\documentclass{aa}  

\usepackage{graphicx}
\usepackage{txfonts}
\usepackage[colorlinks=true,allcolors=blue]{hyperref}
\usepackage{xspace}
\usepackage{threeparttable}
\usepackage{capt-of}

\newcommand{\oiii}{[O\,\textsc{iii}]\xspace}
\newcommand{\oii}{[O\,\textsc{ii}]\xspace}
\newcommand{\nii}{[N\,\textsc{ii}]\xspace}
\newcommand{\sii}{[S\,\textsc{ii}]\xspace}

\newcommand{\ha}{H$\alpha$\xspace}
\newcommand{\hb}{H$\beta$\xspace}
\newcommand{\hg}{H$\gamma$\xspace}

\newcommand{\neiii}{[Ne\,\textsc{iii}]\xspace}
\newcommand{\cii}{[C\,\textsc{ii}]\xspace}
\newcommand{\lya}{Ly$\alpha$\xspace}

\newcommand{\rii}{\oii$\lambda\lambda$3727,30/\hb}
\newcommand{\riii}{\oiii$\lambda$5008/\hb}

\newcommand{\roiioiii}{(\oii$\lambda\lambda$3727,30 + \oiii$\lambda\lambda$4960,5008)/\hb}
\newcommand{\oiiioii}{\oiii$\lambda$5008/\oii$\lambda\lambda$3727,30\xspace}
\newcommand{\neiiioii}{\neiii$\lambda$3870/\oii$\lambda\lambda$3727,30\xspace}
\newcommand{\oiiisii}{\oiii$\lambda$5008/\hb\ / \sii$\lambda\lambda$6718,33/\ha}
\newcommand{\rsoiii}{\oiii$\lambda$5008/\hb\ + \sii$\lambda\lambda$6718,33/\ha}

\newcommand{\yd}[1]{A2744-YD#1\xspace}
\newcommand{\bdf}{BDF-3299\xspace}
\newcommand{\cosmos}{COSMOS24108\xspace}

\begin{document}

   \title{Gas-phase metallicity gradients in galaxies at $z \sim 6-8$}


   \author{G. Venturi
          \inst{\ref{inst:SNS}}\fnmsep\inst{\ref{inst:OAA}}\fnmsep\thanks{\href{mailto:giacomo.venturi1@sns.it}{giacomo.venturi1@sns.it}.}
          \and
          S. Carniani
          \inst{\ref{inst:SNS}}\fnmsep\inst{\ref{inst:OAA}}
          \and
          E. Parlanti
          \inst{\ref{inst:SNS}}
          \and
          M. Kohandel
          \inst{\ref{inst:SNS}}
          \and
          M. Curti
          \inst{\ref{inst:ESO}}
          \and
          A. Pallottini
          \inst{\ref{inst:SNS}}
          \and
          L. Vallini
          \inst{\ref{inst:OAS}}
          \and
          S. Arribas
          \inst{\ref{inst:CAB}}
          \and
          A. J. Bunker
          \inst{\ref{inst:oxford}}
          \and
          A. J. Cameron
          \inst{\ref{inst:oxford}}
          \and
          M. Castellano
          \inst{\ref{inst:OAR}}
          \and
          A. Ferrara
          \inst{\ref{inst:SNS}}
          \and
          A. Fontana
          \inst{\ref{inst:OAR}}
          \and
          S. Gallerani
          \inst{\ref{inst:SNS}}
          \and
          V. Gelli
          \inst{\ref{inst:DAWN}}\fnmsep\inst{\ref{inst:NBI}}
          \and
          R. Maiolino
          \inst{\ref{inst:KICC}}\fnmsep\inst{\ref{inst:UCam}}\fnmsep\inst{\ref{inst:UCL}}
          \and
          E. Ntormousi
          \inst{\ref{inst:SNS}}
          \and
          C. Pacifici
          \inst{\ref{inst:STScI}}
          \and
          L. Pentericci
          \inst{\ref{inst:OAR}}
          \and
          S. Salvadori
          \inst{\ref{inst:unifi}}\fnmsep\inst{\ref{inst:OAA}}
          \and
          E. Vanzella
          \inst{\ref{inst:OAS}}
          }

   \institute{Scuola Normale Superiore, Piazza dei Cavalieri 7, I-56126 Pisa, Italy \label{inst:SNS}
            \and
            INAF - Osservatorio Astrofisico di Arcetri, Largo E. Fermi 5, I-50125 Firenze, Italy \label{inst:OAA}
            \and
            European Southern Observatory, Karl-Schwarzschild-Strasse 2, 85748, Garching, Germany \label{inst:ESO}
            \and
            INAF - Osservatorio di Astrofisica e Scienza dello Spazio, via Gobetti 93/3, I-40129, Bologna, Italy \label{inst:OAS}
            \and
            Centro de Astrobiología (CAB), CSIC-INTA, Cra. de Ajalvir Km. 4, 28850, Torrejón de Ardoz, Madrid, Spain \label{inst:CAB}
            \and
            Department of Physics, University of Oxford, Denys Wilkinson Building, Keble Road, Oxford, OX1 3RH, UK \label{inst:oxford}
            \and
            Cosmic Dawn Center (DAWN), Jagtvej 128, 2200, Copenhagen N, Denmark \label{inst:DAWN}
            \and
            Niels Bohr Institute, University of Copenhagen, Jagtvej 128, 2200, Copenhagen N, Denmark \label{inst:NBI}
            \and
            Kavli Institute for Cosmology, University of Cambridge, Madingley Road, Cambridge, CB3 0HA, UK \label{inst:KICC}
            \and
            Cavendish Laboratory - Astrophysics Group, University of Cambridge, 19 JJ Thomson Avenue, Cambridge, CB3 0HE, UK \label{inst:UCam}
            \and
            Department of Physics and Astronomy, University College London, Gower Street, London WC1E 6BT, UK \label{inst:UCL}
            \and
            INAF - Osservatorio Astronomico di Roma, via di Frascati 33, I-00078 Monte Porzio Catone, Italy \label{inst:OAR}
            \and
            Space Telescope Science Institute, 3700 San Martin Drive, Baltimore, MD 21218, USA \label{inst:STScI}
            \and
            Dipartimento di Fisica e Astronomia, Università degli Studi di Firenze, Largo E. Fermi 1, 50125, Firenze, Italy \label{inst:unifi}
             }

   \date{Received 5 March 2024; accepted 12 August 2024}

 
  \abstract
  {The study of gas-phase metallicity and its spatial distribution at high redshift is crucial to understand the processes that shaped the growth and evolution of galaxies in the early Universe.
  Here we study the spatially resolved metallicity in three systems at $z$ $\sim$ 6--8, namely \yd4, \bdf, and \cosmos, with JWST NIRSpec IFU low-resolution ($R$ $\sim$ 100) spectroscopic observations.
  These are among the highest-$z$ sources in which metallicity gradients have been probed so far.
  Each of these systems hosts several spatial components in the process of merging within a few kiloparsecs, identified from the rest-frame UV and optical stellar continuum and ionised gas emission line maps. 
  The sources have heterogeneous properties, with stellar masses log($M_*/M_\odot$) $\sim$ 7.6--9.3, star formation rates (SFRs) $\sim$ 1--15 $M_\odot$~yr$^{-1}$, and gas-phase metallicities 12+log(O/H) $\sim$ 7.7--8.3, which exhibit a large scatter within each system.
  Their properties are generally consistent with those of the highest-redshift samples to date ($z$ $\sim$ 3--10), though the sources in \yd4 and \cosmos are at the high end of the mass-metallicity relation (MZR) defined by the $z$ $\sim$ 3--10 sources. Moreover, the targets in this work follow the predicted slope of the MZR at $z$ $\sim$ 6--8 from most cosmological simulations.
  The gas-phase metallicity gradients are consistent with being flat in the main sources of each system. Flat metallicity gradients are thought to arise from gas mixing processes on galaxy scales, such as mergers or galactic outflows and supernova winds driven by intense stellar feedback, which wash out any gradient formed in the galaxy.  The existence of flat gradients at $z$ $\sim$ 6--8 sets also important constraints on future cosmological simulations and chemical evolution models, whose predictions on the cosmic evolution of metallicity gradients often differ significantly, especially at high redshift, but are mostly limited to $z$ $\lesssim$ 3 so far.
  }


   \maketitle
%

\section{Introduction}\label{sec:intro}

\begin{figure}
    \centering
    \includegraphics[width=\columnwidth,trim={18cm 0 20cm 0},clip]{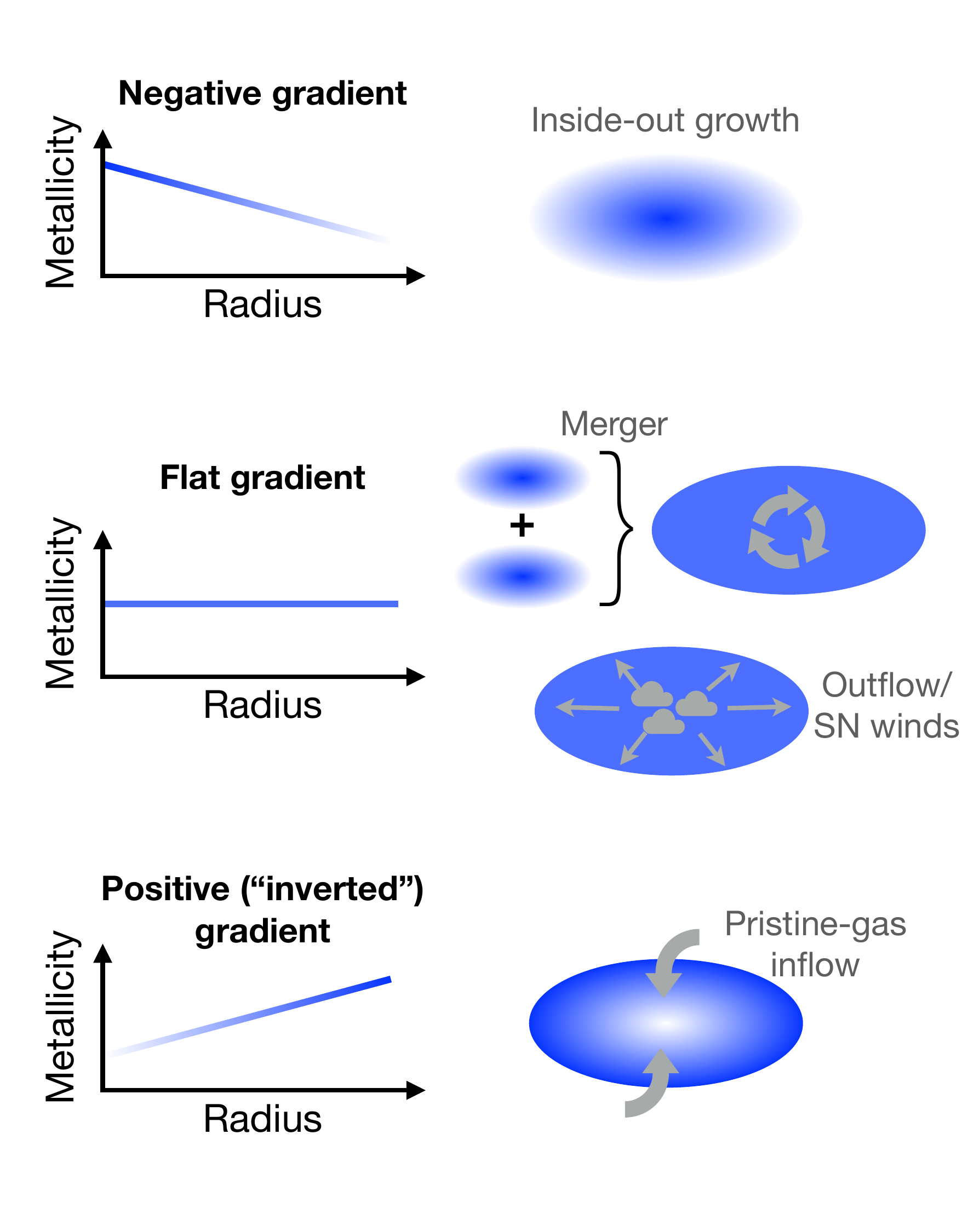}
    \caption{
    Simple sketch summarising the three main classes of radial gas-phase metallicity gradients and possible physical origin.
    Negative (decreasing) gradients can be interpreted as the result of the inside-out growth of a galaxy, in which star formation and the following chemical enrichment start earlier in the inner regions.
    Flat gradients can result from radial mixing processes, such as galaxy mergers as well as SN winds redistributing the metals in the interstellar medium (ISM) and galactic outflows of metal-enriched gas evacuated (or re-accreted) in the outer regions.
    Positive (increasing) inverted gradients may arise from accretion of pristine gas to the central regions of the galaxy; this can also lead to flattened gradients.}
    \label{fig:cartoon}
\end{figure}

Gas-phase metallicity, its scaling relations with other galactic properties, and its spatial distribution are fundamental tools to study and understand the evolution of galaxies \citep[e.g.][]{MaiolinoMannucci2019}.
The interplay between gas accretion, star formation, outflows, and mergers regulates the growth of galaxies and the buildup of their metal content, leading to the correlation between gas-phase metallicity and stellar mass in galaxies known as mass-metallicity relation (MZR; e.g. \citealt{Dave2011, Pallottini2014, SomervilleDave2015}).
The spatial distribution of metals in galaxies, usually described through radial metallicity gradients, bears the imprint of these underlying processes.
Therefore, spatially resolved studies of metallicity in the early Universe, where the formation and primordial growth of galaxies were taking place, are key to better constrain the main mechanisms that have contributed to shape them.

Negative (radially decreasing) gradients are usually interpreted as resulting from an inside-out galaxy formation scenario, in which stars start forming earlier in the inner parts of galaxies and thus have more time to chemically enrich the inner regions than the outer ones \citep[e.g.][]{Samland1997, Prantzos2000, Dave2011, Pilkington2012, Gibson2013, Hemler2021, Tissera2022}.
Flattened gradients could arise from radial mixing of gas and redistribution of metals on galaxy scales, induced by supernova (SN) winds within the galaxy or galactic outflows of metal-enriched material expelled from the galaxy and re-accreted in the outer regions (so-called galactic fountains), driven by intense stellar feedback \citep[e.g.][]{Gibson2013,Ma2017}, as well as by galaxy merging and interactions \citep[e.g.][]{Rupke2010a,Rupke2010b,Rich2012,Torres-Flores2014}.
%
Finally, positive (radially increasing; also called inverted) gradients may be produced by accretion of external pristine gas towards the central regions of the galaxy \citep[e.g.][]{Ceverino2016}; this could also contribute to flatten the metallicity gradient, if not invert it. Particularly strong metal-loaded galactic outflows could also contribute to produce a positive gradient, by moving the metal-rich gas from the central starburst regions to the outskirts \citep[e.g.][]{Tissera2022}.
A schematic cartoon that summarises the above framework is shown in Fig. \ref{fig:cartoon}.

In the local Universe, most spiral galaxies show negative metallicity gradients, with more metal-enriched gas in the inner than in the outer regions of the galaxy (e.g. \citealp{Magrini2009, Magrini2017, Stanghellini2010, Luck2011} for the  Milky Way; \citealp{Zaritsky1994, Magrini2010, Kewley2010, Bresolin2011, Sanchez2014, Berg2015, Ho2015, Belfiore2017} for other local galaxies).
By using metallicity diagnostics that trace past metal enrichment, primarily planetary nebulae (PNe; up to 5--10 Gyr ago, i.e. $z$ $\lesssim$ 2) some studies have suggested that the Milky Way and nearby galaxies had flatter gradients in the past 
which grew progressively steeper with cosmic time \citep{Stanghellini2014, Stanghellini2018, Magrini2016}.
A dependence of gradient slope with mass has been observed in the local Universe, with more massive galaxies exhibiting progressively steeper gradients, while low-mass galaxies having almost flat gradients \citep{Belfiore2017}.
This relation between mass and metallicity gradient may be the result of an evolutionary sequence in mass, with more massive, more evolved galaxies having steeper gradients and low-mass, less evolved ones, analogues of high-$z$ galaxies, having flatter gradients \citep[e.g.][]{MaiolinoMannucci2019}.

Cosmological simulations of galaxy evolution make very different predictions on metallicity gradients at high redshift (up to $z$ $\sim$ 3), which can strongly vary depending on prescriptions on star formation and stellar feedback models and on their relative contribution to gas mixing as compared to mergers \citep[e.g.][]{Gibson2013, Ma2017, Hemler2021, Tissera2022}.
For example, according to the FIRE simulations \citep{Hopkins2014}, mergers and rapid variations of metallicity gradients in high-$z$ galaxies, induced by starburst episodes which drive strong outflows and disrupt the gas disc, would flatten any negative gradients previously developed, especially in low-mass galaxies where feedback mechanisms are more efficient \citep[][]{Ma2017}.
Instead, the TNG50 simulations \citep{Pillepich2019, Nelson2019} which feature less bursty feedback, predict negative gradients at all redshifts, steeper (more negative) with increasing $z$ \citep[][]{Hemler2021}.
The metallicity gradients resulting from the MUGS \citep[`normal' feedback;][]{Stinson2010} and MaGICC \citep[`enhanced' feedback;][]{Brook2012} simulation suites are compared in \cite{Gibson2013}. 
The enhanced feedback (including pre-SN early stellar feedback from massive stars) re-distributes energy and re-cycled ISM material over large scales, through re-accretion in the outer parts of the galaxy of gas expelled via outflows. In this case, relatively flat and temporally invariant abundance gradients are predicted, in contrast to the steeper negative gradients at increasing $z$ emerging from the normal feedback scenario.
Finally, the EAGLE simulations \citep{Schaye2015} find that the median gradient is zero (flat) at all redshifts, but with the scatter around the median increasing with $z$ due to individual galaxies transitioning between steep (either negative or positive) and flatter gradients \citep[][]{Tissera2022}. The authors report that this behaviour results from the higher frequency of major mergers at higher $z$ leading to episodes of enhanced accretion of low-metallicity gas which trigger intense star formation and ejection of metal-enriched gas.

Non-cosmological chemical evolution models have also been employed to study the evolution of metallicity gradients with cosmic time, mostly for the Milky Way. 
\cite{Mott2013}, assuming an inside-out formation of the disc, a constant star formation efficiency (SFE) along the disc, and the presence of radial flows with varying speed, predict an inversion of the gradients (from negative to positive) at $z$ $\gtrsim$ 1, due to a strong infall of primordial gas in the innermost disc regions at early times; only a variable SFE does not lead to an inversion of the gradients at high $z$ in their models.
\cite{Kubryk2015} study the role of the radial motions of gas and stars on the evolution of abundance profiles in the Milky Way disc, finding steep abundance profiles at high $z$ which flatten with time, as a result of the inside-out formation of the disc.
\cite{Molla2019} investigate the role of the growth of the stellar disc, the effect of infall rate and star formation prescriptions, as well as the pre-enrichment of the infall gas, and find a smooth evolution of the gradients with a slight flattening from $z$ = 4 to 1. 
Finally, the first principles-based model of \cite{Sharda2021} finds that the gradient in Milky Way-like galaxies has steepened over time and also predicts the evolution of metallicity gradients with redshift in galaxy samples matched in both stellar masses and abundances, finding that disc galaxies transition from the advection- to the accretion-dominated regime from high to low $z$; in general little evolution of the gradients is predicted for $z$ $\gtrsim$ 1.

Observations at the cosmic noon epoch (i.e., $z$ $\sim$ 1--4) have found heterogeneous results, with both negative, positive, and flat metallicity gradients \citep[e.g.][]{Cresci2010, Yuan2011, Queyrel2012, Swinbank2012, Stott2014, Troncoso2014, Jones2013, Jones2015, Leethochawalit2016, Wuyts2016, Wang2017, Wang2022, Carton2018, Forster-Schreiber2018, Curti2020klever}.
Nevertheless, the majority of high-$z$ measurements is consistent with little or no cosmic evolution of metallicity gradients, which are found to be flat or only moderately negative or positive ($\lesssim$ $|\pm0.1|$ dex~kpc$^{-1}$; see e.g. the compilation in \citealt{Curti2020klever}).

At higher redshift ($z$ $\sim$ 7), \cite{Vallini2024a} also found flat gradients by making use of a physically motivated Bayesian model to derive metallicities from rest-frame far-infrared lines (\oiii 88$\mu$m and \cii 158$\mu$m) from Atacama Large Millimeter/submillimeter Array (ALMA) observations. However, as the authors point out as a caveat, 
there may be systematics in comparing their results with gradients of O/H abundance obtained with the more standard methods based on optical lines, since the adopted model does not take into account the likely enhancement of O/C at low metallicities \citep[see e.g.][]{MaiolinoMannucci2019}, which would result in the model returning higher values of metallicity. Therefore, while their results consistent with flat gradients at $z$ $\sim$ 7 are very significant, they require independent confirmation.

The \textit{James Webb} Space Telescope (JWST) has recently opened up the possibility of measuring metallicity gradients in the first 1--2 Gyr of the Universe by using rest-frame optical emission lines \citep{Wang2022, Rodriguez-del-Pino2023, Arribas2023} which are extensively used and calibrated at lower redshifts \citep[e.g.][]{Curti2017,Curti2020calib}. 
In this work, we study the spatially resolved gas-phase metallicity and investigate the shape of metallicity gradients at high redshift ($z$ $\gtrsim$ 6), by tracing the warm ($T$~$\sim$~$10^4$~K) ionised gas emission.
Specifically, we observed three high-$z$ systems with the Near-InfraRed Spectrograph (NIRSpec) on board  JWST in its Integral Field Unit (IFU) mode with the low-spectral resolution PRISM/CLEAR ($R$\,$\sim$\,100) disperser-filter combination. These are \yd4, part of the proto-cluster A2744-z7p9OD ($z$~$\sim$~7.88), \bdf ($z$~$\sim$~7.11), and \cosmos ($z$~$\sim$~6.36).  
Basic information on the targets is given in Table \ref{tab:table1}.
We trace gas-phase metallicity by making use of strong-line calibrators relying on rest-frame optical and near-UV emission lines.
We also present the basic properties of each source, specifically stellar mass, obtained from spectral energy distribution (SED) fitting, and star formation rate (SFR), from \hb or \ha.
The star formation history of these targets is studied in detail in a separate paper (Kohandel et al., in prep.).

Throughout this work, the reported wavelengths are in vacuum and we adopt a flat $\Lambda$CDM cosmology with $H_0$ $\simeq$ 67.7~km~s$^{-1}$~Mpc$^{-1}$, $\Omega_\mathrm{M}$ $\simeq$ 0.31, and $\Omega_\mathrm{\Lambda}$ $\simeq$ 0.69 \citep{Planck2020}.

\section{Description of observations, data reduction, and analysis}

The observations employed in this work were carried out between October and December 2022 as part of the JWST GO program ID 1893 (PI: S. Carniani; \citealt{Carniani2021jwstprop}). The three sources were observed for 15.102 ks ($\sim$4 h) on-source each (22.059 ks with overheads; $\sim$6 h) with 1.678 ks of background exposure time each (2.830 ks with overheads).
The observations were performed with the PRISM/CLEAR disperser-filter combination and an eight-point dither `medium' cycling pattern (step $\sim$0.5$''$).
The JWST NIRSpec $\sim$3$''\times3''$ IFU PRISM observations simultaneously span the spectral range 0.6--5.2~$\mu$m with a spectral resolution ranging between $R$ $\sim$ 30--330.

\begin{table*}
\centering
\caption{Basic information on the observed targets.}
\begin{tabular}{c c c c c}
\hline\hline
Target name & RA [hh:mm:ss.ss] & Dec [dd:mm:ss.ss]  &$z$\tablefootmark{a} &Scale [kpc arcsec$^{-1}$]\\
\hline
\yd4 & +00:14:24.90 & --30:22:56.10  & 7.879&3.512\tablefootmark{b}\\
\bdf & +22:28:12.31 & --35:10:00.39  & 7.114&5.289\\
\cosmos & +10:00:47.33 & +02:28:43.14  & 6.361&5.647\\
\hline
\end{tabular}
\label{tab:table1}
\tablefoot{
\tablefoottext{a}{From this work.}
\tablefoottext{b}{The reported scale is corrected for the lensing magnification factor of $\sim$2 \citep{Morishita2023, Bergamini2023}.}
}
\end{table*}

\subsection{Data reduction}

We retrieved the raw data from the MAST archive and we ran the three stages of the pipeline using version 1.11 with CRDS (calibration reference data system) context `jwst\_1094.pmap'. First, at stage 1 `calwebb\_detector1', the pipeline applied the detector-level corrections (e.g. check for saturation, dark exposure subtraction, flagging of bad pixels and cosmic-ray persistences) and performed ramp fitting for individual exposures. We then calibrated the count rate images by executing stage 2 `calwebb\_spec2' of the pipeline, which corrects for flat field and performs the wavelength calibration. The background was subtracted from each exposure during this stage by using the observations of the dedicated background for each target. We processed the background targets up to stage 2 of the pipeline, then we applied the background step for the science target. In the background step, the pipeline subtracts the background exposure from each target exposure in the detector space.
Finally, each calibrated exposure was combined in stage 3 `calwebb\_spec3' by using `drizzle' weighting and a spaxel size of 0.05\arcsec\ to obtain the final cubes. During stage 3, we applied the outlier rejection step built in the pipeline and then a sigma clipping to remove any residual outliers in the final cube.

\subsection{Data analysis}\label{ssec:data_anal}

\begin{table}[]
    \centering
    \caption{Emission line diagnostic ratios used in this work together with their compact notation.}
    \begin{tabular}{cc} 
        \hline\hline
         Name&  Emission line ratio\\ 
        \hline
         R$_2$& 
         \rii\\ 
         R$_3$&\riii\\
         O$_3$O$_2$&\oiiioii\\
         $\hat{\mathrm{R}}$&0.47 R$_2$ + 0.88 R$_3$\\
         Ne$_3$O$_2$&\neiiioii\\
         RS$_{32}$&\rsoiii\\
         O$_3$S$_2$&\oiiisii\\
        \hline
 \end{tabular}
    \label{tab:ratio_notation}
\end{table}

In this section, we describe the analysis of the NIRSpec R100 IFU data. 
In brief, we obtained the emission line fluxes from emission-line modelling of the spectra, which were used to infer the gas-phase metallicity. This was done on both an integrated basis (to get the integrated metallicity of each target) and a spatially resolved one, in this case both spaxel-by-spaxel (to produce maps) and in concentric radial annuli (to obtain radial gradients). 
From the integrated spectra, we also obtained the stellar mass ($M_*$) of each source from SED fitting and their SFR from the emission-line modelling.
We provide more details in the following.

\subsubsection{Emission line fitting}
The main goal of this work is to measure radial gas-phase metallicity gradients. To do so, we infer the oxygen abundance relative to hydrogen (12+log(O/H)), a proxy of gas-phase metallicity, by making use of the optical and near-UV strong-line diagnostic ratios reported in Table \ref{tab:ratio_notation}. 
We adopt the new diagnostic ratio $\hat{\mathrm{R}}$ = 0.47 R$_2$ + 0.88 R$_3$, first introduced in \cite{Laseter2024}, in place of the more traditional R$_{23}$ = \roiioiii, since the former is more suited for high-$z$ galaxies, while the latter is a projection mostly driven by low-metallicity local analogues.
%
We adopt the best-fit polynomial calibrations from \cite{Curti2017,Curti2020calib}, slightly revisited in \cite{Curti2023_ero, Curti2023_jades} to better probe the low-O/H regime, to infer the gas metallicity from the combination of the above ratios. 
We tested other gas-phase metallicity calibrations \citep[e.g.][]{Nakajima2022, Sanders2024}. These gave 12+log(O/H) values similar to the Curti et al. ones within 0.1 dex, consistent with the uncertainties, and virtually no difference in the metallicity radial gradients.

The full width at half maximum (FWHM) of the point spread function (PSF) of JWST depends on wavelength. Specifically, the PSF FWHM of NIRSpec IFU ranges from around 0.09$''$ at 1 $\mu$m up to around 0.16$''$ at 4.5 $\mu$m, and the variation with wavelength is stronger in the direction perpendicular to the IFU slices than along them \citep{Deugenio2023a}.
Since the aim of this work is to obtain spatially resolved metallicities from emission line ratios sometimes far in wavelength (see Table \ref{tab:ratio_notation}), we applied a wavelength-dependent spatial smoothing to the data cube prior to the line fitting with the aim of achieving the same spatial resolution at all wavelengths. We adopted the PSF FWHM curves from \cite{Deugenio2023a}, empirically calibrated by matching NIRSpec IFU observations with NIRCam ones of the same target reported in their Eqs. 3 and 4, for the along- and across-slice cases, respectively.
The smoothing was done by convolving the cube with a wavelength-dependent 2D Gaussian kernel in order to obtain a smoothed cube having the same spatial resolution at each wavelength, specifically the PSF FWHM reported in \cite{Deugenio2023a} at the wavelength of the highest-wavelength line of interest for the along-slice case (which has the largest FWHM among the two cases). This corresponds to FWHM$_\mathrm{goal}$ $\sim$ 0.16$''$ (at \oiii) for \yd4, $\sim$ 0.14$''$ (at \oiii) for \bdf, and $\sim$ 0.18$''$ (at \ha-\sii) for \cosmos. Specifically, the 2D Gaussian kernel to be used for the convolution, at a given wavelength $\lambda$, was defined as follows: FWHM$^2_\mathrm{kern;dir}$($\lambda$) = FWHM$^2_\mathrm{goal}$($\lambda$) -- FWHM$^2_\mathrm{mod;dir}$($\lambda$), where FWHM$_\mathrm{mod;dir}$ is the model FWHM from \cite{Deugenio2023a} and `dir' indicates the direction, either along or across the slices.

In general, we employed the original unsmoothed data cube for the flux maps of continuum and emission lines and for integrated measurements from circular apertures, while we adopted the smoothed data cube for the maps and radial profiles involving line ratios and metallicity, which would have otherwise been affected by the wavelength-dependence of the PSF.
In case of \yd4, the highest-$z$ target of our sample, we adopted the smoothed data cube for every map, to obtain visually clearer maps as compared to the more noisy ones from the unsmoothed cube. 

The data analysis consisted in modelling the rest-frame optical and near-UV emission lines available in the observed spectral range. 
The emission lines are spectrally unresolved in the PRISM spectra (the spectral resolution is $\sigma_\mathrm{res}$ $\sim$ 1250 km~s$^{-1}$ at $\sim$3 $\mu$m and $\sigma_\mathrm{res}$ $\sim$ 450 km~s$^{-1}$ at $\sim$5 $\mu$m), therefore a single Gaussian function per line was used.
We fitted the \oiii-\hb, the \oii-\neiii-\hg, and the \ha-\nii-\sii line complexes separately. The \nii6550,85 was included to allow for a better modelling of the \ha line profile, which was otherwise showing a marked residual in its redward wing. This was done only when fitting integrated spectra from circular apertures or concentric annuli for radial profiles, not in the lower-S/N case of spaxel-by-spaxel fitting, for which no residual wing indicative of \nii was detected above the noise.
For each line complex, the velocity was tied to be the same for all the lines, while we allowed the line width to vary to match the PRISM spectral resolution, given that all lines are spectrally unresolved as mentioned. 
We fixed the flux ratios \oiii$\lambda$5008/$\lambda$4960 and \nii $\lambda$6585/$\lambda$6550 to their theoretical value of 3 \citep{StoreyZeippen2000}.
We included an underlying first-order polynomial to model the continuum. We accounted for the Balmer break of the continuum at around 3645 \AA\ rest-frame, when needed, by employing three (first-order) polynomials, one modelling the jump and the other two on each side of it.

The line fluxes used to infer the metallicity or the SFR were first corrected for dust extinction, when possible (see below), using the \hg/\hb ratio (for \yd4 and \bdf) or \ha/\hb (for \cosmos), only when both lines had S/N$>$3.
We adopted a \cite{Calzetti2000} reddening curve, with an $R_V$ = 4.05, suitable for high-$z$ low-metallicity (12+log(O/H) $<$ 8.5) star-forming galaxies at the wavelengths of interest here of $\gtrsim$3000~\AA\ (see e.g. \citealt{Shivaei2020}). We assumed the theoretical extinction-free \hg/\hb and \ha/\hb ratios of 0.466 and 2.87, respectively, valid for case-B recombination and an electron temperature of $T_\mathrm{e}$ $\sim$ 10$^4$~K \citep{Osterbrock:2006a}, typical of the warm ionised gas emitting rest-frame optical lines.

We selected the line ratio diagnostics for the metallicity estimation in an adaptive way, based on the line S/N. Specifically, a certain ratio (e.g. \neiii/\oii) was employed only when all the lines involved in it had S/N$>$3. In this way, for each fitted spectrum we only selected the sub-sample of line ratio diagnostics whose line fluxes exceeded the S/N threshold, instead of using all the ratios (even those involving S/N$<$3 lines) or none of them (when just one or a few of the lines were below the S/N threshold).
Finally, given a set of diagnostic ratios, the best-fit metallicity was obtained by minimising the chi-squared defined simultaneously by the different observed ratios and their relative calibration curves, weighted by the observed uncertainties and the intrinsic dispersion of the calibration (added in quadrature; see \citealt{Curti2020calib}).

We first modelled the emission lines in integrated spectra extracted from circular apertures centred on each sub-source present in the three studied systems. From these, we obtained the integrated gas-phase metallicity (as described above) and SFR of each sub-source. The SFR was obtained from the extinction-corrected flux of \ha, using the relation from \cite{Kennicutt2012}. In \yd4 and \bdf, for which \ha was not available, the SFR was obtained from the extinction-corrected flux of \hb, converted to \ha by adopting the theoretical \ha/\hb ratio of 2.87.
We employed circular apertures of 0.15\arcsec\ radius for \yd4 and \cosmos, and of 0.1\arcsec\ for \bdf, due to the spatial vicinity of the sources in this latter system.

The emission line modelling was then performed both on a spaxel-by-spaxel basis (for both the unsmoothed and the smoothed data cubes) and in concentric radial annuli centred on each source (for the smoothed data cube). The latter was done to increase the S/N on the emission lines and get more robust estimates of metallicity with the goal of obtaining metallicity gradients.
For this annular line modelling, we extracted integrated spectra at each radius by collapsing the spaxels within concentric circular annuli. The annuli, having radial width of 1 spaxel (0.05$''$) each, were centred on each of the main spatial components detected in each targeted system, with a variable maximum aperture radius depending on the component extension (as described in the next sections separately for each target).

The extinction-correction of emission line fluxes used to infer metallicity was done only for the case of integrated spectra from circular apertures and from concentric radial annuli. This was not possible for the spaxel-by-spaxel case because, among the useful Balmer lines tracing extinction, \hg is detected in almost no spaxels in \yd4 and \bdf. Therefore, the maps of metallicity were obtained without accounting for possible extinction. For \cosmos, also including \ha in the spectral range, \hb is detected in a large enough number of spaxels to attempt for an extinction correction of the line fluxes for the metallicity. However, the resulting map is very similar to the non-extinction-corrected one, only more noisy, therefore we report the metallicity map obtained without the extinction correction. 

\begin{table*}
\caption{Properties of the individual sources.}
    \centering
    \def\arraystretch{1.2}
    \begin{tabular}{ccccccccccc}
    \hline\hline
        Source& $z$ & $\log (M_*/M_\odot)$ & $F_{\rm \oii}$ & $F_{\rm \neiii}$ & $F_{\rm H\beta}$ & $F_{\rm \oiii}$ & $F_{\rm H\alpha}$ & $F_{\rm \sii}$ & SFR & 12+log(O/H) \\
        {\small (1)}  & {\small (2)}     & {\small (3)}  & {\small (4)} & {\small (5)} & {\small (6)} & {\small (7)} & {\small (8)} & {\small (9)} & {\small (10)} & {\small (11)}\\
        
    \hline
    \hline
        \\[-2ex]
        \multicolumn{10}{c}{{\large \yd4 }}\\
        
           YD4 & 7.879 &    8.69$^{+0.18}_{-0.17}$ & 72$\pm$3 & $<$21 & 36$\pm$3 & 180$\pm$4 & -- & -- & 2.2$^{+0.2}_{-0.2}$ &  8.19$_{-0.13}^{+0.09}$\\
           YD6 & 7.880 &    8.40$^{+0.27}_{-0.13}$& 26$\pm$3 & $<$20 & $<$17 & 44$\pm$3 & -- & -- & $<$1.0 & 8.27$_{-0.20}^{+0.16}$ \\
           YD6-\oiii-E & 7.881 &    7.83$^{+0.46}_{-0.22}$& $<$24 & $<$24 & $<$14 & 57$\pm$3 & -- & -- & $<$0.87 & \\
           YD4-\oiii-W & 7.877 &    7.82$^{+0.41}_{-0.48}$& $<$14 & $<$15 & $<$19 & 46$\pm3$ & -- & -- & $<$1.1 & \\
           ZD1 & 7.872 &    7.58$^{+0.18}_{-0.37}$& $<$13 & $<$13 & $<$18 & 27$\pm$3 & -- & -- & $<$1.1 & \\
           YD1 & 7.883 &    8.58$^{+0.28}_{-0.19}$& 107$\pm$4 & $<$24 & 47$\pm$3 & 346$\pm$3 & -- & -- & 2.85$^{+0.18}_{-0.18}$& 8.11$_{-0.07}^{+0.05}$  \\
           YD1-E & 7.882 &    8.44$^{+0.43}_{-0.23}$& 54$\pm$3 & $<$20 & 34$\pm$2 & 270$\pm$2 & -- & -- & 2.02$^{+0.13}_{-0.13}$& 7.98$_{-0.09}^{+0.07}$  \\
           s1 & 7.878 &    8.01$^{+0.28}_{-0.47}$& $<$24 & $<$25 & 18$\pm$3 & 151$\pm$3 & -- & -- & 1.09$^{+0.16}_{-0.16}$& 7.72$_{-0.20}^{+0.20}$  \\
           \hline
            \\[-2ex]
        \multicolumn{10}{c}{{\large \bdf }}\\

          a & 7.114 &   7.90$^{+0.25}_{-0.13}$ & 24$\pm$2 & $<$13 & 36$\pm$1 & 195$\pm$3 & -- & -- & 3.47$^{+0.13}_{-0.13}$ & 7.68$_{-0.09}^{+0.13}$ \\
          b & 7.114 &   8.21$^{+0.30}_{-0.11}$ & 28$\pm$2 & $<$11 & 44$\pm$2 & 285$\pm$4 & -- & -- & 4.2$^{+0.2}_{-0.2}$ & 7.69$_{-0.06}^{+0.10}$ \\
          c & 7.112 &   7.62$^{+0.44}_{-0.14}$ & $<$13 & $<$12 & 15$\pm$2 & 93$\pm$4 & -- & -- & 1.5$^{+0.2}_{-0.2}$ & 7.80$_{-0.38}^{+0.15}$ \\
          \hline 
          \\[-2ex]
        \multicolumn{10}{c}{{\large \cosmos}}\\

         a & 	6.361 &   9.29$^{+0.09}_{-0.08}$ & 179$\pm$5 & 48$\pm$6 & 53$\pm$5 & 358$\pm$5 & 219$\pm$4 & 34$\pm$4 & 14.4$^{+1.3}_{-1.3}$ & 8.20$_{-0.06}^{+0.05}$ \\
        b & 6.358 &   8.81$^{+0.11}_{-0.17}$ & 150$\pm$6 & 47$\pm$6 & 72$\pm$4 & 470$\pm$4 & 245$\pm$3 & $<$18 & 9.8$^{+0.3}_{-0.3}$ & 8.11$_{-0.07}^{+0.04}$ \\
          c & 6.359 &   8.89$^{+0.10}_{-0.06}$ & 67$\pm$5 & $<$36 & 28$\pm$4 & 92$\pm$4 & 68$\pm$4 & $<$22 &  1.74$^{+0.09}_{-0.09}$  & 8.33$_{-0.10}^{+0.08}$ \\
          \oiii-Ea & 6.362 &   8.53$^{+0.30}_{-0.22}$ & 62$\pm$3 & 36$\pm$3 & 41$\pm$3 & 296$\pm$3 & 183$\pm$3 & $<$21 & 14.5$^{+1.3}_{-1.3}$ & 8.01$_{-0.12}^{+0.08}$ \\
          \oiii-Eb & 6.356 &   8.40$^{+0.72}_{-0.32}$ & 44$\pm$4 & $<$24 & 37$\pm$3 & 244$\pm$3 & 113$\pm$3 & $<$22 & 3.43$^{+0.11}_{-0.11}$ & 7.95$_{-0.17}^{+0.14}$ \\
    \hline
    \end{tabular}
    \tablefoot{(1) Source name. (2) Redshift from \oiii$\lambda$5008. (3) Stellar mass from SED fitting with \textsc{bagpipes}. (4-9) \oii$\lambda\lambda$3727,30, \neiii$\lambda$3870, \hb, \oiii$\lambda$5008, \ha, and \sii$\lambda$6718,33 measured fluxes (in units of 10$^{-20}$ erg~s$^{-1}$~cm$^{-2}$), (10) SFR (in $M_\odot$~yr$^{-1}$), and (10) gas-phase metallicity from emission line Gaussian fitting. For the sources in the \yd4 system, the reported $M_*$ and SFR are corrected for the lensing magnification factor of $\sim$2 \citep{Morishita2023, Bergamini2023}; the line fluxes are instead the observed ones. All the properties are obtained from integrated spectra extracted from circular apertures (having radius of 0.15\arcsec\ for \yd4 and \cosmos and of 0.1\arcsec\ for \bdf) centred on each source. For the line fluxes and SFR, we also report the $3 \sigma$ upper limits in case of no detection.}
    \label{tab:integr_meas}
\end{table*}

\subsubsection{SED fitting}
With the goal of obtaining the stellar mass of each target, we fitted the spatially integrated spectra extracted from the data cube by using the SED fitting code \textsc{bagpipes} \citep{Carnall2018}. We adopted the stellar population models by \cite{Bruzual2003} and included the nebular emission with \textsc{cloudy} \citep{Ferland2017} with  the ionisation parameter
($-3.0 < \log U < 0.0$) as a free parameter and adopting a ${\rm [C/O]=[C/O]}_{\odot}$. We assumed a \cite{Kroupa2001} initial mass function truncated at 0.01 and 100 M$_\odot$ and a \cite{Calzetti2000} attenuation curve. Finally, we used a non-parametric star-formation history model with continuity priors (see \citealt{Leja2019}) and with four time bins:  $0<t<10$ Myr,  $10<t<50$ Myr,  $50<t<100$ Myr, and  $100<t<300$ Myr.

\section{\yd4}

\begin{figure*}
    \centering
    \includegraphics[scale=0.27,trim={3cm 0 1.5cm 0},clip]{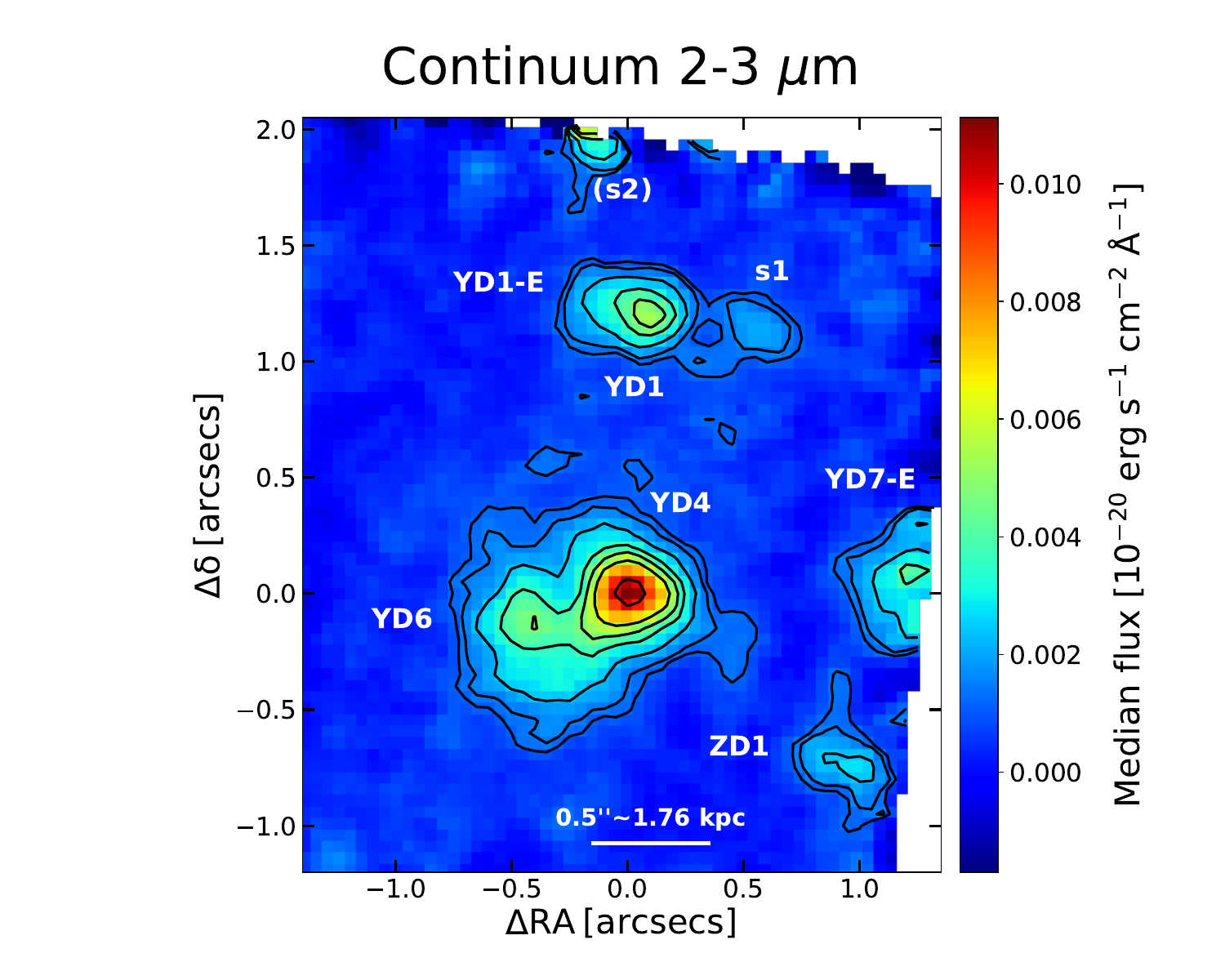}
    \includegraphics[scale=0.27,trim={3cm 0 1.5cm 0},clip]{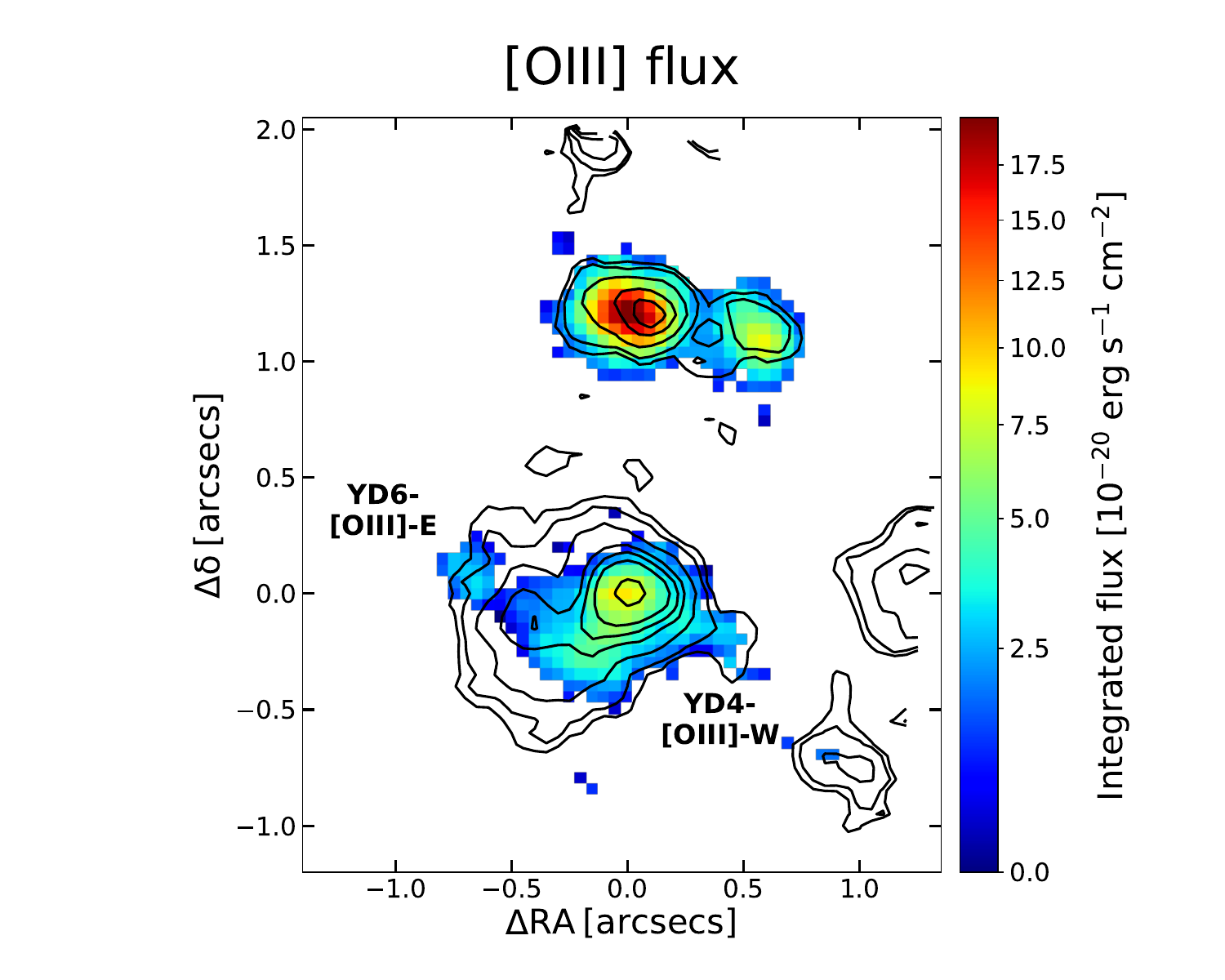}
    \includegraphics[scale=0.27,trim={3cm 0 1.5cm 0},clip]{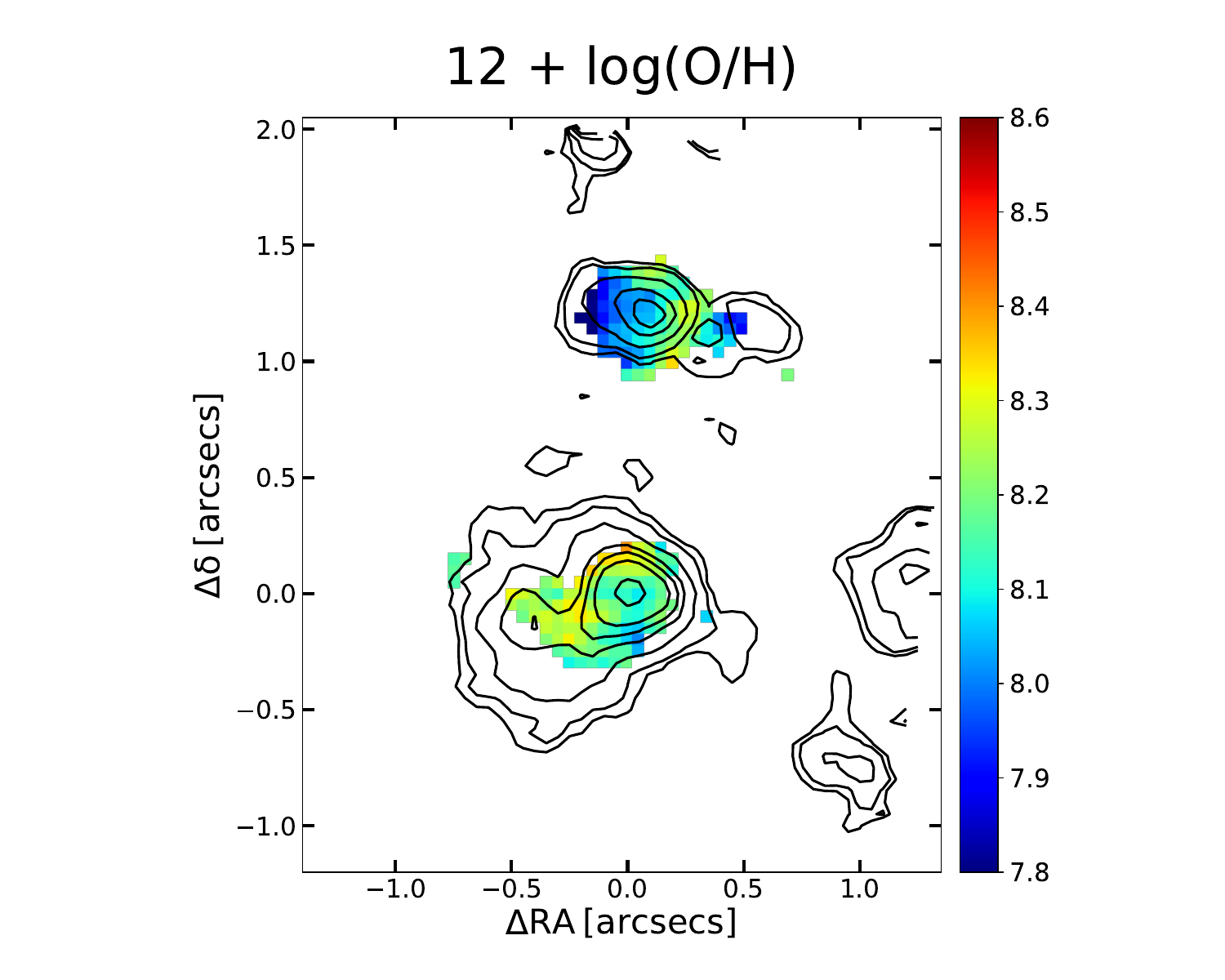}
    \includegraphics[width=0.8\textwidth,trim={0 0.4cm 0 0.4cm},clip]{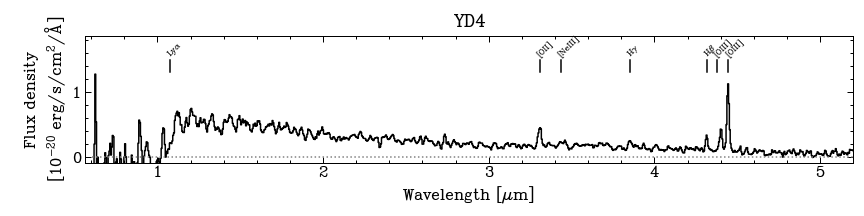}
    \caption{Maps for \yd4 from JWST NIRSpec IFU. All the maps are obtained from the (wavelength-dependent) spatially smoothed data cube (details in Sect. \ref{ssec:data_anal}). No extinction correction is applied. A cut of S/N$>3$ on the peak flux of each line is applied. Median stellar continuum emission in the observed spectral range 2--3 $\mu$m (0.23--0.34 $\mu$m rest-frame; left); \oiii integrated flux (centre); 
    map of oxygen abundance, 12+log(O/H) (right).
    Contours mark the continuum from first panel.
    NIRSpec PRISM/CLEAR spectrum extracted from a circular aperture with radius of 0.15\arcsec\ centred at the location of the target YD4 (bottom).}
    \label{fig:yd4_maps}
\end{figure*}

A2744-YD4 is part of the proto-cluster A2744-z7p9OD at $z$~$\sim$~7.883 located behind the strong lensing cluster Abell 2744. The proto-cluster includes 22 sources at 7 < $z$ < 9 identified through combined deep HST and \textit{Spitzer} IRAC photometry \citep{Laporte2014,Zheng2014,Atek2015,Ishigaki2016} as part of the Hubble Frontier Fields program \citep{Lotz2017}. Seven of them were recently spectroscopically confirmed to be at $z$ = 7.88 through JWST NIRSpec MSA spectroscopy \citep{Morishita2023}. This makes A2744-z7p9OD the most distant proto-cluster known so far.
The system is highly over-dense, with an excess of surface number density from the field average $\delta$ = $(n-\overline{n})/\overline{n}$ $\sim$ 130$^{+66}_{-51}$ \citep{Ishigaki2016}.

The $\sim$3$''\times3''$ IFU FOV of our observations includes five sources of the proto-cluster known as the `quintet', namely YD1, YD4, YD6, YD7, and ZD1, as well as a new source identified by \cite{Hashimoto2023} and named s1 (Fig. \ref{fig:yd4_maps}, top-left). We only partially cover the East component of YD7 (named YD7-E), while its emission toward to West lies outside the FOV.
The lensing magnification factor of the source in the FOV is $\mu$ $\simeq$ 2.0 \citep{Morishita2023, Bergamini2023}. 
YD4, at the centre of our IFU observation, was reported to be at $z$ = 8.38 based on Ly$\alpha$ and ALMA \cii\ 158 $\mu$m and \oiii\ 88 $\mu$m line emission \citep{Laporte2017,Laporte2019,Carniani2020}. Recently, \cite{Morishita2023} measured $z$ = 7.88 for YD4 based on the detection of high-S/N \oiii 4959,5007 $\AA$ and \hb emission lines in high-spectral resolution ($R$2700) NIRSpec MSA observations, thus ruling out the previous redshift measurement. 

The top-left panel of Fig.~\ref{fig:yd4_maps} shows the map of the continuum emission in the range 2--3 $\mu$m observed wavelength ($\sim$0.23--0.34~$\mu$m rest-frame).
In addition to the sources named YD4, YD6, YD1, s1, YD7-E, and ZD1, the continuum map also shows the presence of a source at the northernmost part of the FOV, which is also present in the NIRCam image but is not labelled in \cite{Hashimoto2023}; we name this source s2. However, given that this is at the edge of the FOV, where many artefacts are present in the NIRSpec IFU data, we consider this source as tentative and report its label in parenthesis in the figure.
Moreover, in the map of the continuum obtained from the unsmoothed data cube integrated over the spectral range 1.2--2 $\mu$m (Fig.~\ref{fig:yd4_maps_appdx} in the Appendix\footnote{The Appendix is available at \url{https://doi.org/10.5281/zenodo.13327942}.}, top-left), where the spatial resolution is the highest, the eastern tail of YD1 appears as a separate peaked spatial component. We label this extra source as YD1-E.

Fig. \ref{fig:yd4_maps}, bottom panel,  illustrates the integrated spectrum (aperture radius = 0.15$''$) associated with the main component of the \yd4 system, that is, YD4 itself. The rest-frame optical lines \oii, \hb, and \oiii are detected with high S/N ($\gtrsim$6 on the peak).
The spectroscopic redshift based on Lyman break and optical lines ($z$ $\sim$ 7.88) is consistent with that found by \cite{Morishita2023} and \cite{Hashimoto2023} from the NIRSpec $R$2700 data.
We find a stellar mass of log($M_*/M_\odot$) $\sim$ 8.7 from the SED fitting and a SFR $\sim$ 2 $M_\odot$~yr$^{-1}$ from \hb for YD4 (Table \ref{tab:integr_meas}).
These were corrected for the magnification factor of 2 due to the lensing. The gas-phase metallicity, relying on line ratios, is instead not affected by it.
For the other spatial components in the system, we find log($M_*/M_\odot$) $\sim$ 7.5--8.6 and SFR $\sim$ 1--3 $M_\odot$~yr$^{-1}$.

The \oiii flux map resulting from our spectral emission-line modelling is displayed in Fig. \ref{fig:yd4_maps}, central panel (the maps of \hb and \oii are shown in Fig.~\ref{fig:yd4_maps_appdx}\footnotemark[\value{footnote}]).
From the \oiii map, we identify two new emission regions which are weak or absent in continuum, which we label as YD6-[OIII]-E and YD4-[OIII]-W. 
Overall, the \oiii\ maps obtained from the PRISM data presented here are deeper and reveal fainter and more extended features than those from high-resolution grating data presented in \cite{Hashimoto2023} (who do not report the detection of any other emission line due to the lower S/N of their data).
We can count nine sources in total in the $\sim$3$''\times3''$ NIRSpec IFU FOV, among those detected in continuum and in line emission, namely YD4, YD6, YD6-[OIII]-E, YD4-[OIII]-W, YD7-E, ZD1, YD1, YD1-E, and s1, and possibly a tenth source, s2. 
All these sources lie at $z$ $\sim$ 7.88, either confirmed spectroscopically (YD4, YD6, YD6-[OIII]-E, YD4-[OIII]-W, YD1, YD1-E, and s1; e.g. \citealt{Morishita2023}, \citealt{Hashimoto2023}, and this work) or from photometry (ZD1), except for YD7-E (and the tentative source s2) whose redshift is not assessed.
The spectra of all targets, but YD4, are reported in  Figs.~\ref{fig:spectra_yd4_1} and \ref{fig:spectra_yd4_2} in the Appendix\footnote{The Appendix is available at \url{https://doi.org/10.5281/zenodo.13327942}.}. 


In the top-right panel of Fig. \ref{fig:yd4_maps}, we show the map of oxygen abundance, 12 + $\log$\,(O/H), a proxy for gas-phase metallicity. 
The metallicity is inferred by making use of the strong-line diagnostics reported in Table \ref{tab:ratio_notation} as described in Sect. \ref{ssec:data_anal} (the maps of the \riii\ and \oiiioii\ emission line ratios are shown in Fig. \ref{fig:yd4_maps_appdx}\footnotemark[\value{footnote}]).
The ratios involving lines which are not close in wavelength (all but \oiii\ and \hb) may be affected by extinction. Unfortunately, the only robustly detected Balmer line at the spaxel level is \hb, and only in a handful of spaxels, therefore estimating the dust extinction spaxel-by-spaxel is not possible. Based on this, the abundance map should not be taken as a robust measurement; moreover, the map is very noisy. 
Nevertheless, the map seems to suggest that, in the northern system, YD1 is embedded between two sources at lower metallicity, s1 and YD1-E. 
No clear pattern of metallicity shows up in the southern system (the YD4 one) from the map.

\section{\bdf}

\begin{figure*}
    \centering
    \includegraphics[scale=0.27,trim={1.5cm 0 1.5cm 0},clip]{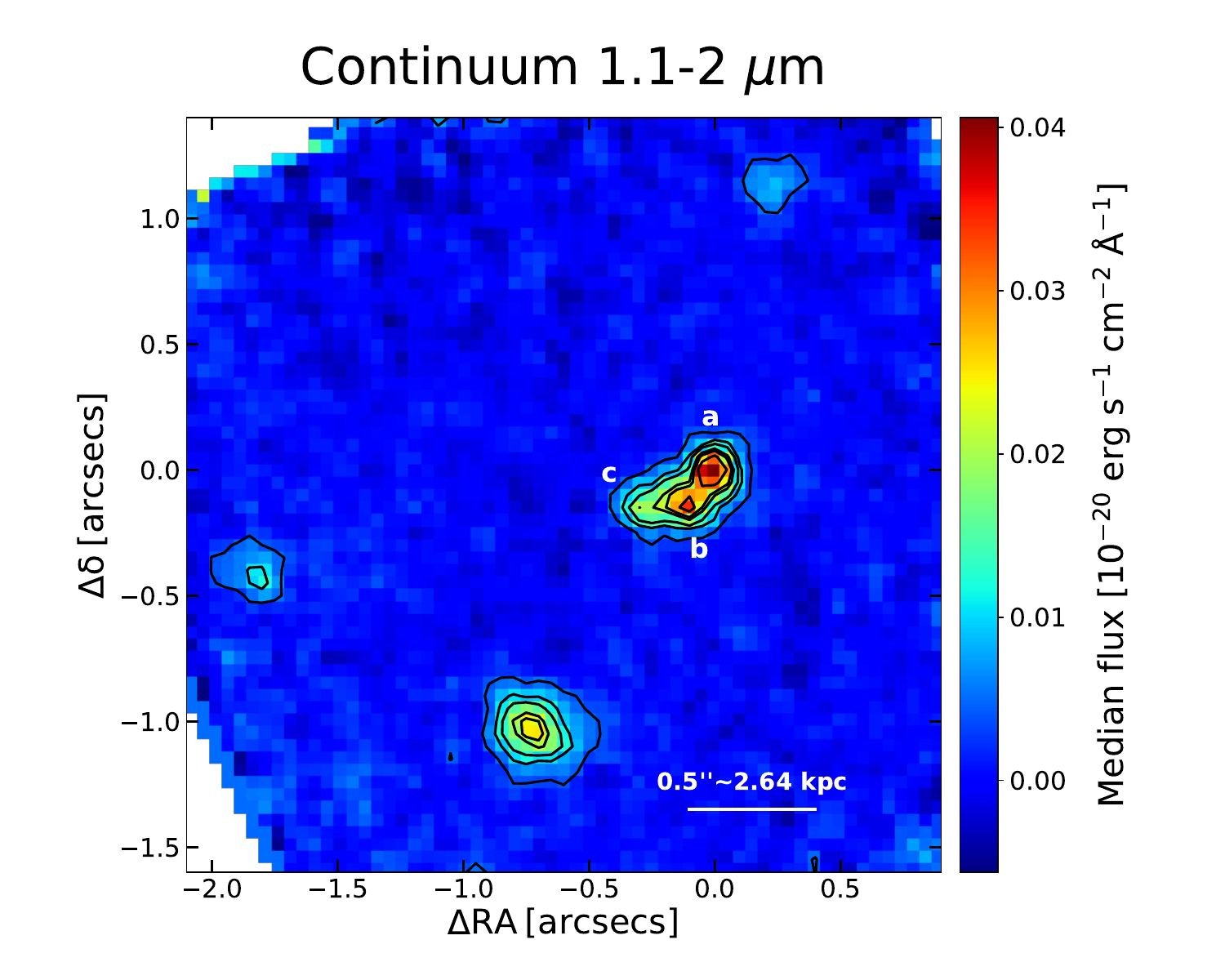}
    \includegraphics[scale=0.27,trim={1.5cm 0 1.5cm 0},clip]{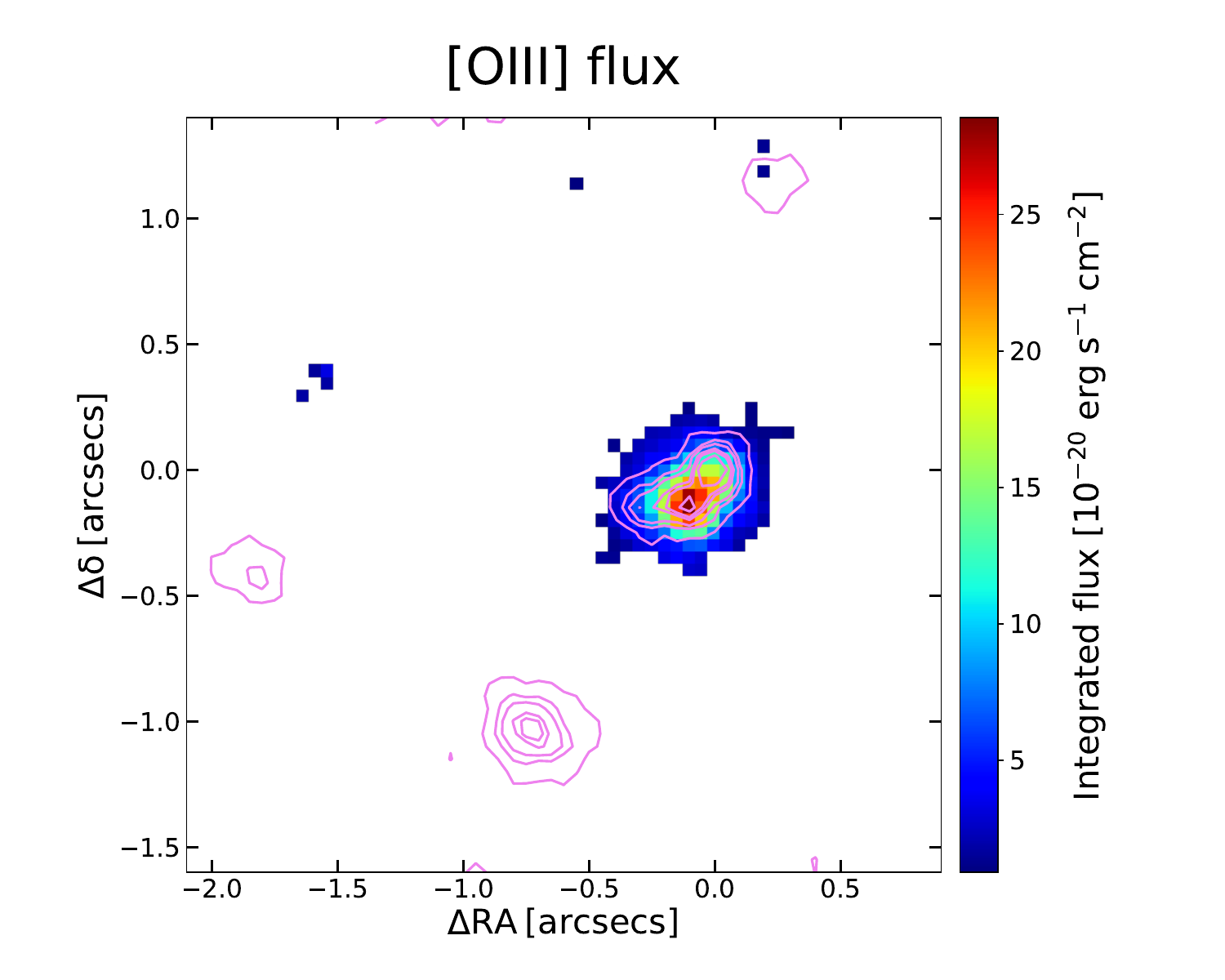}
    \includegraphics[scale=0.27,trim={1.5cm 0 2cm 0},clip]{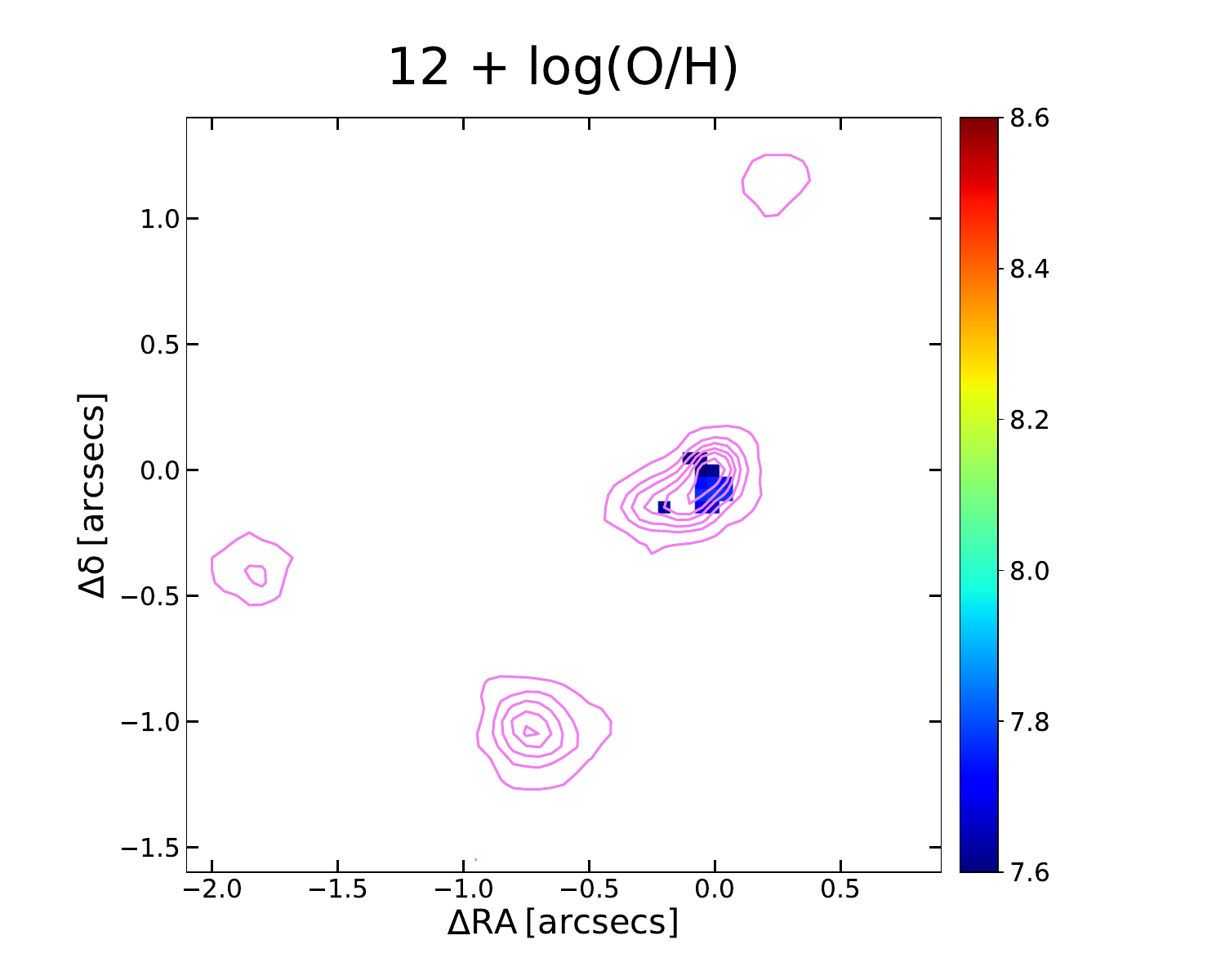}
    \includegraphics[width=0.8\textwidth,trim={0 0.4cm 0 0.4cm},clip]{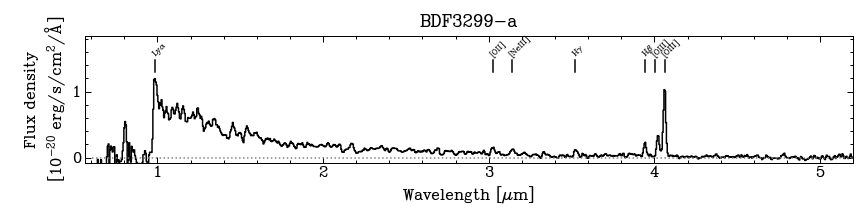}
    \caption{Maps for \bdf from JWST NIRSpec IFU. The flux maps are obtained from the original data cube, while the line ratio and metallicity maps from the spatially smoothed data cube (details in Sect. \ref{ssec:data_anal}). Median stellar continuum emission in the observed spectral range 1.1--2 $\mu$m ($\sim$0.14--0.25 $\mu$m rest-frame; top-left). The rest is as in Fig. \ref{fig:yd4_maps}. The PRISM/CLEAR spectrum (bottom) is extracted from a circular aperture with radius of 0.1\arcsec\ centred at the location of the most luminous component in continuum, \bdf-a.}
    \label{fig:bdf_maps}
\end{figure*}


\bdf is a spectroscopically confirmed star-forming galaxy (SFR$_\mathrm{UV}$ $\sim$ 6 $M_\odot$~yr$^{-1}$, SFR$_\mathrm{dust}$ $\lesssim$ 12 $M_\odot$/yr) at $z$ $\sim$ 7.109 \citep{Castellano2010, Vanzella2011, Maiolino2015, Carniani2017}, located in an over-density of galaxies \citep{Castellano2016}. 
We report the maps for \bdf from the NIRSpec IFU observations in Fig. \ref{fig:bdf_maps}. 

The median continuum in the observed spectral range 1.1--2 $\mu$m ($\sim$0.14--0.25 $\mu$m rest-frame) shows that \bdf is composed of three spatial components. The two brighter ones are located at the NW (the strongest of the two in the continuum) and at the centre of the system, respectively; a third, fainter one resides in the E part. We label these sources as \bdf-a, \bdf-b, and \bdf-c, respectively, in decreasing order of continuum brightness, as labelled in Fig. \ref{fig:bdf_maps}, left panel.
The other continuum emitters in the FOV are lower-redshift sources.

Fig. \ref{fig:bdf_maps}, bottom panel, shows the integrated spectrum (aperture radius = 0.1$''$)  associated with the brightest component in continuum emission, \bdf-a. 
The rest-frame optical lines \oii, \hb, and \oiii are detected with high S/N ($\gtrsim$6 on the peak).
From the lines, we estimate a redshift of 7.114, slightly higher but roughly consistent with the previously reported one from \lya \citep[7.109;][]{Vanzella2011} and \cii \cite[7.107;][]{Carniani2017}.
We obtain a stellar mass of log($M_*/M_\odot$) $\sim$ 7.9 from the SED fitting and a SFR $\sim$ 3.5 $M_\odot$~yr$^{-1}$ from \hb for \bdf-a (Table \ref{tab:integr_meas}).
The spectra of the other spatial components in the system, namely \bdf-b and c, are reported in Fig.~\ref{fig:spectra_bdf} in the Appendix\footnotemark[\value{footnote}]).
From these, we find log($M_*/M_\odot$) $\sim$ 7.6--8.2 and SFR $\sim$ 1.5--4 $M_\odot$~yr$^{-1}$. 

The \oiii emission (Fig. \ref{fig:bdf_maps}, central panel) peaks on \bdf-b rather than on \bdf-a (the strongest in continuum emission).
The line ratio maps are quite noisy and only few spaxels are above the S/N threshold of 3 (Fig.~\ref{fig:bdf_maps_appdx}). Therefore, the same applies to the metallicity map, which generally shows low values, of 12+log(O/H) $\lesssim$ 7.8, in the few spaxels where it can be estimated (Fig. \ref{fig:bdf_maps}, right).


\section{COSMOS-24108}

\begin{figure*}
    \centering
    \includegraphics[scale=0.27,trim={1.5cm 0 1.5cm 0},clip]{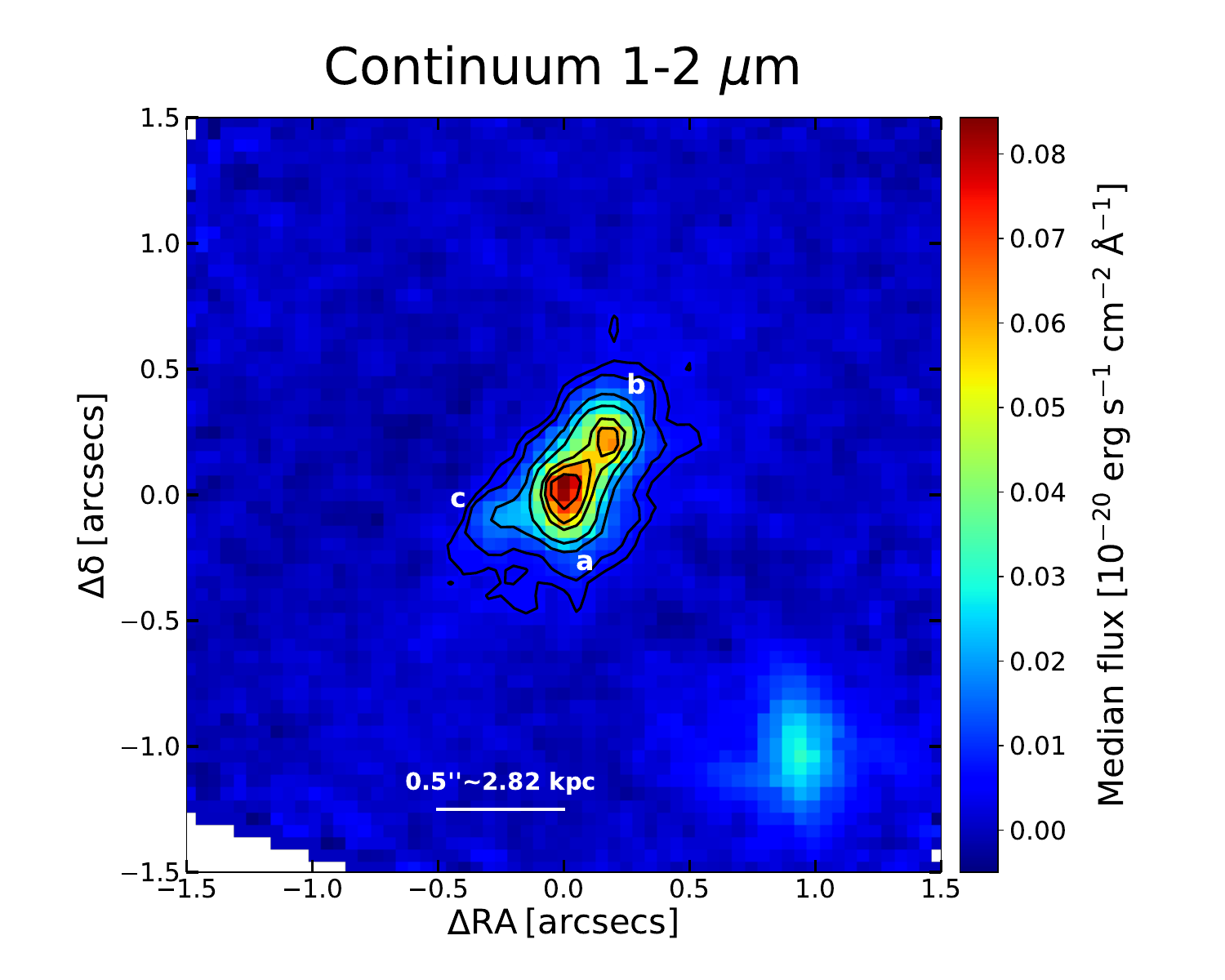}
    \includegraphics[scale=0.27,trim={1.5cm 0 1.5cm 0},clip]{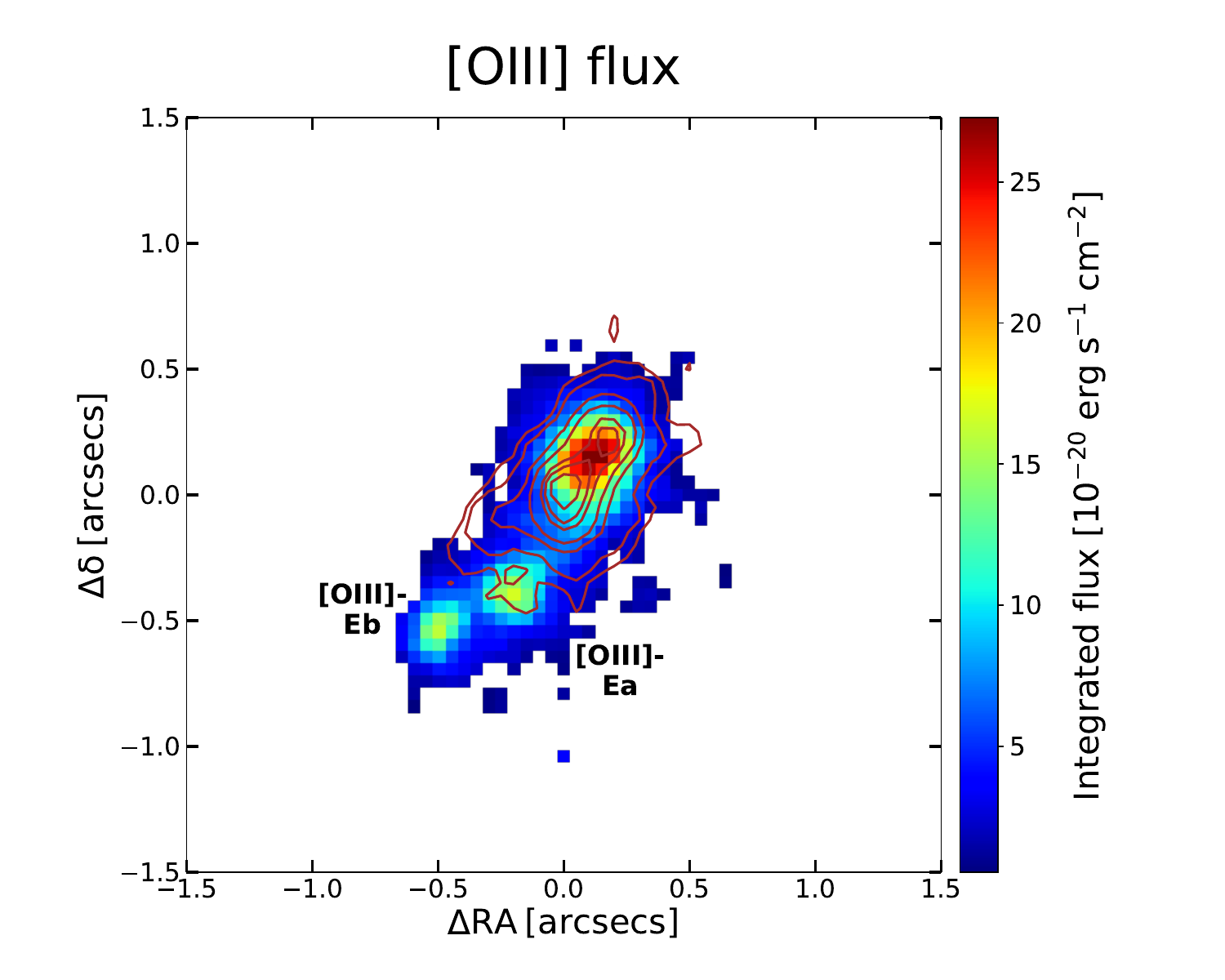}
    \includegraphics[scale=0.27,trim={1.5cm 0 1.5cm 0},clip]{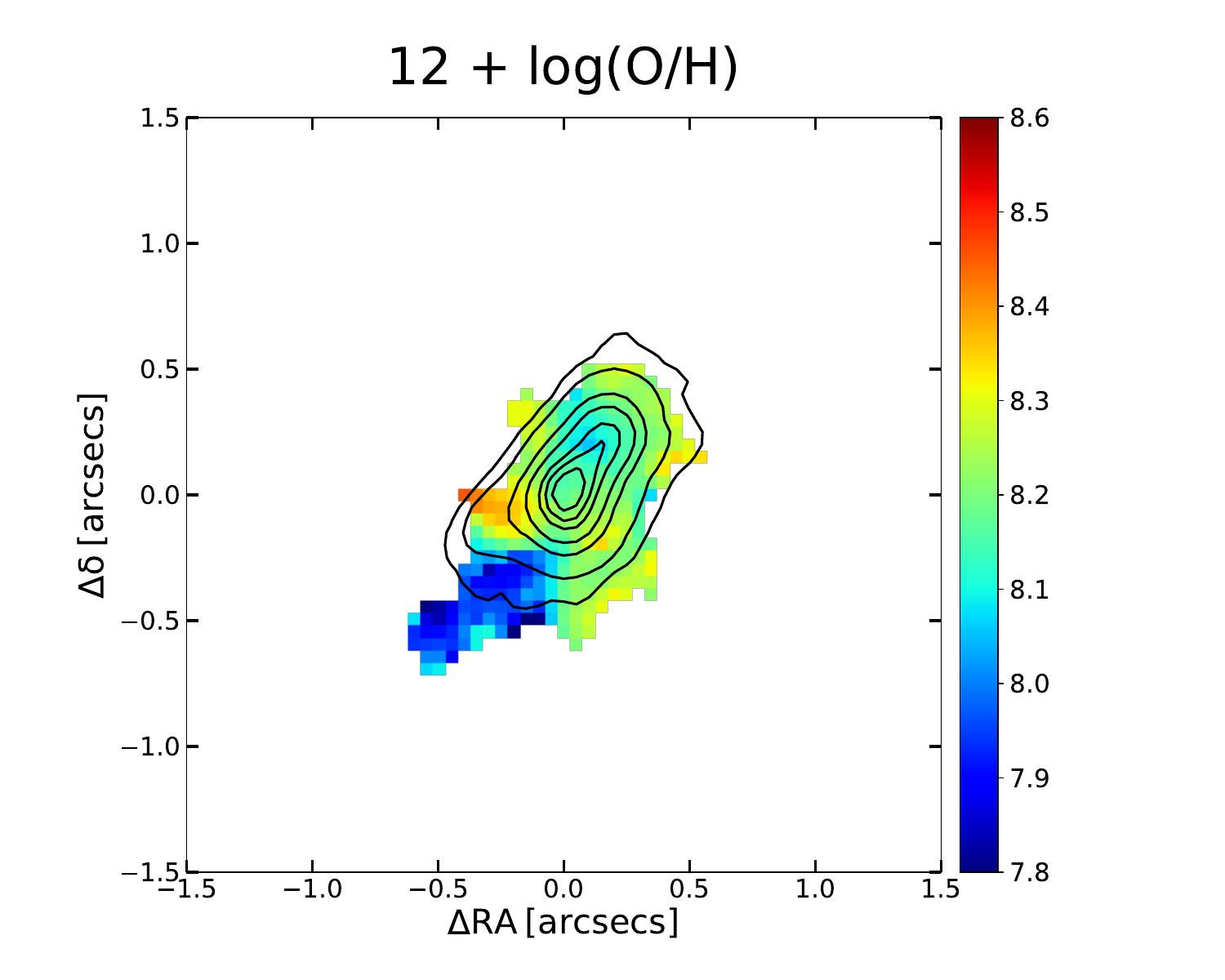}
    \includegraphics[width=0.8\textwidth,trim={0 0.4cm 0 0.4cm},clip]{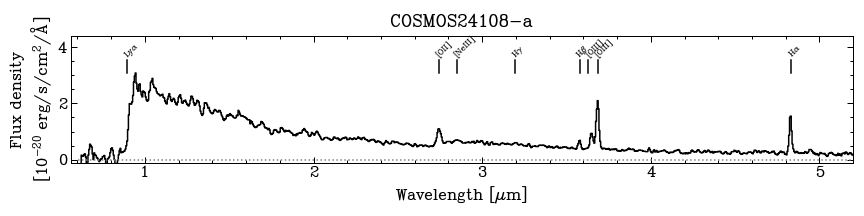}
    \caption{Maps for \cosmos from JWST NIRSpec IFU. Same as in Fig.~\ref{fig:bdf_maps}. The continuum map (left) is the median in the 1--2 $\mu$m observed range ($\sim$0.14--0.27~$\mu$m rest-frame). The PRISM/CLEAR spectrum (bottom) is extracted from a circular aperture with radius of 0.15\arcsec\ centred at the location of the most luminous component in continuum, \cosmos-a.}
    \label{fig:cosmos_maps}
\end{figure*}

COSMOS-24108 ($z$~$\sim$~6.36; SFR$_\mathrm{UV}$ $\sim$ 29 $M_\odot$/yr, SFR$_\mathrm{dust}$ $\lesssim$ 6.2 $M_\odot$/yr) shows two, or possibly three, spatial components in H-band HST rest-frame UV imaging \citep{Pentericci2016}.
In Fig.~\ref{fig:cosmos_maps}, left panel, we show the 1--2 $\mu$m observed continuum ($\sim$0.14--0.27 $\mu$m rest-frame) from the NIRSpec IFU data.
This shows two main spatial components, and a third, fainter one in the SE part of the system, consistent with those seen with HST in a similar spectral band. 
We label these three spatial components as \cosmos-a, b, and c, in order of continuum brightness.
The continuum at redder wavelengths, between 3--3.5 $\mu$m ($\sim$0.39--0.46 $\mu$m rest-frame; Fig.~\ref{fig:cosmos_maps_appdx} in the Appendix\footnotemark[\value{footnote}]),
is dominated by the southernmost of the two main components, and only an extended emission towards the northernmost source is present at these wavelengths, instead of a more clearly separate spatial component as at lower wavelengths. Therefore, the northernmost source has bluer continuum than the southernmost one.
In the SW corner of the FOV, another galaxy at a lower redshift 
is also present.

Fig. \ref{fig:cosmos_maps}, bottom panel, shows the integrated spectrum (aperture radius = 0.15$''$) associated with the brightest component in continuum emission, \cosmos-a. The rest-frame optical lines \oii, \hb, \oiii, \ha, and \sii are detected with high S/N ($\gtrsim$4 on the peak).
We obtain a stellar mass of log($M_*/M_\odot$) $\sim$ 9.3 from SED fitting and a SFR $\sim$ 15 $M_\odot$~yr$^{-1}$ from \ha for \cosmos-a (Table \ref{tab:integr_meas}).
The integrated spectra of the other spatial components in the system are shown in Fig.~\ref{fig:spectra_cosmos} in the Appendix\footnotemark[\value{footnote}]. For these, we find log($M_*/M_\odot$) $\sim$ 8.4--8.9 and SFR $\sim$ 2--15 $M_\odot$~yr$^{-1}$. 


The \oiii ionised gas line emission (Fig. \ref{fig:cosmos_maps}, central panel; \hb, \oii, and \ha maps are reported in Fig. \ref{fig:cosmos_maps_appdx}\footnotemark[\value{footnote}]) is much more extended than the continuum, revealing two additional bright clumps to the SE of the system, one of the two also tentatively detected in continuum (see contours).
We label these as \cosmos-[OIII]-Ea and [OIII]-Eb.
The \ha\ map (Fig.~\ref{fig:cosmos_maps_appdx}, bottom-left) shows a bridge of gas (visible also in \oiii, though weaker relative the rest of the emission) connecting the main system to the closer of these two clumps, \cosmos-[OIII]-Ea. 
By comparing the ionised gas line emission with the 1--2 $\mu$m continuum, we see that the former peaks in between the two main continuum components, with a preferential extension towards the northernmost, weaker continuum component (\cosmos-b) rather than to the southernmost, brighter one (\cosmos-a).

Fig.~\ref{fig:cosmos_maps}, right panel, reports the map of gas metallicity (the \riii and \oiiioii line ratio maps are shown in Fig. \ref{fig:cosmos_maps_appdx}\footnotemark[\value{footnote}]). 
The metallicity exhibits differences of up to $\sim$0.5--0.6 dex among the different sources in the system. The northernmost of the two main continuum sources, \cosmos-b, appears to have lower metallicity (12 + $\log$\,(O/H) $\sim$ 8.0--8.1) than the southernmost one, \cosmos-a ($\sim$ 8.2), while the SE minor continuum component, \cosmos-c, has larger metallicity ($\sim$ 8.4). Towards the two ionised gas clumps to the SE, \cosmos-[OIII]-Ea and \cosmos-[OIII]-Eb, the metallicity is the lowest, with values of $\sim$ 7.8--7.9.

\begin{figure*}
    \centering
    \includegraphics[width=0.49\textwidth,trim={0.6cm 0 2.5cm 0},clip]{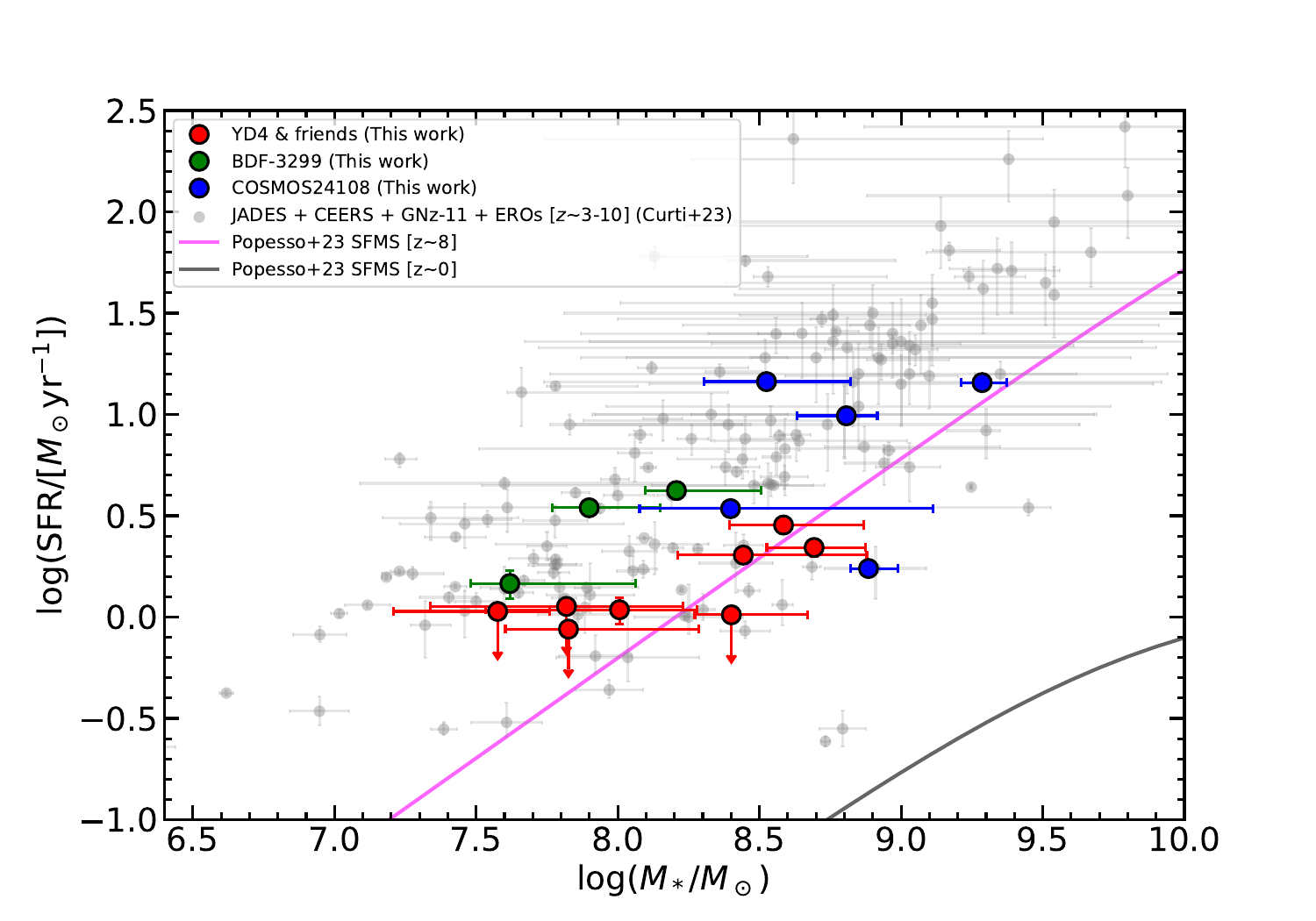}
    \includegraphics[width=0.49\textwidth,trim={0.6cm 0 2.5cm 0},clip]{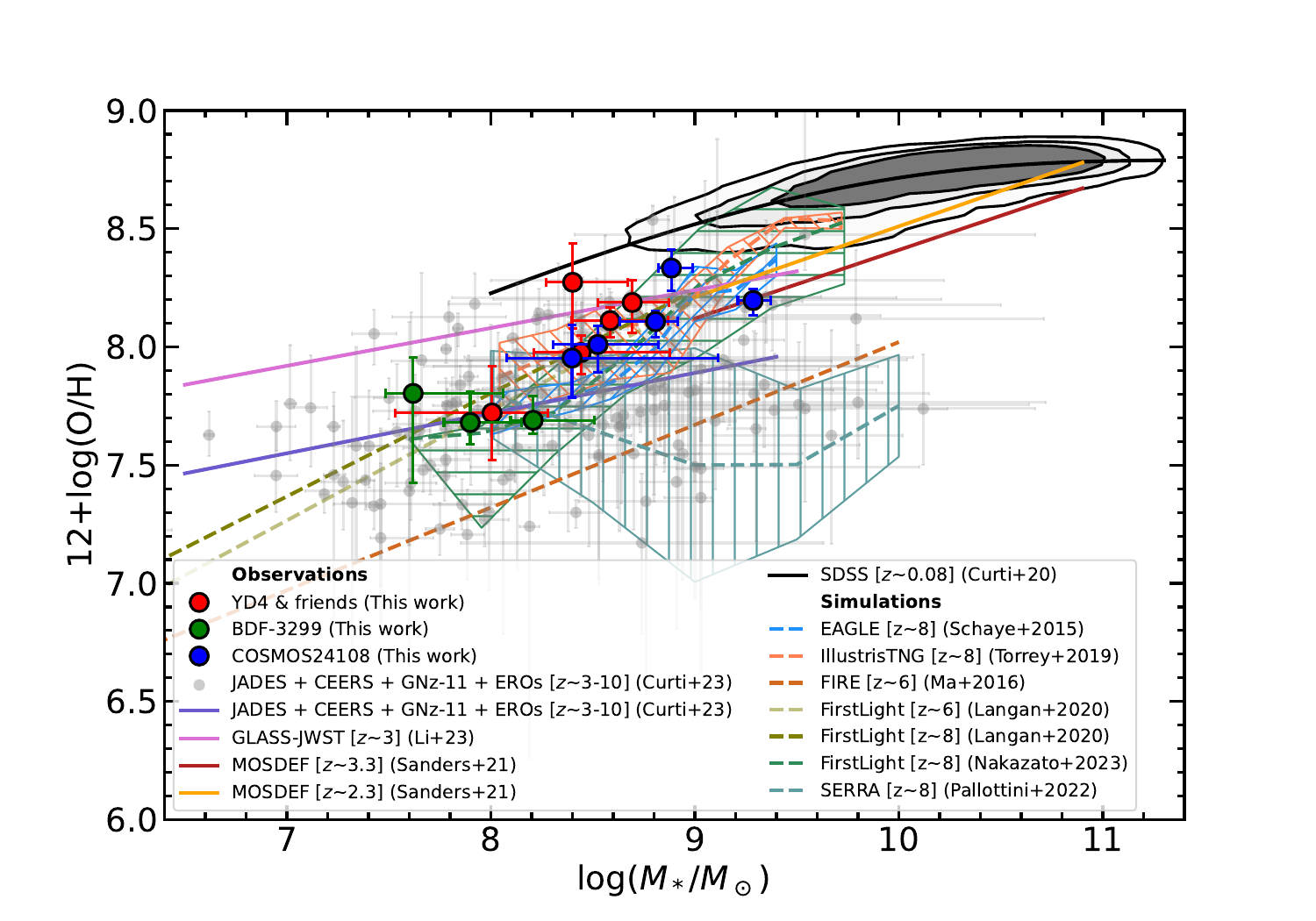}
    \caption{
    Star formation rate versus stellar mass (left) and gas-phase metallicity versus stellar mass (right) for the single spatial components in the systems from this work (red, green, and blue circles). 
    In grey circles we show for comparison the compilation of $z$ $\sim$ 3--10 galaxies from JADES (\citealt{Curti2023_jades}; including GNz-11, \citealt{Bunker2023}), CEERS \citep[as re-computed for consistency in \citealt{Curti2023_jades}]{Nakajima2023}, and EROs \citep{Curti2023_ero, Laseter2024}. Left: The main sequences of star formation (SFMS) at z $\sim$ 0 and z $\sim$ 8 from \cite{Popesso2023} are displayed for reference. Right: Best-fit mass-metallicity relations (MZRs) from \cite{Curti2023_jades} for these $z$ $\sim$ 3--10 targets, \cite{Li2023} at $z$ $\sim$ 3 (GLASS-JWST), \cite{Sanders2021} at $z$ $\sim$ 3.3 and 2.3 (MOSDEF), and \cite{Curti2020calib} at $z$ $\sim$ 0.08 (SDSS).    
    The high-$z$ MZRs predicted by EAGLE \citep[$z$ $\sim$ 8;][]{Schaye2015}, FIRE \citep[$z$ $\sim$ 6;][]{Ma2016}, IllustrisTNG \citep[$z$ $\sim$ 8;][]{Torrey2019}, FirstLight ($z$ $\sim$ 6 and $z$ $\sim$ 8, for $\log(M_*/M_\odot)$ $\lesssim$ 9 at $z$ $\sim$ 6; \citealt{Langan2020}; and $z$ $\sim$ 8, for $\log(M_*/M_\odot)$ $\gtrsim$ 9 at $z$ $\sim$ 6; \citealt{Nakazato2023}), and SERRA \citep[$z$ $\sim$ 8;][]{Pallottini2022} cosmological simulations are also displayed.}
    \label{fig:MZR}
\end{figure*}

\section{Integrated mass-metallicity relation}

We obtained the properties of each sub-source belonging to each target, by extracting integrated spectra from circular apertures of radius of 0.15$''$ for \yd4 and \cosmos and of 0.1$''$ for \bdf centred on each sub-source (Table \ref{tab:integr_meas}).
We find emission line ratios of log(\oiii/\hb) $\sim$ 0.4--0.9 for \yd4, log(\oiii/\hb) $\sim$ 0.7--0.8 for \bdf, and log(\oiii/\hb) $\sim$ 0.5--0.9 and log(\sii/\ha) $\lesssim$--0.8 for \cosmos. These are consistent with the range of values found for $z$ $\sim$ 6--8 star-forming galaxies in the JADES survey \citep{Cameron2023}, of log(\oiii/\hb) $\sim$ 0.5--0.8 and log(\sii/\ha) $\lesssim$ --0.8 (when only considering detections and not upper limits), in a range of stellar masses (log($M_*/M_\odot$) $\sim$ 6.5--9.0) and SFRs ($\sim$0.1--30) comprising those of the sources in this work. In some cases, like \cosmos-a with  log(\oiii/\hb) $\sim$ 0.8 and log(\sii/\ha) $\sim$ --0.8, the ratios are at the high end of the values reported in \citep{Cameron2023}.

We inferred the gas-phase metallicity and the SFR from emission-line fitting and the stellar mass from SED fitting, as described in Sect. \ref{ssec:data_anal}.
In Fig.~\ref{fig:MZR}, we display the metallicity versus SFR and stellar mass diagrams for the spatial components identified in the images (Table~\ref{tab:integr_meas}).
We compare these with the values reported by \cite{Curti2023_jades} for the $z$ $\sim$ 3--10 galaxies from the JADES (including GNz-11; \citealt{Bunker2023}), CEERS \citep[][re-computed by \citealt{Curti2023_jades} for consistency]{Nakajima2023}, and EROs \citep{Curti2023_ero, Laseter2024} samples (grey circles). The best-fit relation for this $z$ $\sim$ 3--10 compilation obtained by \cite{Curti2023_jades} is also shown.
We also report for reference the best-fit mass-metallicity relations (MZRs) for galaxies at $z$ $\sim$ 0.08 from SDSS \citep{Curti2020calib}, $z$ $\sim$ 2--3 from MOSDEF \citep{Sanders2021}, and $z$ $\sim$ 3 from GLASS-JWST \citep{Li2023}. 

Our targets are generally compatible with the values found for the $z$ $\sim$ 3--10 galaxies, especially the sources in \bdf which sit on the MZR defined by this high-$z$ collection. We note that most of the sources in \yd4 and \cosmos ($z$ $\sim$ 8 and 6, respectively) are at the high end of the JADES, CEERS, and EROs points, where the $z$ $\sim$ 2--3 MZRs from MOSDEF and GLASS-JWST lie.
This suggests that the \yd4 and \cosmos systems may comprise more evolved sources as compared to the majority of sources at the same redshifts and be instead more similar to galaxies at cosmic noon.

We further display the predictions for the high-$z$ MZR from a number of cosmological simulations, specifically EAGLE \citep[$z$ $\sim$ 8;][]{Schaye2015}, FIRE \citep[$z$ $\sim$ 6;][]{Ma2016}, IllustrisTNG \citep[$z$ $\sim$ 8;][]{Torrey2019}, FirstLight (from \citealt{Langan2020} at $z$ $\sim$ 6 and $z$ $\sim$ 8 and \citealt{Nakazato2023} at $z$ $\sim$ 8, for $\log(M_*/M_\odot)$ $\lesssim$ 9 and $\gtrsim$ 9, respectively, at $z$ $\sim$ 6), and SERRA \citep[$z$ $\sim$ 8;][]{Pallottini2022}\footnote{The MZR from the SERRA simulation \citep{Pallottini2022} was originally presented in \citet{Curti2023_ero} and was computed using 202 simulated galaxies at $z=7.7$. In the present work, the MZR from SERRA is recomputed from an extended sample of 245 objects \citet{PallottiniFerrara2023}. This causes some differences at the low-mass end of the reported MZR with respect to that in \citet{Curti2023_ero}.}. 
We note that the sources analysed in this work, both as a whole and within the individual \yd4 and \cosmos systems, are better aligned with simulations predicting steeper MZR slopes than the median of the observations at $z$ $\sim$ 3--10. 

\section{Gas-phase metallicity gradients}

\begin{figure*}
    \centering
    \includegraphics[scale=0.265,trim={1cm 0 1cm 0},clip]{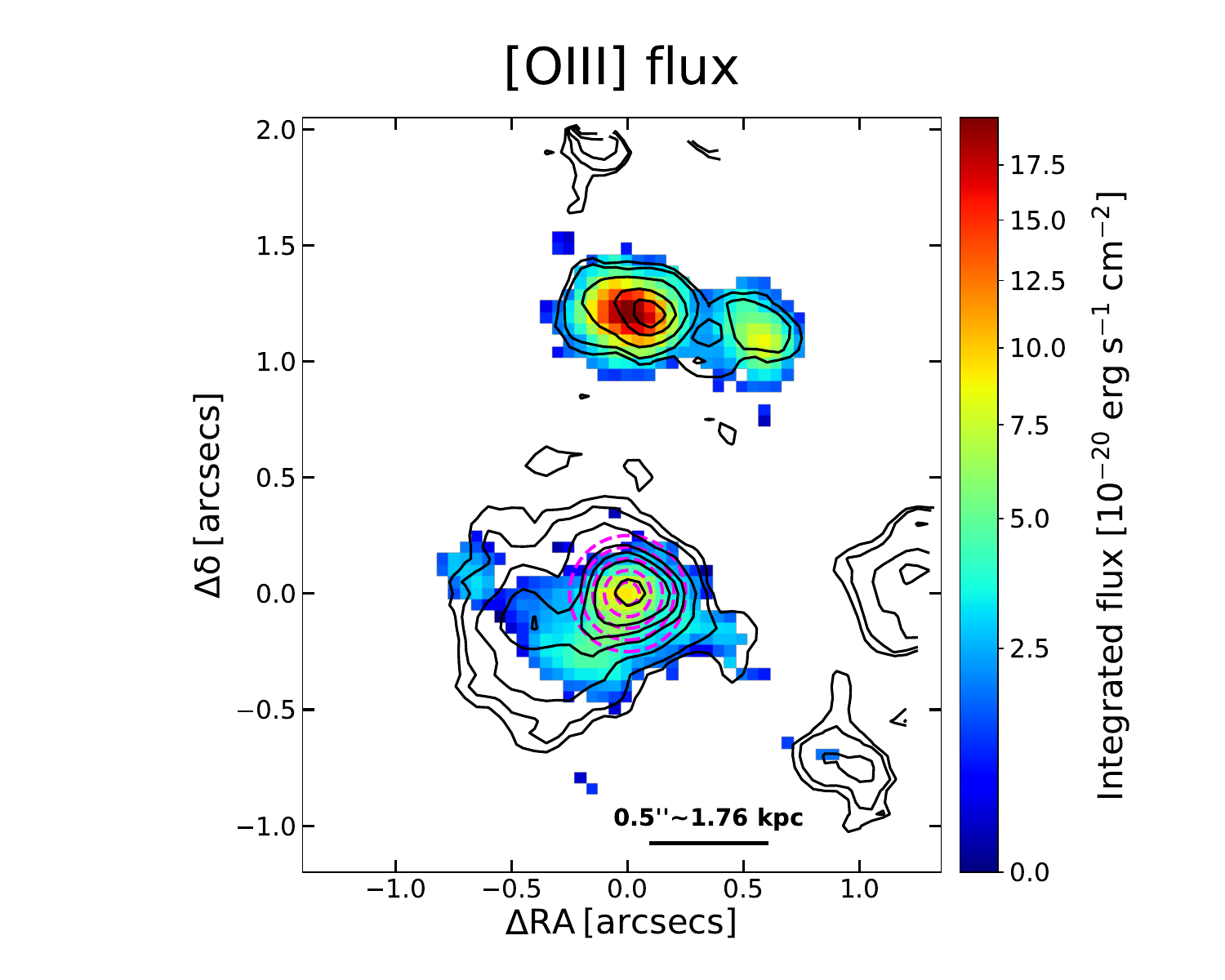}
    \includegraphics[scale=0.29,trim={1cm 0 1cm 0},clip]{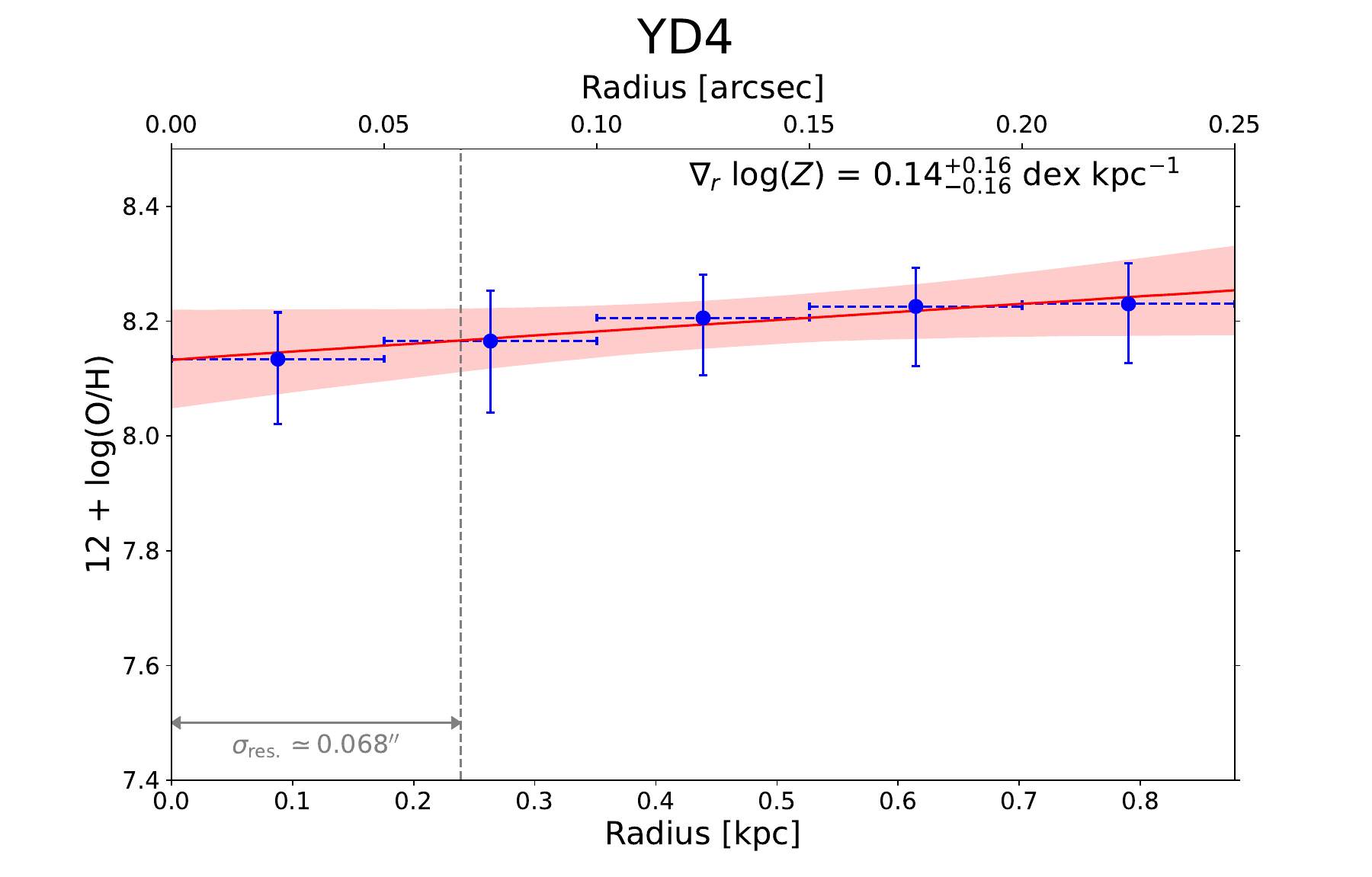}\\
    \includegraphics[scale=0.265,trim={1cm 0 1cm 0},clip]{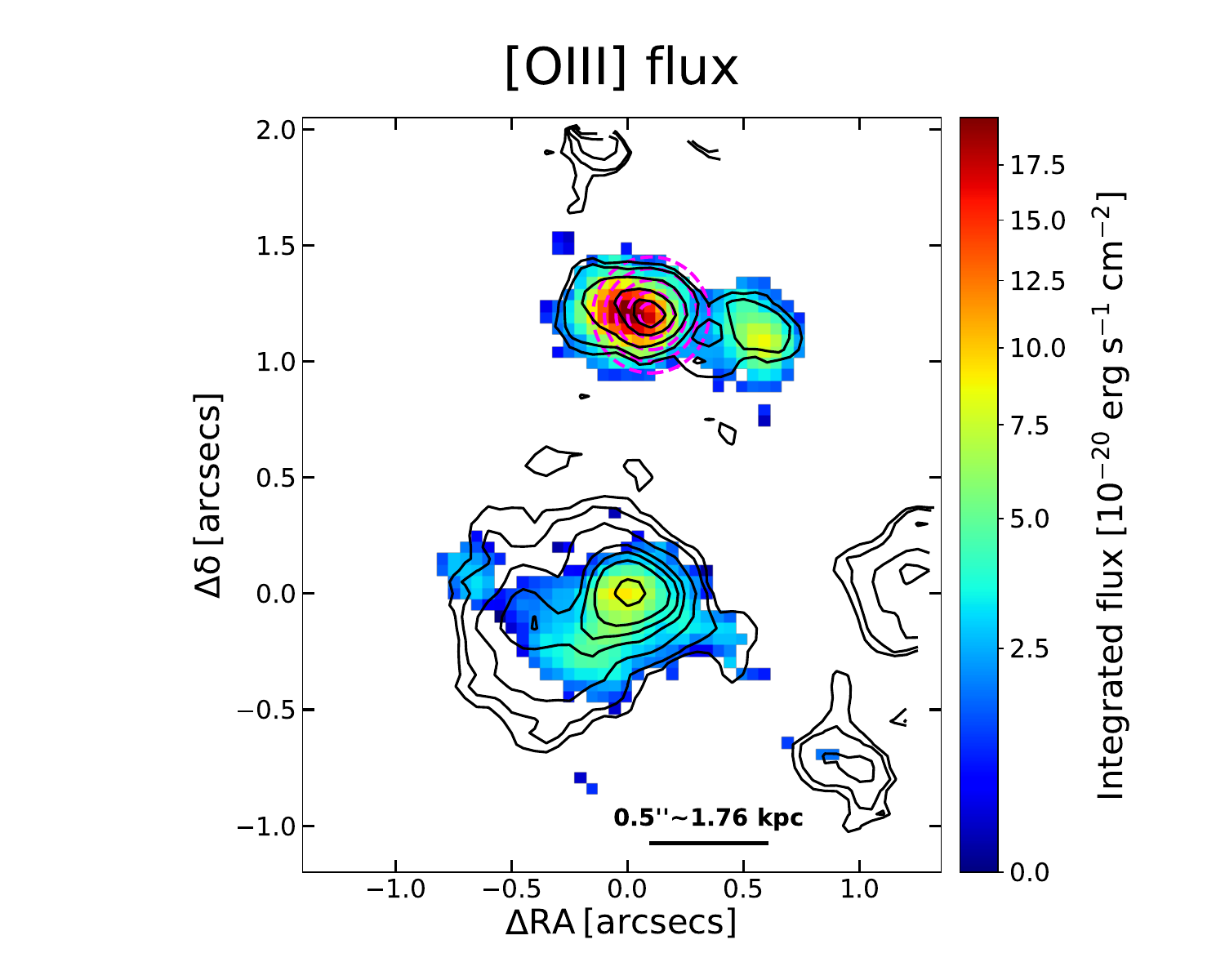}
    \includegraphics[scale=0.29,trim={1cm 0 1cm 0},clip]{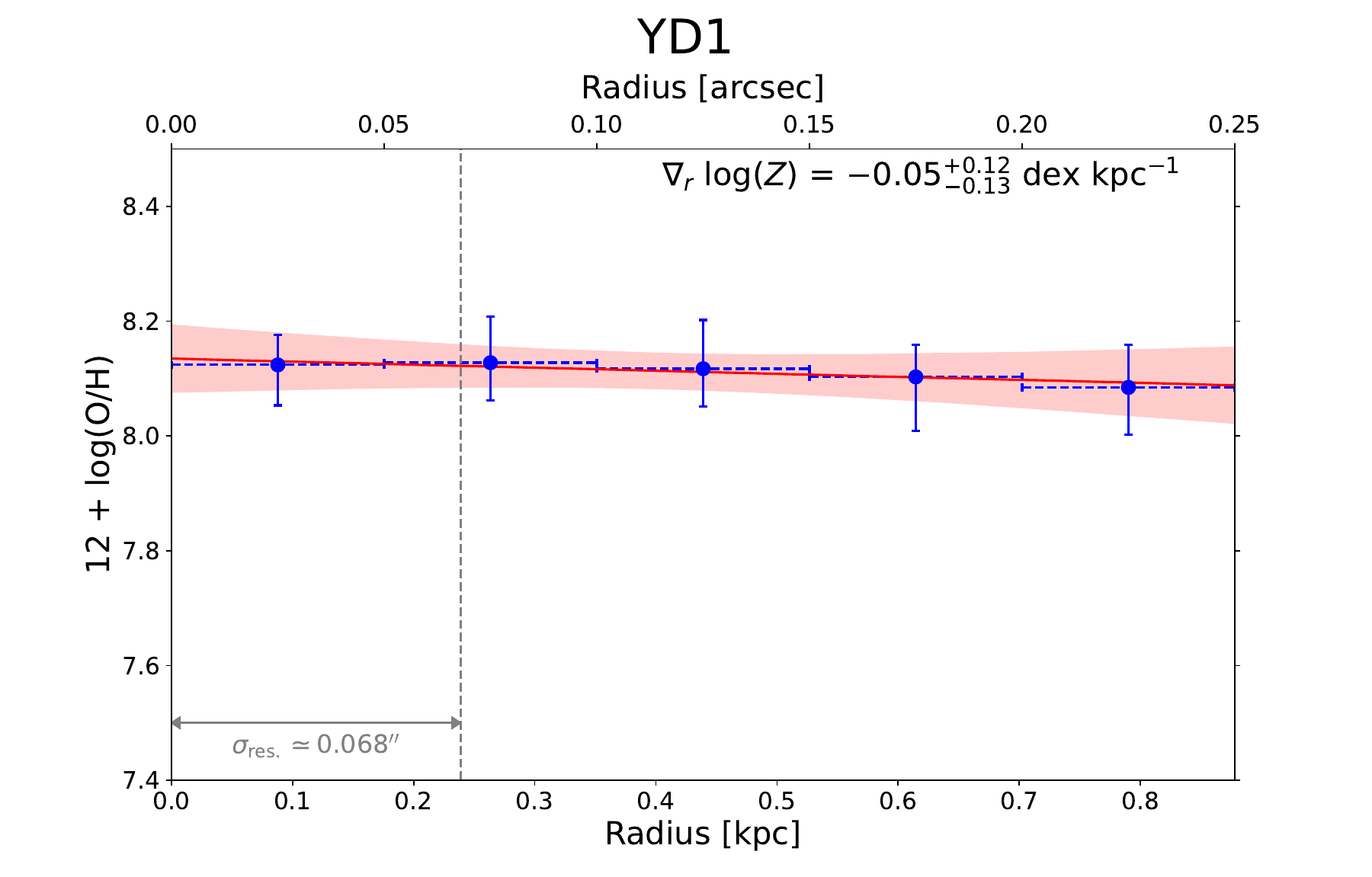}\\
    \includegraphics[scale=0.265,trim={1cm 0 1cm 0},clip]{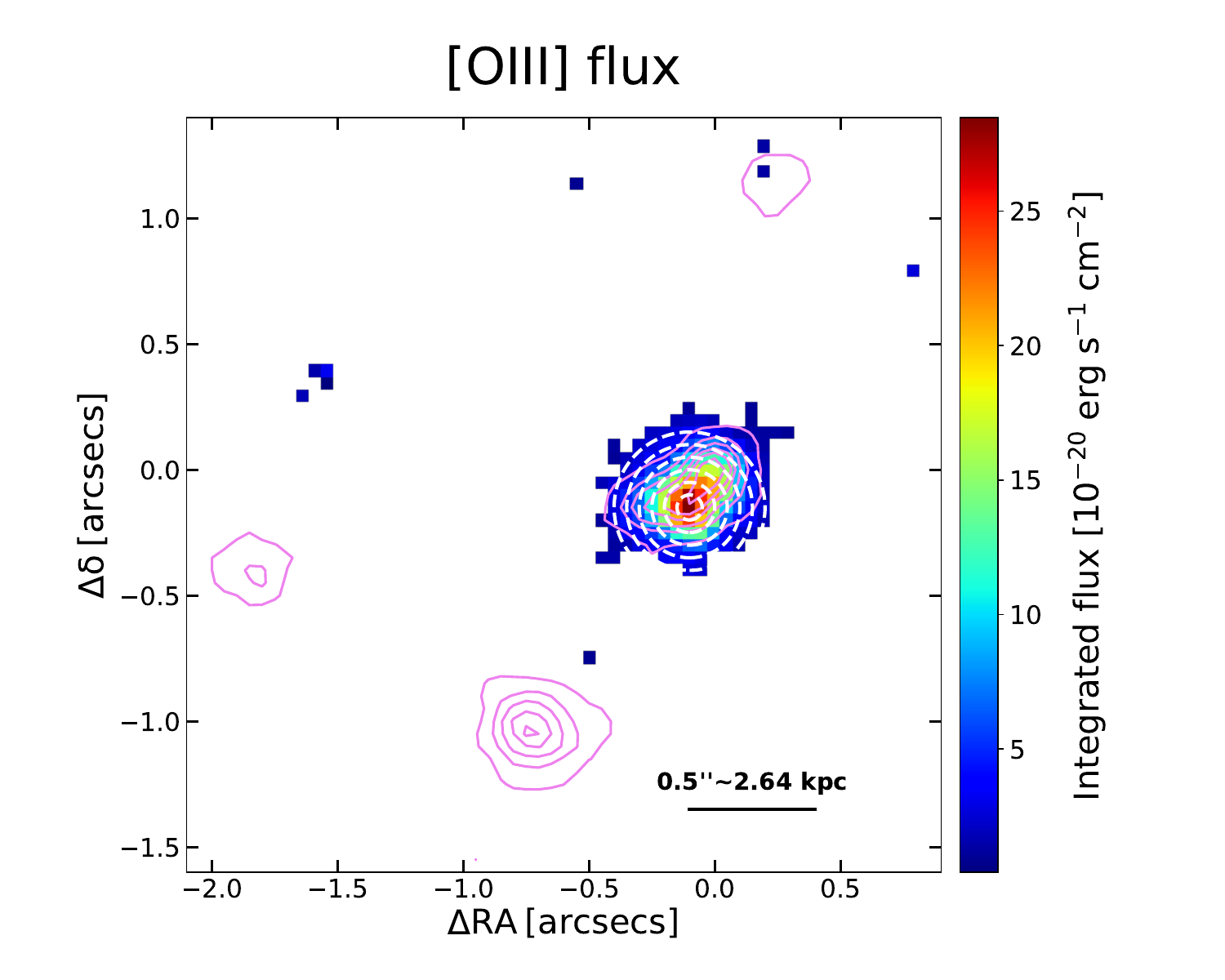}
    \includegraphics[scale=0.29,trim={1cm 0 1cm 0},clip]{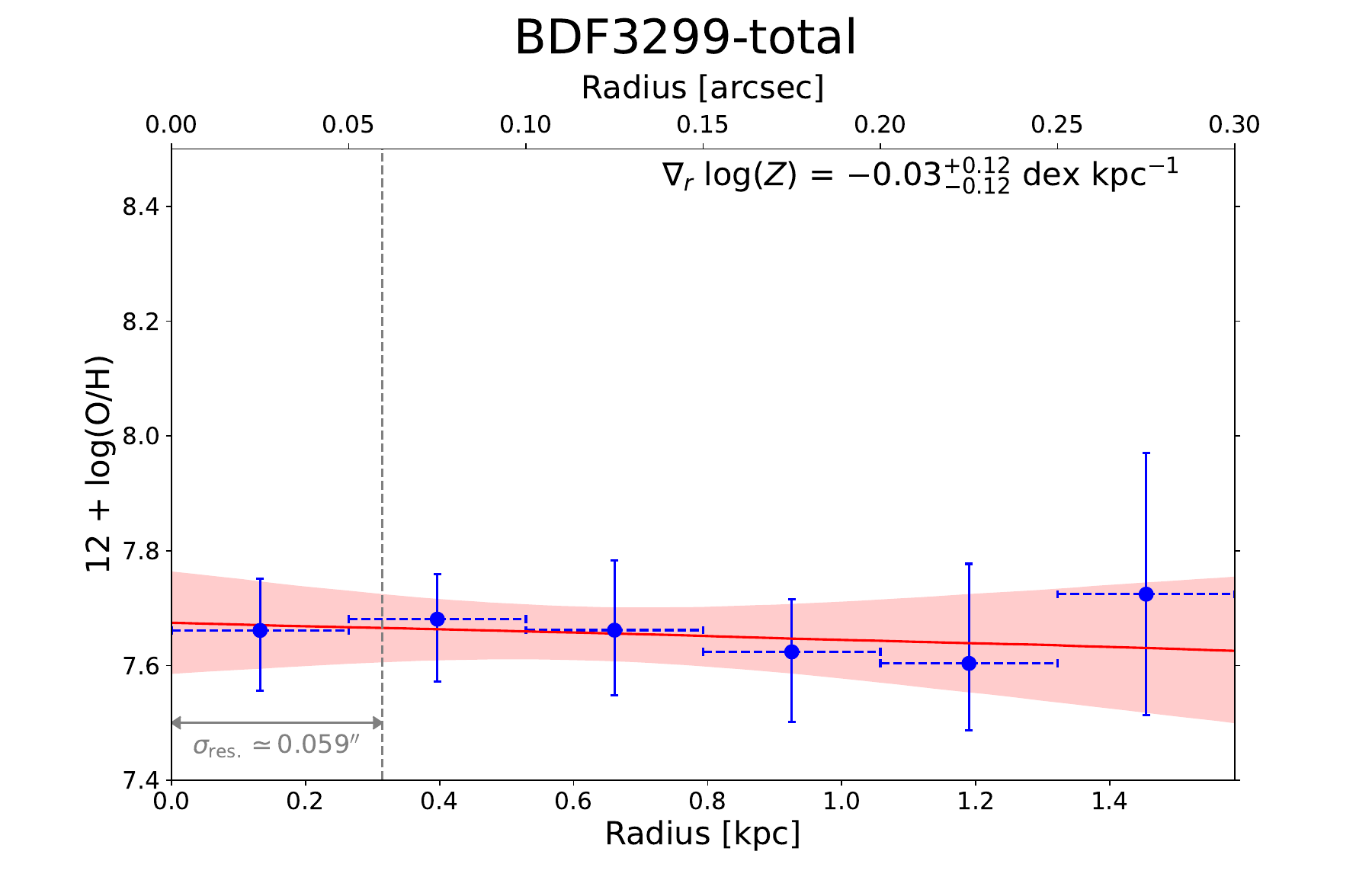}\\
    \includegraphics[scale=0.265,trim={1cm 0.5cm 1cm 0},clip]{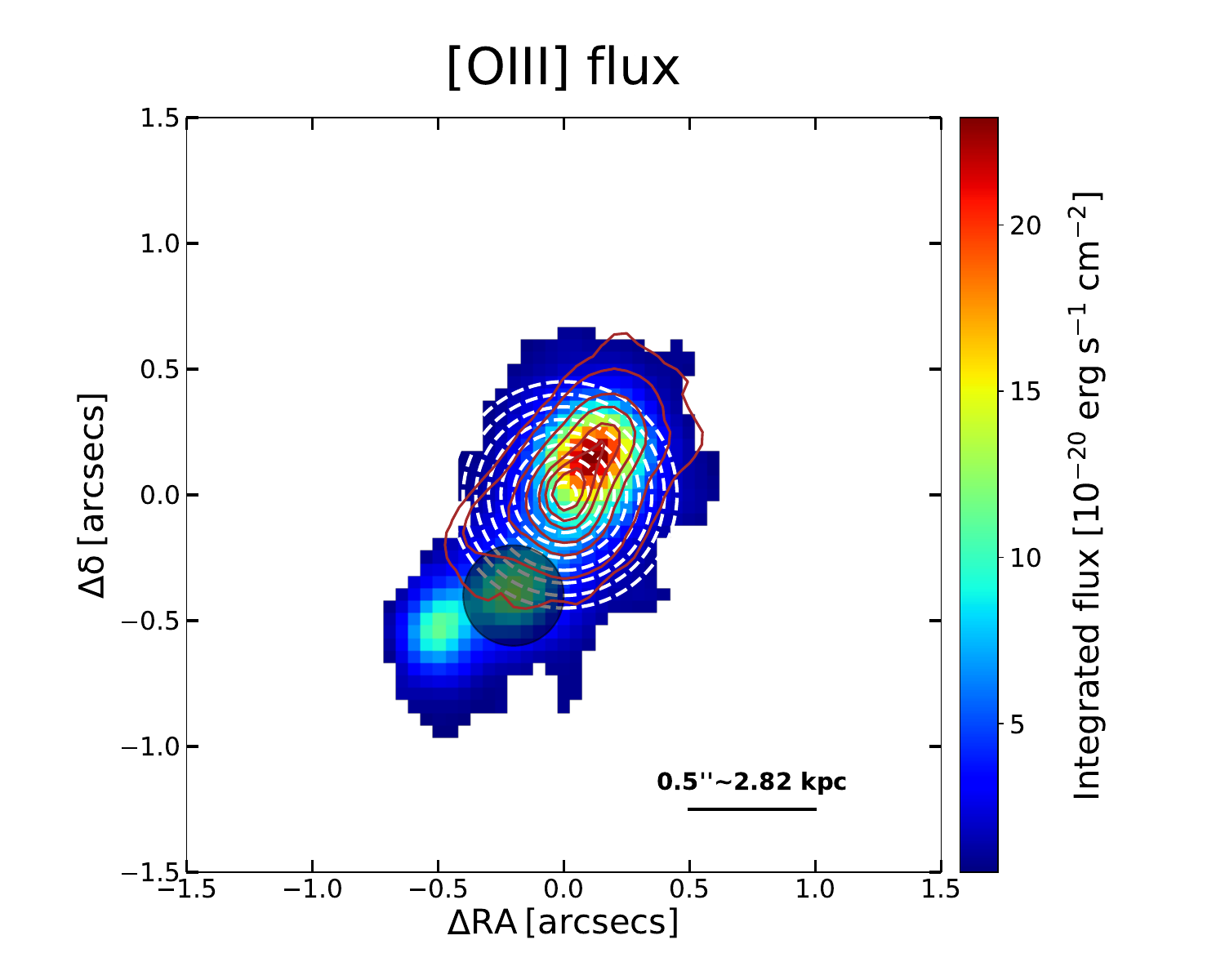}
    \includegraphics[scale=0.29,trim={1cm 0.5cm 1cm 0},clip]{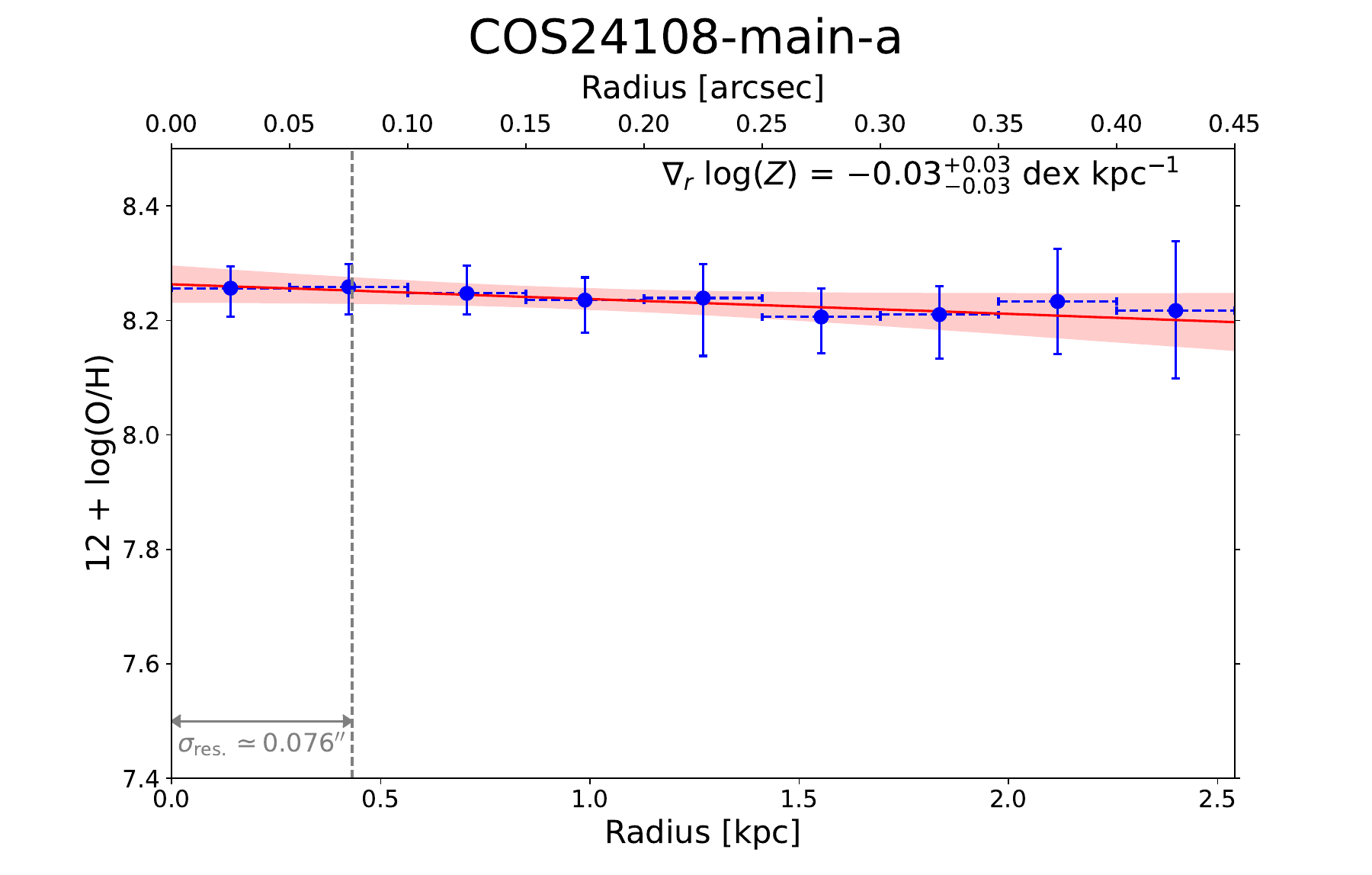}
    \caption{Metallicity gradients for the main sources in \yd4 (top two rows), \bdf (mid-bottom), and \cosmos (bottom). Radial annuli have radial width of 1 spaxel (0.05\arcsec). The wavelength-dependent spatially smoothed cube was used (details in Sect. \ref{ssec:data_anal}). Metallicity gradients (right) and \oiii maps (left; also from smoothed cube), where the concentric annuli are reported on the source (the shaded region in the \cosmos map was masked out from the gradient; see text). The vertical grey dashed line marks the spatial resolution ($\sigma_\mathrm{res.}$). The fluxes of emission lines far in wavelength are corrected for extinction based on \hg/\hb in \yd4 and \bdf, while in \cosmos \ha/\hb is used for that and also \sii and \ha are used to infer the metallicity. A cut of S/N$>3$ on the peak flux of each line is applied.}
    \label{fig:met_grad}
\end{figure*}

\begin{table}
\caption{Metallicity gradients inferred for the main sources in this work.}
    \centering
    \def\arraystretch{1.2}
    \begin{tabular}{ccc}
    \hline\hline
         Source name&  $\nabla_\mathrm{r} \, \log(Z)$ [dex kpc$^{-1}$] & [dex $R_\mathrm{e}^{-1}$]\\
    \hline
         \yd4&  0.14$_{-0.16}^{+0.16}$ & 0.10$_{-0.12}^{+0.12}$\\
         A2744-YD1&  --0.05$_{-0.13}^{+0.12}$ & --0.04$_{-0.9}^{+0.8}$\\
         \bdf&  --0.03$_{-0.12}^{+0.12}$ & --0.02$_{-0.09}^{+0.09}$\\
         \cosmos& --0.03$_{-0.03}^{+0.03}$ & --0.04$_{-0.04}^{+0.04}$\\
    \hline
    \end{tabular}
    \label{tab:metgrad}
\end{table}

To more robustly constrain the spatially resolved metallicity, we extracted the line fluxes from spatially integrated spectra from concentric radial annuli, as described in Sect. \ref{ssec:data_anal}.
This allowed us to increase the S/N of the emission lines at large radii and determine the dust extinction from the faint Balmer lines. 
We thus determined the gas-phase metallicity as a function of the distance from the centre of each source, as defined in the following.

The resulting metallicity gradients are reported in Fig. \ref{fig:met_grad}.
In the system \yd4, we determined the metallicity gradients only for the two brightest sources YD4 and YD1 (centred around their continuum peak) since the metallicity measurements for the other sources suffered from large uncertainties due to low S/N of emission lines.
For \bdf, since we could have only extracted two radial points for each spatial component (\bdf-a, b, and c) due to the compactness of this system, we obtained the metallicity gradient profile for the whole \bdf system, centred on the brightest source in line emission, \bdf-b. 
For the \cosmos system, we centred the gradient around the continuum emission peak, corresponding to the spatial component \cosmos-a (Fig.~\ref{fig:cosmos_maps}, left panel). We masked the region surrounding the low-metallicity \oiii clump \cosmos-Ea (shaded circle in Fig.~\ref{fig:met_grad}), since it has a markedly low metallicity (see Fig.~\ref{fig:cosmos_maps}, right panel) and can be clearly spatially separated from the rest of the system. 
Given that the line emission peaks between the two main continuum components, \cosmos-a and \cosmos-b, and not on either of them (see Fig. \ref{fig:cosmos_maps}, central panel), we also extracted the gradient centred around the \oiii peak, instead of around the continuum. This is reported in Fig.~\ref{fig:cosmos_metgrad_clumps} in the Appendix\footnote{The Appendix is available at \url{https://doi.org/10.5281/zenodo.13327942}.}. The two gradients are not statistically different within the uncertainties.
In Fig.~\ref{fig:cosmos_metgrad_clumps} we also report the metallicity gradient from two additional smaller apertures centred on the emission line clumps to the SE of the main system, \cosmos-Ea and \cosmos-Eb.

For all the sources, \yd4, YD1, \bdf, and \cosmos, the metallicity gradients span a range between $\nabla_r \log(Z)$ $\sim$ $-0.05~{\rm dex~kpc^{-1}}$ and $\nabla_r \log(Z)$ $\sim$ $0.15~{\rm dex~kpc^{-1}}$ (Table~\ref{tab:metgrad}).  We note that, within the uncertainties, all metallicity gradients in our sample are flat. 

We also estimate the gradients relative to the effective (half-light) radius, $R_\mathrm{e}$, as sometimes done in observational (e.g. \citealt{Belfiore2017}) and theoretical studies (e.g. \citealt{Molla2019}); in this case, the gradient should not depend on other properties of galaxies (as suggested by e.g. \citealt{Garnett1997}).
We estimated the effective radius with a 2D Gaussian modelling of the continuum emission for the whole \bdf and \cosmos systems (including all the spatial components in continuum), and for the single YD4 and YD1 components, for consistency with the way metallicity gradients were calculated; for the same reason, we employed the wavelength-dependent spatially smoothed data cube, as done for the gradients. We stress that, therefore, the $R_\mathrm{e}$ thus obtained should not be considered as the intrinsic ones (i.e. de-convolved for the spatial resolution of the observations), and, in the case of \bdf and \cosmos, they are not the $R_\mathrm{e}$ of the single spatial clumps.
The obtained $R_\mathrm{e}$ are $\sim$ 0.74, 0.70, 0.75, and 1.2 kpc for YD4, YD1, \bdf, and \cosmos, respectively.
The resulting gradients are reported in Table~\ref{tab:metgrad}.
These are flat, and in any case $\gtrsim$ --0.1~dex~$R_\mathrm{e}^{-1}$ when considering the uncertainties; in the local Universe, such gradients are typical of low-mass galaxies ($\log(M_* / M_\odot)$ $\lesssim$ 9.5--10; \citealt{Belfiore2017}).

\section{Discussion}\label{sec:disc}

\subsection{Driving mechanisms of the observed gradients}
We find flat gradients within the uncertainties for the sources in \yd4, \bdf, and \cosmos analysed in this work.
As introduced in Sect. \ref{sec:intro} and sketched in Fig.~\ref{fig:cartoon}, negative (radially decreasing) gradients are expected as a consequence of the inside-out galaxy formation scenario, while inverted (radially increasing) gradients may arise as a consequence of pristine gas accretion towards the central regions.
Possible explanations for flat radial gradients are radial gas mixing processes, occurring as a consequence of mergers, SN-driven winds and/or large-scale gas circulation (galactic fountains, i.e. metal-loaded outflows expelled from the galaxy and re-accreted in the external regions), 
which redistribute the gas in the galaxy and wash out any pre-existing radial gradient of metallicity.

All the targets in this work show multiple sources within a few kpc and disturbed morphologies. This indicates that they are experiencing, or have experienced, interactions and galaxy merging processes, which are expected to be more frequent in the early Universe according to cosmological simulations \citep[see e.g.][]{Kohandel2020, PallottiniFerrara2023}. Specifically, galaxies at $z$ $\sim$ 6 with stellar masses $\gtrsim$ 5$\times$10$^8$ $M_\odot$ living in dense environments, similar to those studied in this work (see Table~\ref{tab:integr_meas}), are expected to have already experienced multiple merger events \citep{Gelli2020}.
On the other hand, while in the low-spectral resolution ($R$ $\sim$ 100) NIRSpec PRISM data analysed in this work the emission lines are spectrally unresolved, \oiii and \hb do not show evidence for asymmetric wings or any complex line profile indicative of outflows in the higher resolution ($R$ $\sim$ 2700) NIRSpec IFU or MSA data from \cite{Hashimoto2023} and \cite{Morishita2023}, respectively, for the case of YD4, YD1, and s1 in the \yd4 system (YD4 having both IFU and MSA $R$ $\sim$ 2700 data).
Therefore, this indicates that strong outflows are not occurring at present in this system.
However, in principle we cannot exclude that the redistribution of metals across the galaxy may also be the result of past intense SN winds mixing the ISM or re-accretion in the outer regions of metal-enriched gas expelled by past galactic outflows, whose trace is absent in present-day spectra.
All in all, mergers and/or possibly either past galactic outflows or SN wind mixing seem the most likely mechanism driving the observed flat metallicity gradients. 

We stress that, even if we did our best in extracting the metallicity gradient for each separate spatial component, complications arise because in some cases these sub-sources are very close to each other and may contaminate each other's metallicity (as appears to be the case of YD1 and YD1-E; Fig. \ref{fig:yd4_maps}) or there is confusion as to which source the detected gas belongs (as in the case of \cosmos-a and \cosmos-b, where gas is located between the two sources). Moreover, the sources are not settled in radially symmetric metallicity distributions. All these aspects can contribute to yielding a flat gradient.
Nevertheless, the spatial vicinity of the sources and the displacement of gas and stars are the result of the ongoing merging processes. Therefore, ultimately, mergers appear to be the most likely cause for the flatness of the metallicity gradients.

\subsection{Metallicity gradients across cosmic time}

\begin{figure*}
    \centering
    \includegraphics[width=\textwidth]{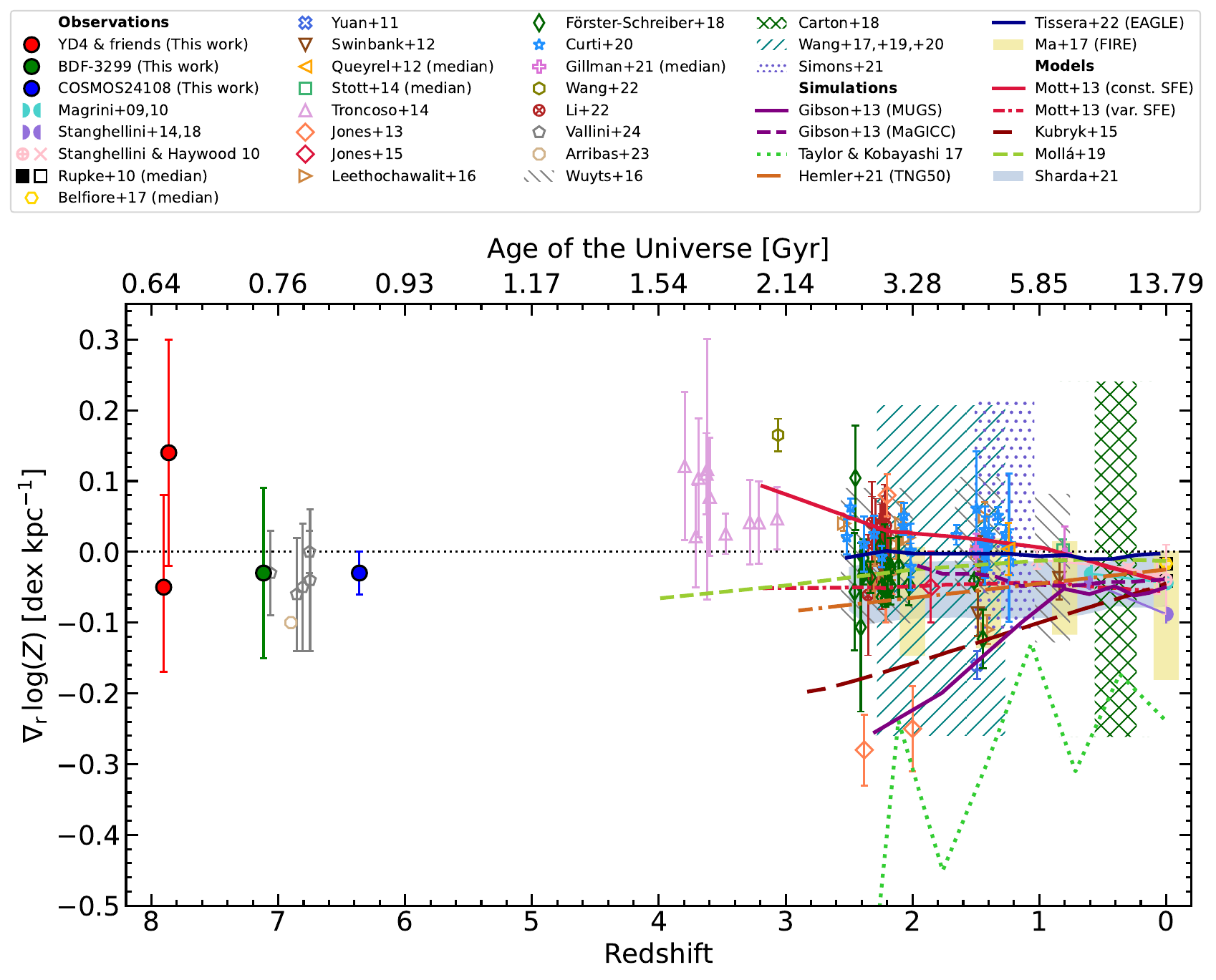}
    \caption{Gas-phase metallicity gradient (in dex~kpc$^{-1}$) as a function of redshift (bottom $x$ axis) and cosmic time (upper $x$ axis). The targets analysed in this work are marked with red (\yd4 and YD1), green (\bdf), and blue (\cosmos). The other symbols are a compilation of gas-phase metallicity gradients from literature. Specifically, at $z$ $\sim$ 0, we report the measurements from \citet[full and empty squares for isolated and merging  galaxies, respectively]{Rupke2010c} and \cite[MaNGA, median]{Belfiore2017}.
    The evolution of metallicity gradients with cosmic time for the Milky Way, M33, and M81 from H II regions ($z$ $\sim$ 0) and PNe (up to a few Gyr ago), from \cite{Magrini2009, Magrini2010}, \cite{Stanghellini2010, Stanghellini2018}, and \cite{Stanghellini2014} is shown.
    We report the gradients estimated at various $z$ from \citet{Yuan2011}, \citet[HiZELS]{Swinbank2012}, \citet[MASSIV, median]{Queyrel2012}, \citet[KMOS-HiZELS, median]{Stott2014}, \citet[AMAZE + LSD]{Troncoso2014}, \citet[GLASS]{Jones2013, Jones2015}, \citet[CASSOWARY]{Leethochawalit2016}, \citet[KMOS$^\mathrm{3D}$]{Wuyts2016}, \citet{Carton2018}, \citet[GLASS + GLASS JWST]{Wang2017, Wang2019, Wang2020, Wang2022}, \citet[SINS/zC-SINF]{Forster-Schreiber2018}, \citet[KLEVER]{Curti2020klever}, \citet[KROSS + KGES, median]{Gillman2021}, \citet[CLEAR]{Simons2021}, \cite[MAMMOTH-Grism]{Li2022}, \citet{Vallini2024a}, and \citet[GA-NIFS]{Arribas2023}. Hatched regions are reported in case of large samples. 
    The predictions for the evolution of metallicity gradients from cosmological simulations are also shown, specifically for MUGS (enhanced feedback) and MaGICC (normal feedback) from \cite{Gibson2013}, FIRE \citep{Ma2017}, \cite{Taylor2017}, TNG50 \citep{Hemler2021}, and EAGLE \citep{Tissera2022}, together with the predicted evolution from the chemical evolution models from \citet[for both constant and variable SFE]{Mott2013}, \cite{Kubryk2015}, \cite{Molla2019}, and \cite{Sharda2021}.
    \label{fig:metgrad_vs_z}
    }
\end{figure*}

In Fig. \ref{fig:metgrad_vs_z}, we report the metallicity gradients from a compilation of observational studies, from $z$ = 0 up to $z$ $\sim$ 8 (this study), namely the highest redshift probed so far. 
We report the gradients in dex~kpc$^{-1}$, given that most values reported in literature are in these units rather then in dex~$R_\mathrm{e}^{-1}$.
The flat gradients within the uncertainties that we find at $z$ $\sim$ 6--8 are compatible with the other currently available measured gradients at these redshifts, namely \cite{Vallini2024a} from FIR lines with ALMA and \cite{Arribas2023} from rest-frame optical lines with JWST/NIRSpec. 
Specifically, the results from \cite{Vallini2024a} showcase the potential for synergies between spatially resolved observations with JWST and ALMA, though the spatial resolutions may be quite different ($\sigma$ $\sim$ 0.2--0.4 kpc in our case versus $\sim$ 1.5 kpc in the case of \citealt{Vallini2024a}).

We also report the predictions for the evolution of metallicity gradients with redshift from cosmological simulations.
Some of them predict steeper negative gradients with increasing redshift (MUGS, with normal feedback, \citealt{Gibson2013}; TNG50, \citealt{Hemler2021}), in some cases extremely steep \citep[for which the case of a $\log (M_*/M_\odot)$ = 12 galaxy at the center of a cluster with no AGN is shown]{Taylor2017}, while others predict median flatter gradients at higher $z$ though with increasing scatter (MAGICC, with enhanced feedback, \citealt{Gibson2013}; FIRE, \citealt{Ma2017}; EAGLE, \citealt{Tissera2022}).
Predictions from chemical evolution models, also displayed in Fig.~\ref{fig:metgrad_vs_z}, are very heterogeneous as well, with little to no evolution with redshift (\citealt{Mott2013} for the Milky Way, with variable SFE, and \citealt{Sharda2021}, for which the $\log (M_*/M_\odot)$ = 11.1 case is shown) and slightly to steeply increasing gradients with $z$ (\citealt{Molla2019} and \citealt{Kubryk2015}, respectively, both for the Milky Way), as well as gradients inverting at z $\gtrsim$ 1 (\citealt{Mott2013} with constant SFE).

Unfortunately, predictions for the cosmic evolution of metallicity gradients from cosmological simulations and models are limited to $z$ $\lesssim$ 3--4 so far. 
If we assumed that the trend with redshift follows a simple linear extrapolation to z $\sim$ 8 of the negative trends from TNG50 \citep{Hemler2021} and MUGS \citep{Gibson2013}, as well as the model by \cite{Kubryk2015},  we would get metallicity gradients 
that are incompatible with our estimates considering the uncertainties (Fig. \ref{fig:metgrad_vs_z}), which would instead be more in line with flat gradient predictions at high $z$ (MaGICC, \citealt{Gibson2013}; FIRE, \citealt{Ma2017}; EAGLE, \citealt{Tissera2022}; \citealt{Mott2013} with variable SFE; \citealt{Sharda2021}).
However, this very rough linear extrapolation is most likely wrong, given that metals are expected to start forming from uniformly distributed, pristine gas and therefore the gradients should converge to zero at some point in the past.
The only case which extends beyond $z$ $\sim$ 3--4 is the simulations of \cite{Taylor2017}, which reach up to $z$ $\sim$ 6, where gradients even steeper than --1.5 dex~kpc$^{-1}$ (not shown in the figure to avoid an excessive shrink of the $y$ axis) are predicted, way steeper than those observed at z $\sim$ 6--8; however, even in this case, the redshifts up to 8 probed by the current observations are not explored.

All in all, we cannot draw any clear conclusion on what cosmological simulations and models best reproduce the observed gradients at high $z$. Instead, we stress the need for updated metallicity gradient predictions from cosmological simulations and chemical models at $z$ $\gtrsim$ 3 in order to match the redshifts reached by current observations.


We point out that the merger fraction increases with redshift (e.g. theoretically \citealt{Fakhouri2010, RodriguezGomez2017, OLeary2021}; observationally \citealt{Duncan2019}), therefore galaxies at high redshift often have disturbed morphologies due to interactions and do not have radially symmetric gas and chemical distributions (see e.g. Figs. \ref{fig:yd4_maps}, \ref{fig:bdf_maps}, and \ref{fig:cosmos_maps}; also e.g. observations in \citealt{DeGraaf2023, Arribas2023, Rodriguez-del-Pino2023} and simulations in \citealt{Pallottini2017,Pallottini2019}).
Specifically, the metallicity maps in our work show in some cases non-azimuthally symmetric structures which are averaged out when producing radial gradients (as it is e.g. the case of YD1, having the highest values to the W and the lowest to the E of its spatial emission peak).
Therefore, radial gradients may not be the optimal means to describe the distribution of metals in high-$z$ galaxies and other alternative methods to quantify it (such as the metallicity scatter in the galaxy, as suggested by \citealt{MaiolinoMannucci2019}) should be sought and included in predictions from simulations. 

In summary, the flat radial gradients and the asymmetric spatial distribution of metallicity observed in some cases seem to support the merging scenario, though ISM mixing due to past SN winds or re-accretion in the outer regions of metal-enriched gas from past galactic outflows (galactic fountains) are also a viable mechanism to explain the flatness of the gradients.
Spatially asymmetric low-metallicity accretion from the circumgalactic medium (CGM) and intergalactic medium (IGM) could also contribute to produce non-azimuthally symmetric metallicities as suggested by \cite{Arribas2023}.

\section{Conclusions}

In this work, we have presented new JWST NIRSpec IFU observations at low spectral resolution ($R$\,$\sim$\,100; PRISM/CLEAR) of three high-$z$ systems, namely \yd4 in the proto-cluster A2744-z7p9OD ($z$~$\sim$~7.88), \bdf ($z$~$\sim$~7.11), and \cosmos ($z$~$\sim$~6.36). At these redshifts, the NIRSpec PRISM spectra (spanning $\sim$0.6--5.2 $\mu$m) cover the rest-frame UV and optical spectral ranges and include the main optical emission lines from warm ($T$~$\sim$~$10^4$~K) ionised gas, \oii, \hg, \hb, \oiii, and, for \cosmos, also \ha and \sii. We modelled the ionised gas lines in the spectra with Gaussian functions and mapped their emission.
The main goal of this work is to study the spatially resolved gas-phase metallicity and investigate the shape of metallicity gradients at high redshift ($z$ $\gtrsim$ 6).

The targets have very disturbed morphologies in both the stellar continuum and ionised gas line emission. We identify several spatial components concentrated within a few kiloparsecs in all the three systems.
We obtained the main integrated properties of each spatial component in each target, specifically stellar mass from SED fitting, SFR from \ha or \hb, and gas-phase metallicity by means of flux ratios of rest-frame optical emission lines which we modelled with Gaussian functions.
We found log($M_*/M_\odot$) $\sim$ 7.6--9.3, SFRs $\sim$ 1--15 $M_\odot$~yr$^{-1}$, and gas-phase metallicities 12+log(O/H) $\sim$ 7.7--8.3 by extracting integrated spectra from circular apertures centred on each spatial component. 

We compared the stellar masses and gas-phase metallicities of the targets with the mass-metallicity relations (MZRs) inferred observationally at different redshifts and with those predicted by cosmological simulations.
In general, the sources in the systems studied in this work are consistent with the distribution of the highest-$z$ galaxies to date ($z$ $\sim$ 3--10) from JADES, CEERS, and EROs samples in the mass-metallicity plane. Nevertheless, most of the sources in \yd4 and \cosmos lie at the upper end of this $z$ $\sim$ 3--10 distribution, in terms of metallicity at a given stellar mass, and are closer to the best-fit MZRs measured at $z$ $\sim$ 2--3 (e.g. MOSDEF and GLASS-JWST surveys) than to the MZR at $z$ $\sim$ 3--10. 
Relative to the MZRs from cosmological simulations, the sources in the three systems studied in this work are in good agreement with the slopes predicted at $z$ $\sim$ 6--8 by most simulations.

We inferred the gas-phase metallicity radial gradients by extracting integrated spectra from concentric radial annuli centred on the main sources of each system for which the S/N allowed for a robust metallicity estimate. 
The gas-phase metallicity gradients are flat within the uncertainties. Flat gradients can be associated with processes which mix the gas and the metals on galaxy scales and therefore wash out any gradient which may have formed in the galaxy, such as mergers, SN wind mixing, and re-accretion of metal-loaded galactic outflows in the outer regions. 

These are among the very few measurements of spatially resolved gas metallicity at these high redshifts ($z$ $\sim$ 6--8). In particular, YD4 (at $z$ $\sim$ 8) constitutes the highest-$z$ source in which a metallicity gradient has been probed so far.

Cosmological simulations and chemical evolution models make very different predictions regarding the cosmic evolution of metallicity gradients. 
Some of them predict steeper negative gradients with increasing redshift (e.g. MUGS and TNG50), while others predict median flatter gradients at higher $z$ (e.g. MaGICC, FIRE, and EAGLE).
Unfortunately, these predictions are generally limited to $z$ $\lesssim$ 3, therefore no conclusions can be drawn on what simulations best reproduce the observed mainly flat gradients at $z$ $\sim$ 6--8 found in this work.

All in all, the results of this work in terms of the MZR and gas-phase metallicity gradients at $z$ $\sim$ 6--8 provide important constraints to guide future cosmological simulations and models. In particular, they urge for specific predictions on the cosmic evolution of metallicity gradients and metallicity maps out to redshift of 8, given that most of such predictions are currently limited to $z$ $\lesssim$ 3.

Moreover, galaxies at high $z$ tend to have more irregular morphologies and may not have azimuthally symmetric chemical distributions, as a result of frequent mergers and asymmetric low-mass gas infall.
Radial gradients, which average out any azimuthal information, may then not be the optimal means to quantify the distribution of metals in high-$z$ galaxies. Therefore, other alternative quantitative tracers of the spatially resolved metallicity should be considered as part of predictions from simulations, to be compared with observations.


\begin{acknowledgements}
We acknowledge support from European Union’s HE ERC Starting Grant No. 101040227 - WINGS (G.V., S.C.), INAF Minigrant `Reionization and fundamental cosmology with high-redshift galaxies' (M.C.), `FirstGalaxies' Advanced Grant from the European Research Council (ERC) under the European Union’s Horizon 2020 research and innovation program, Grant agreement No. 789056 (A.J.B.), ERC Advanced Grant INTERSTELLAR H2020/740120 (A.F.), grant PID2021-127718NB-I00 funded by the Spanish Ministry of Science and Innovation/State Agency of Research, MICIN/AEI/ 10.13039/501100011033 (S.A.).
This work is based on observations made with the NASA/ESA/CSA James Webb Space Telescope. The data were obtained from the Mikulski Archive for Space Telescopes at the Space Telescope Science Institute, which is operated by the Association of Universities for Research in Astronomy, Inc., under NASA contract NAS 5-03127 for JWST. These observations are associated with program \#1893.
This research has made use of NASA’s Astrophysics Data System Bibliographic Services.
This research has made use of the Python packages NumPy \citep{numpy2020}, SciPy \citep{scipy2020},  IPython \citep{ipython2007}, Matplotlib \citep{matplotlib2007}, emcee \citep{emcee:2013}, and Astropy (\url{http://www.astropy.org}), a community-developed core Python package for Astronomy \citep{astropy:2013, astropy:2018}.
This research has made use of the FITS files visualisation tool QFitsView (\url{https://www.mpe.mpg.de/~ott/QFitsView/}).
\end{acknowledgements}

\bibliographystyle{aa} 
\bibliography{bibliography_jwstmet}

\begin{appendix}


\section{Additional figures for \yd4}

\includegraphics[width=0.5\textwidth]{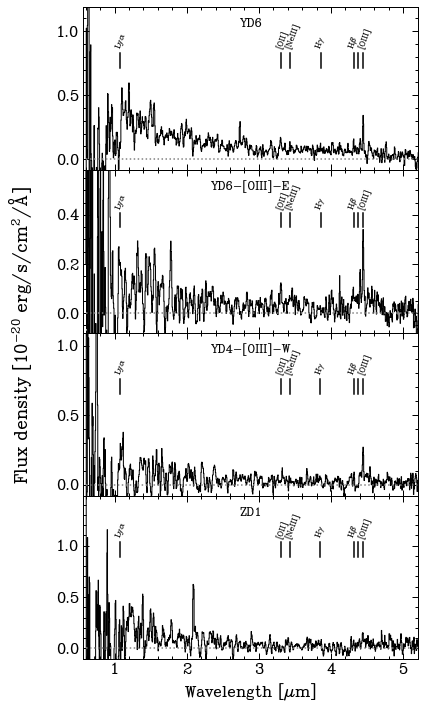}
\captionof{figure}{NIRSpec PRISM/CLEAR spectra extracted from a circular aperture with radius of 0.15\arcsec\ centred at the location of the targets YD6, YD6-\oiii-E, YD4-\oiii-W, and ZD1.} 
\vfill
\label{fig:spectra_yd4_1}
\includegraphics[width=0.5\textwidth]{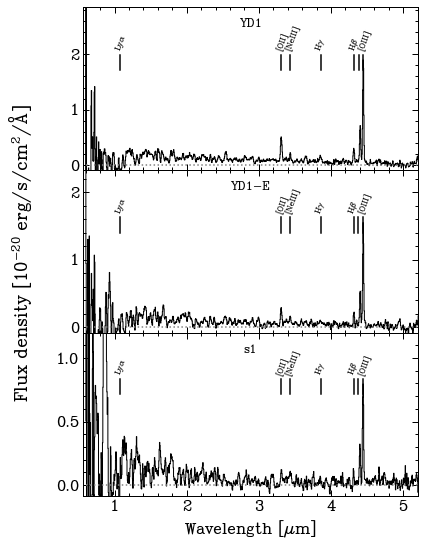}
\captionof{figure}{NIRSpec PRISM/CLEAR spectra extracted from a circular aperture with radius of 0.15\arcsec\ centred at the location of the targets YD1, YD1-E, and s1.} 
\label{fig:spectra_yd4_2}

\noindent\begin{minipage}{\textwidth}
    \centering
    \includegraphics[scale=0.27,trim={3cm 0 1cm 0},clip]{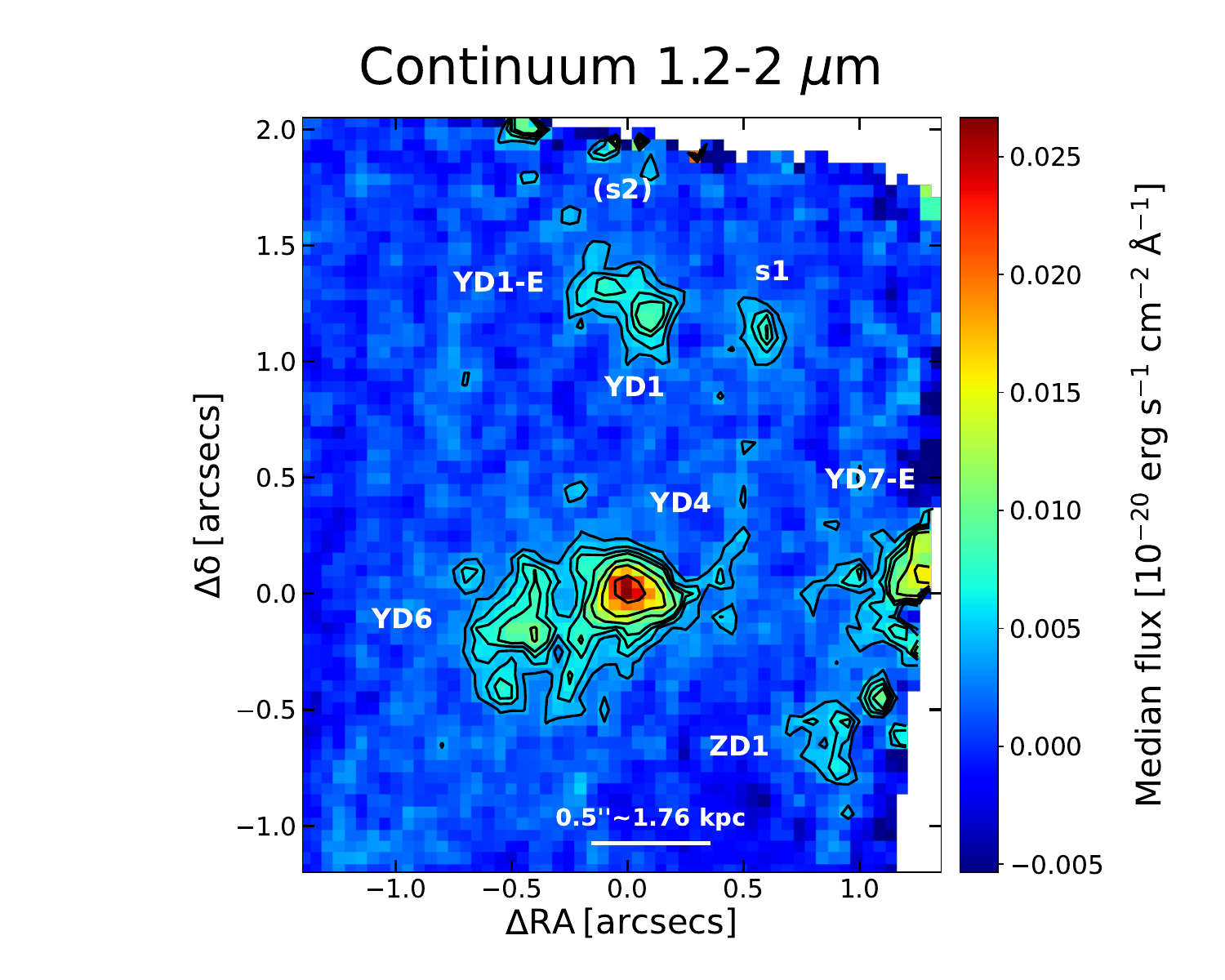}
    \includegraphics[scale=0.27,trim={3cm 0 1.5cm 0},clip]{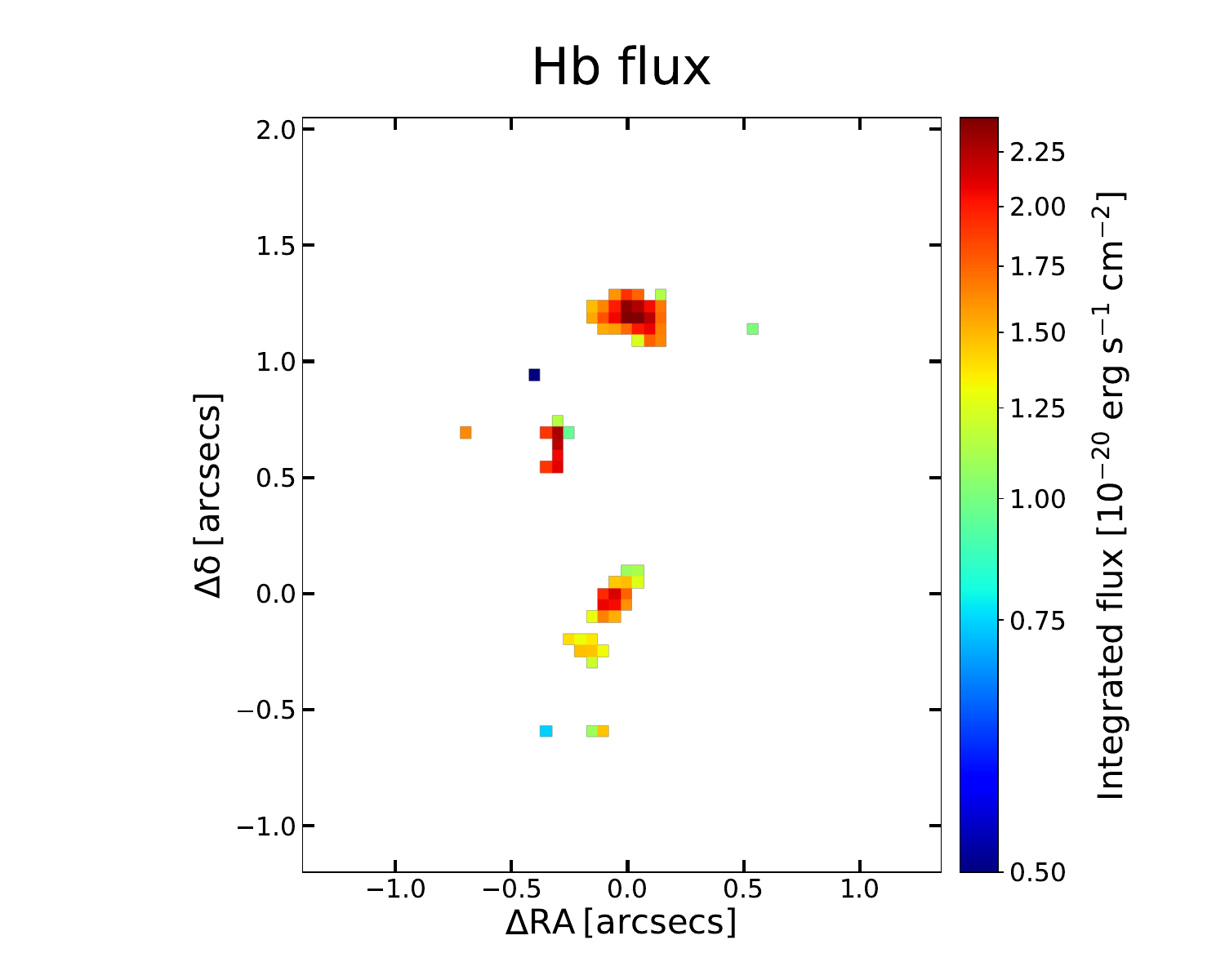}
    \includegraphics[scale=0.27,trim={3cm 0 1.5cm 0},clip]{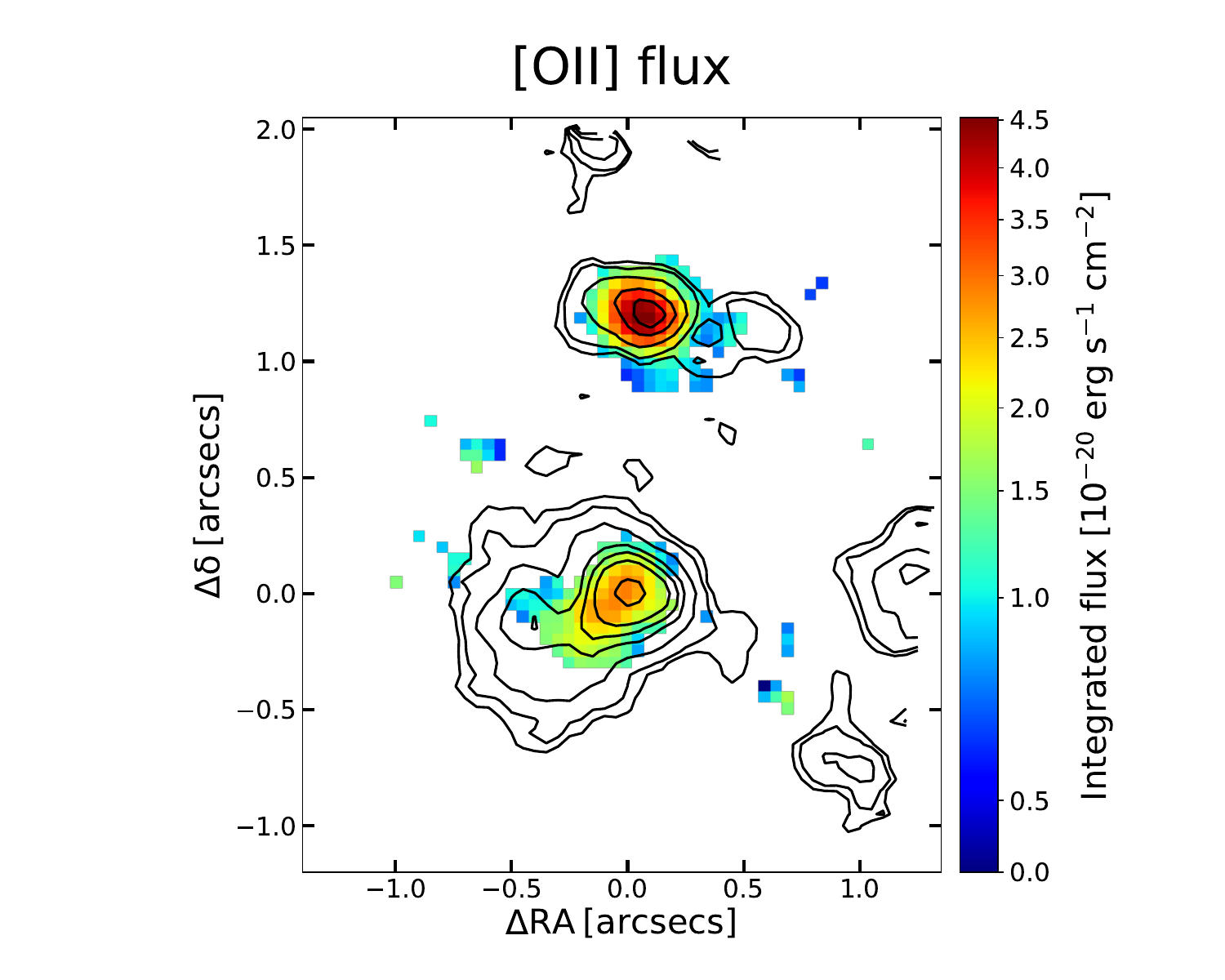}\\
    \includegraphics[scale=0.27,trim={3cm 0 1cm 0},clip]{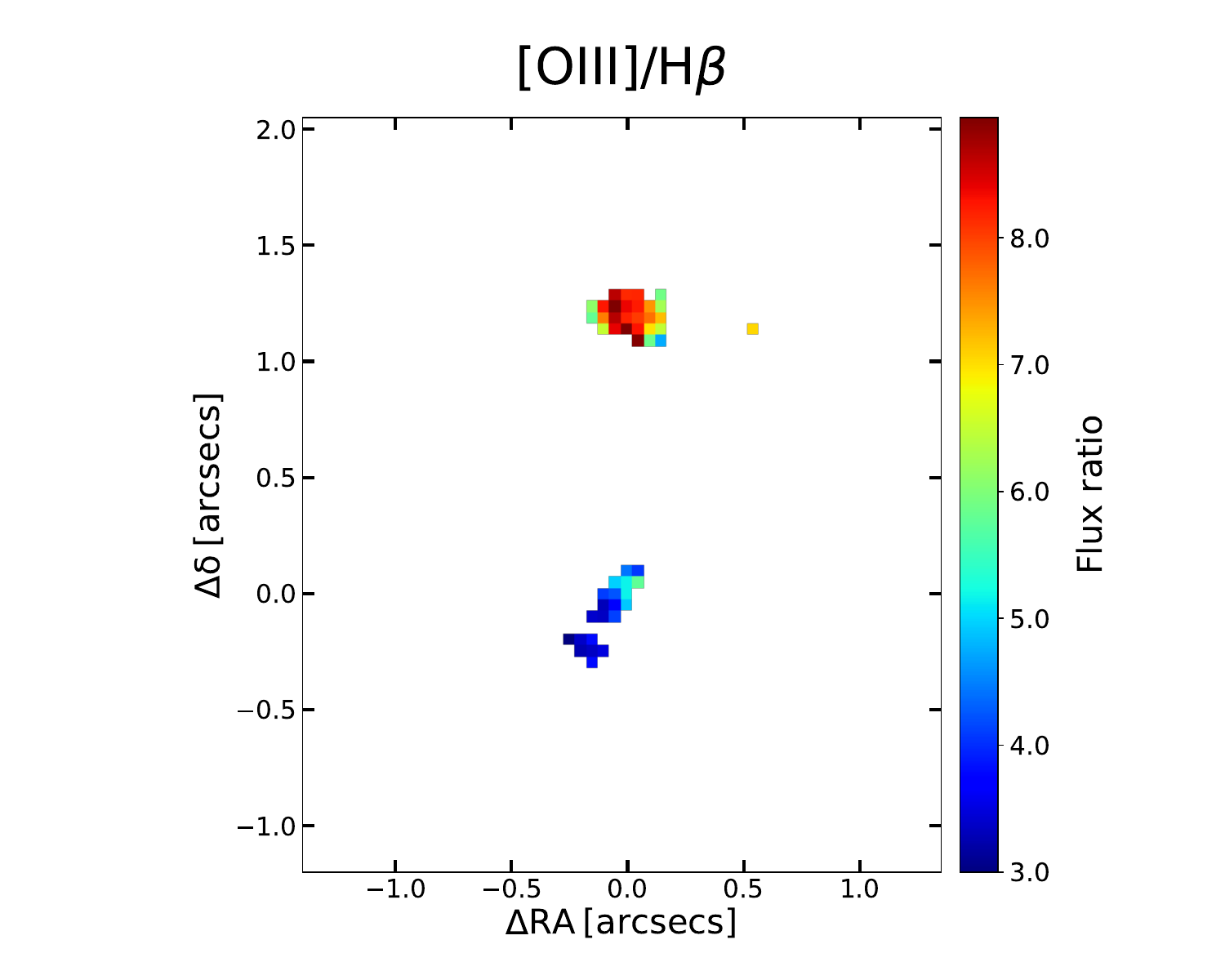}
    \includegraphics[scale=0.27,trim={3cm 0 1.5cm 0},clip]{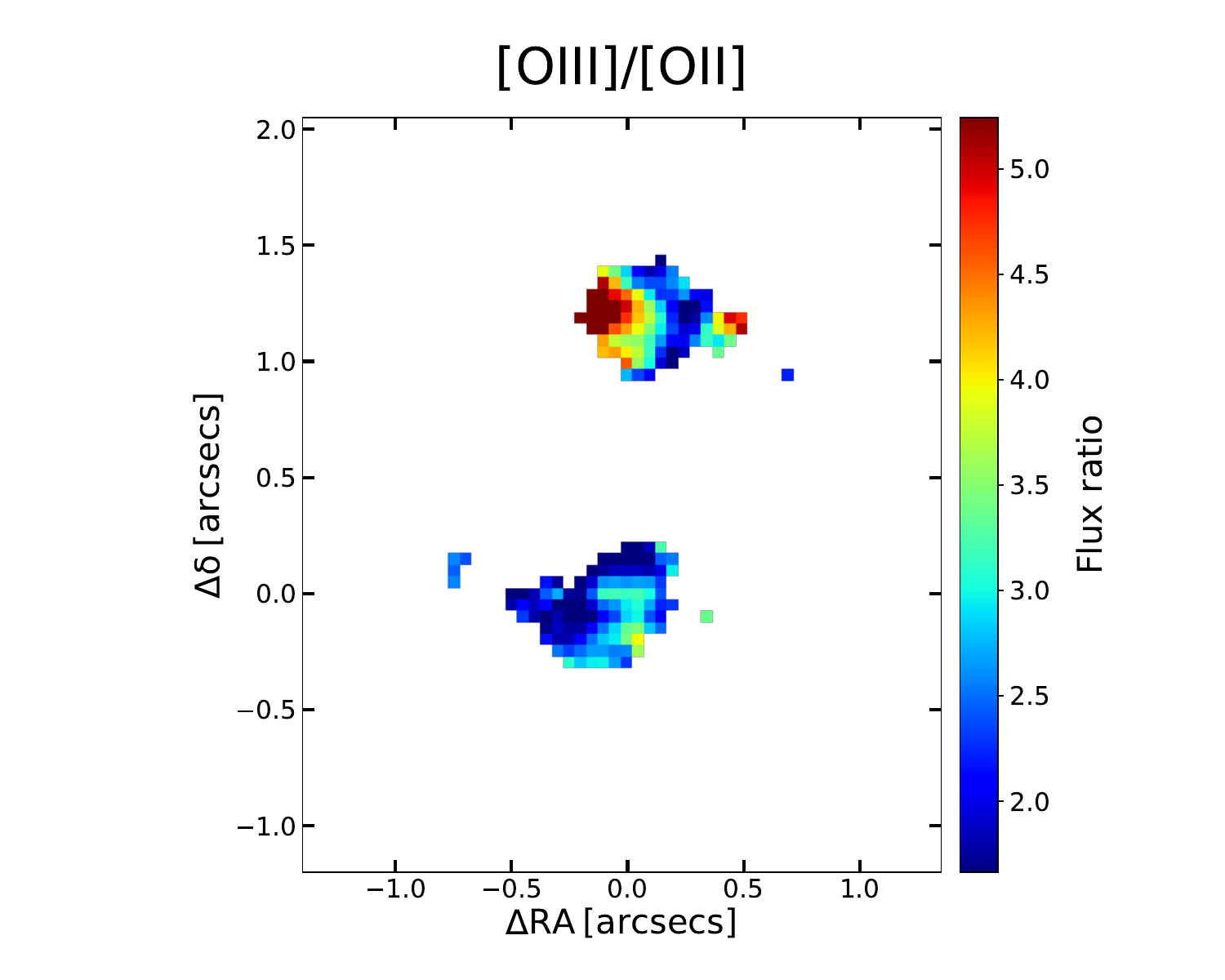}
    \captionof{figure}{Additional maps for \yd4. Highest-resolution continuum map of \yd4 from the original unsmoothed data cube in observed spectral range 1.2-2 $\mu$m, where the spatial resolution is the highest (top-left).
    All the other maps are obtained from the (wavelength-dependent) spatially smoothed data cube, as in Fig. \ref{fig:yd4_maps} (details in Sect. \ref{ssec:data_anal}). First row: \hb (centre) and \oii (right) integrated fluxes. Second row: \oiii/\hb (left) and \oiii/\oii (right) line flux ratios. Contours mark the 2--3 $\mu$m (observed-frame) continuum. 
    }
    \label{fig:yd4_maps_appdx}
\end{minipage}

\section{Additional figures for \bdf}

\includegraphics[width=0.5\textwidth]{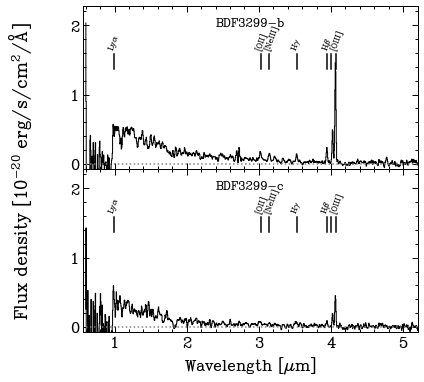}
\captionof{figure}{NIRSpec PRISM/CLEAR spectra extracted from a circular aperture with radius of 0.1\arcsec\ centred at the location of the targets \bdf-b and \bdf-c.} 
\label{fig:spectra_bdf}

\clearpage\noindent\begin{minipage}{\textwidth}
    \centering
    \includegraphics[scale=0.27,trim={1.5cm 0 1.5cm 0},clip]{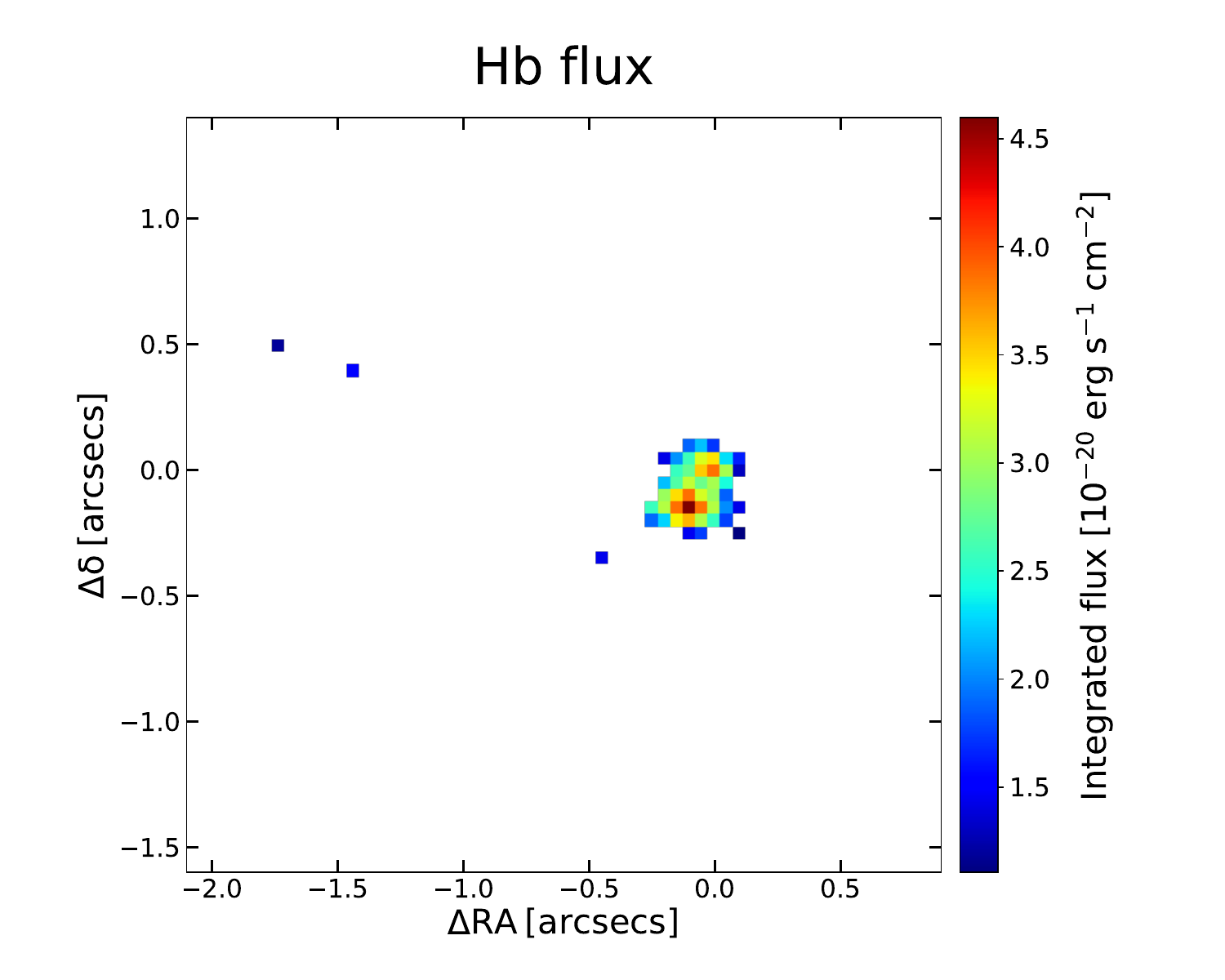}
    \includegraphics[scale=0.27,trim={1.5cm 0 1.5cm 0},clip]{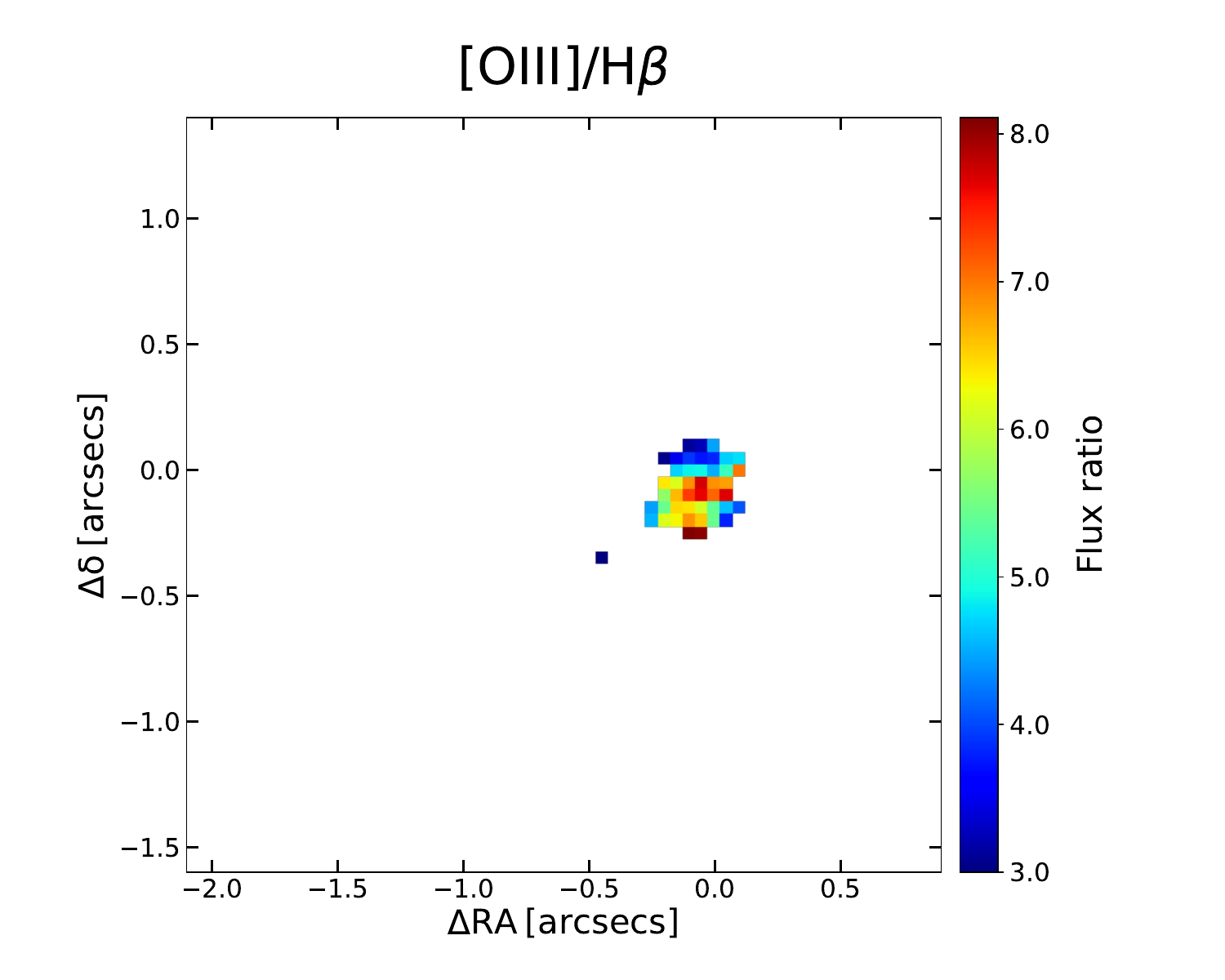}
    \includegraphics[scale=0.27,trim={1.5cm 0 1.5cm 0},clip]{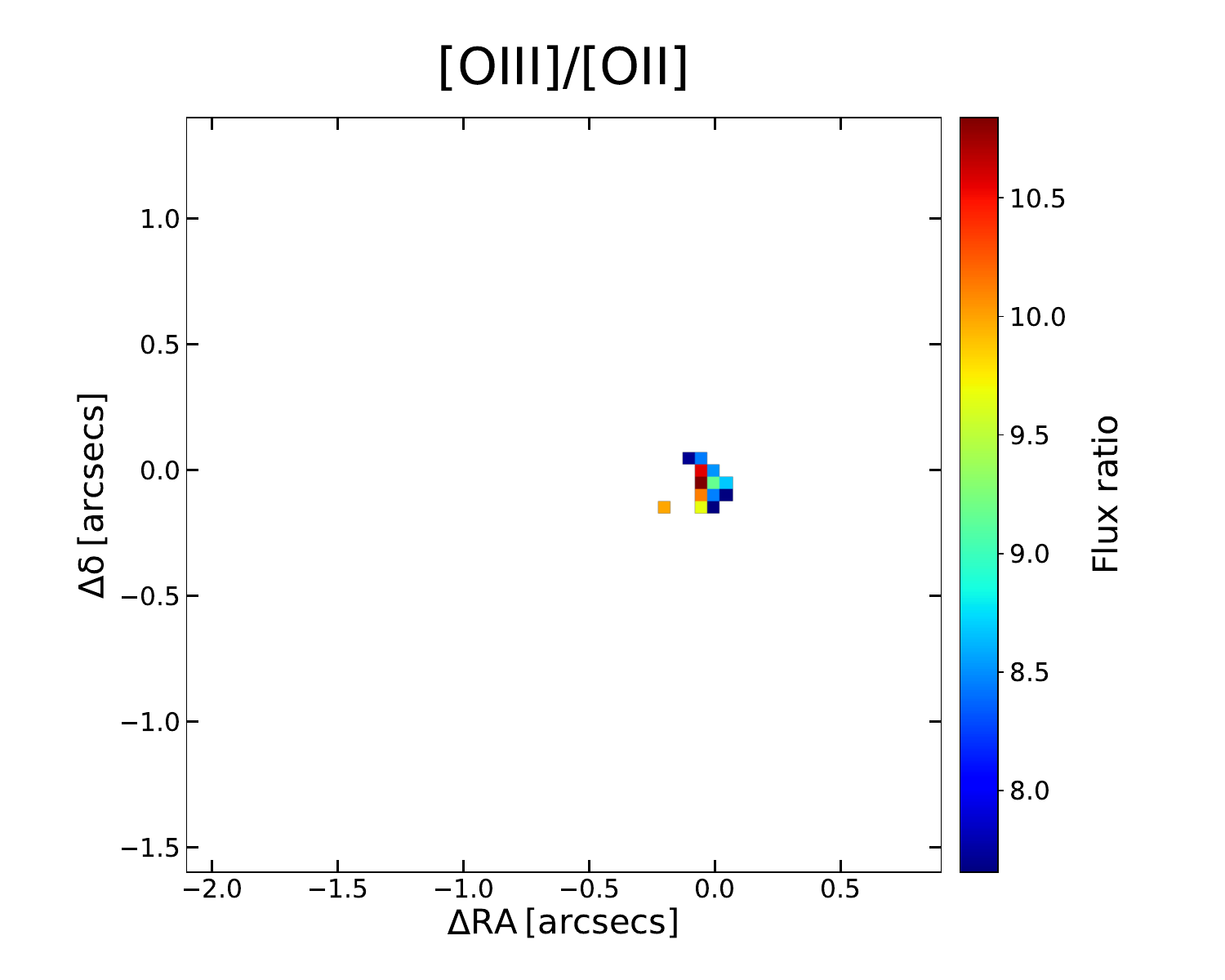}
    \captionof{figure}{Additional maps for \bdf. As in Fig. \ref{fig:bdf_maps}, the flux maps are obtained from the original data cube, while the line ratio maps from the spatially smoothed data cube (details in Sect. \ref{ssec:data_anal}). Top row: maps of \hb integrated flux (left), \oiii/\hb (centre) and \oiii/\oii (right) line flux ratios. 
    }
    \label{fig:bdf_maps_appdx}
\end{minipage}

\section{Additional figures for \cosmos}

\includegraphics[width=0.5\textwidth]{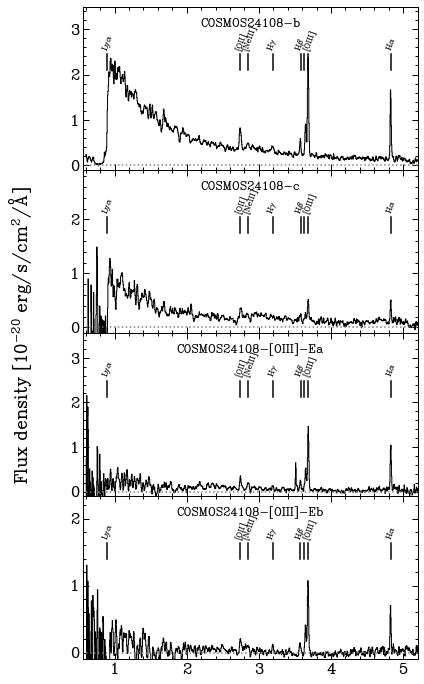}
\captionof{figure}{NIRSpec PRISM/CLEAR spectra extracted from a circular aperture with radius of 0.15\arcsec\ centred at the location of the targets \cosmos-b, \cosmos-c, \cosmos-[OIII]-Ea, and \cosmos-[OIII]-Eb.} 
\label{fig:spectra_cosmos}

\clearpage\noindent\begin{minipage}{\textwidth}
    \centering
    \includegraphics[scale=0.27,trim={1.5cm 0 1.5cm 0},clip]{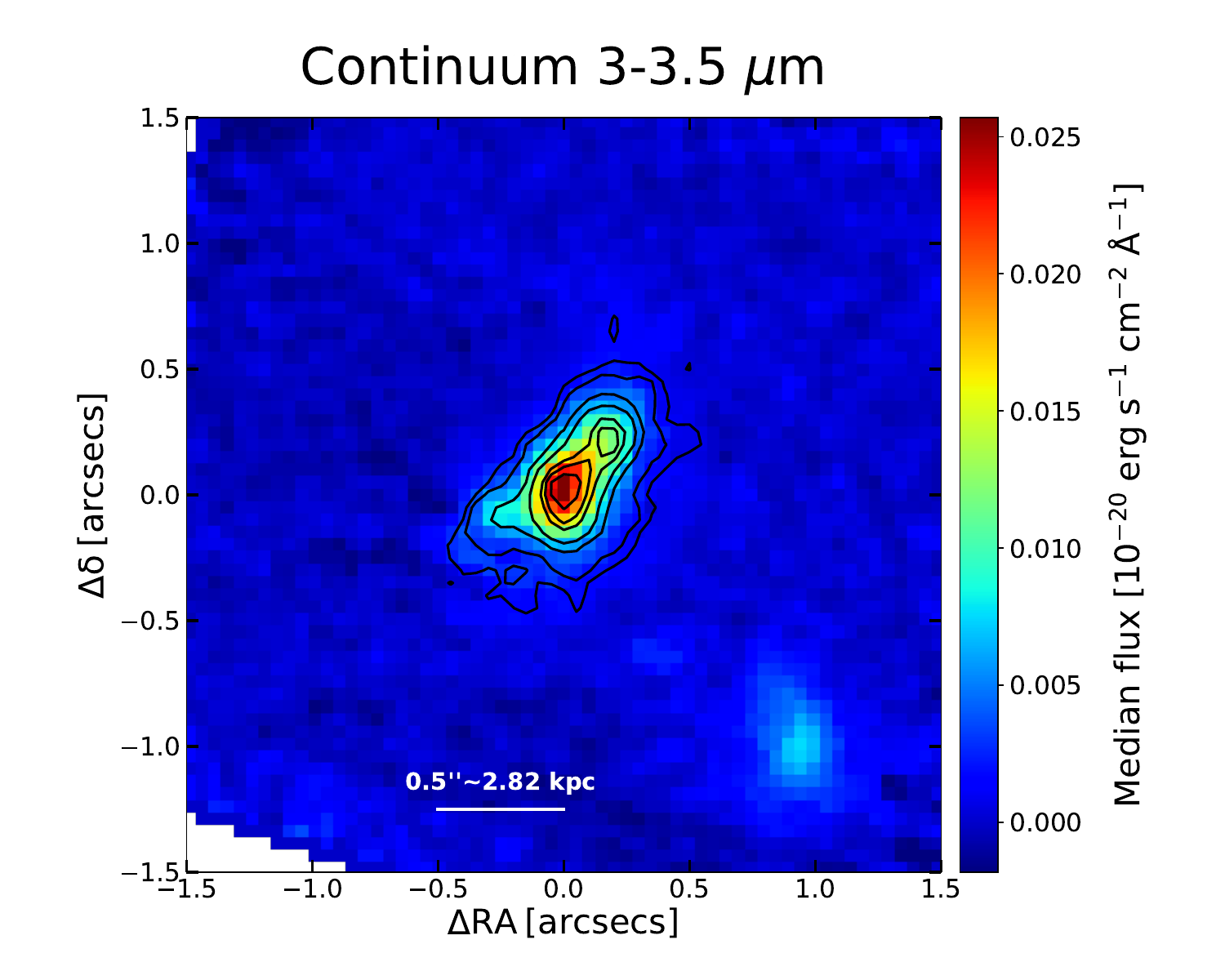}
    \includegraphics[scale=0.27,trim={1.5cm 0 1.5cm 0},clip]{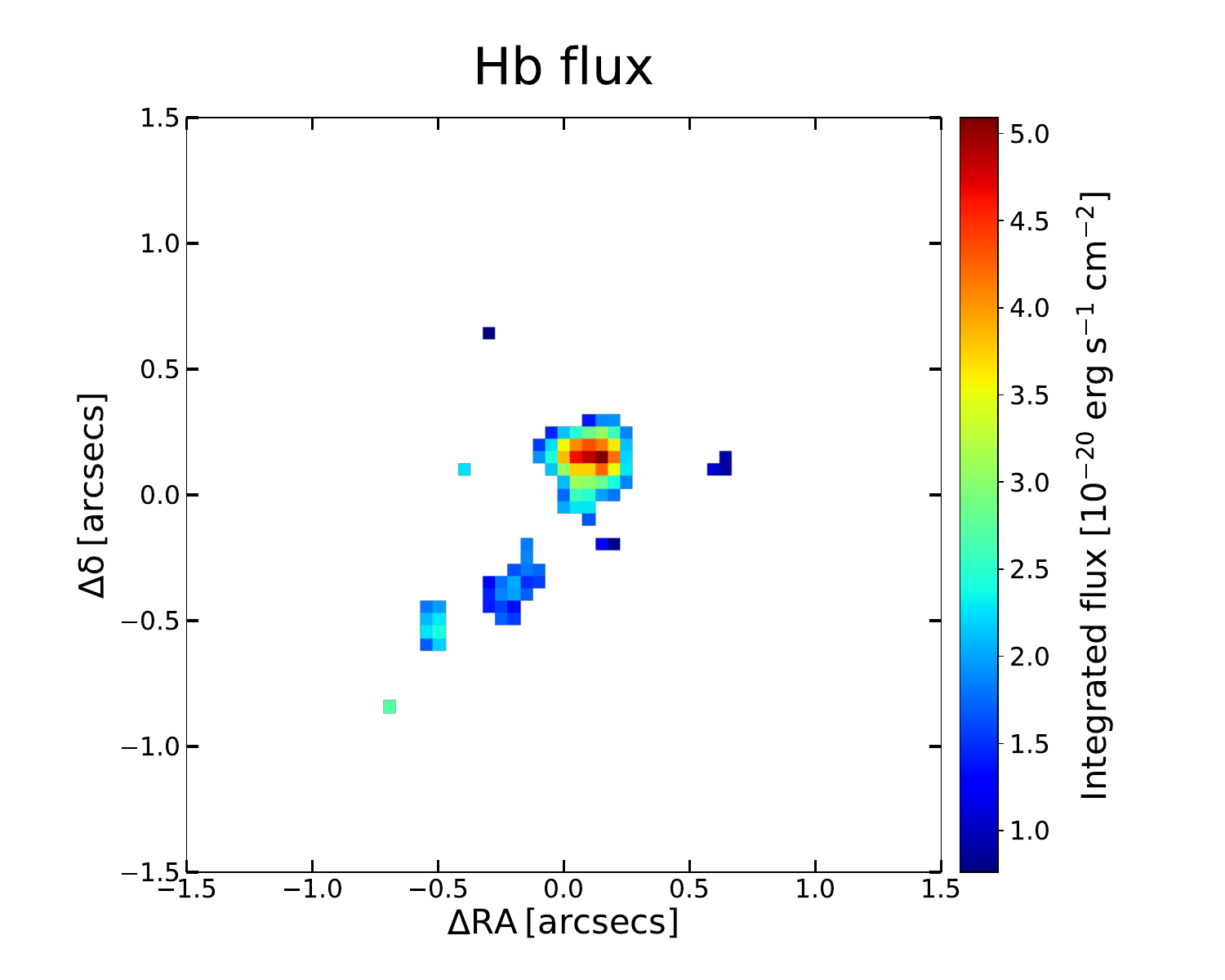}
    \includegraphics[scale=0.27,trim={1.5cm 0 1.5cm 0},clip]{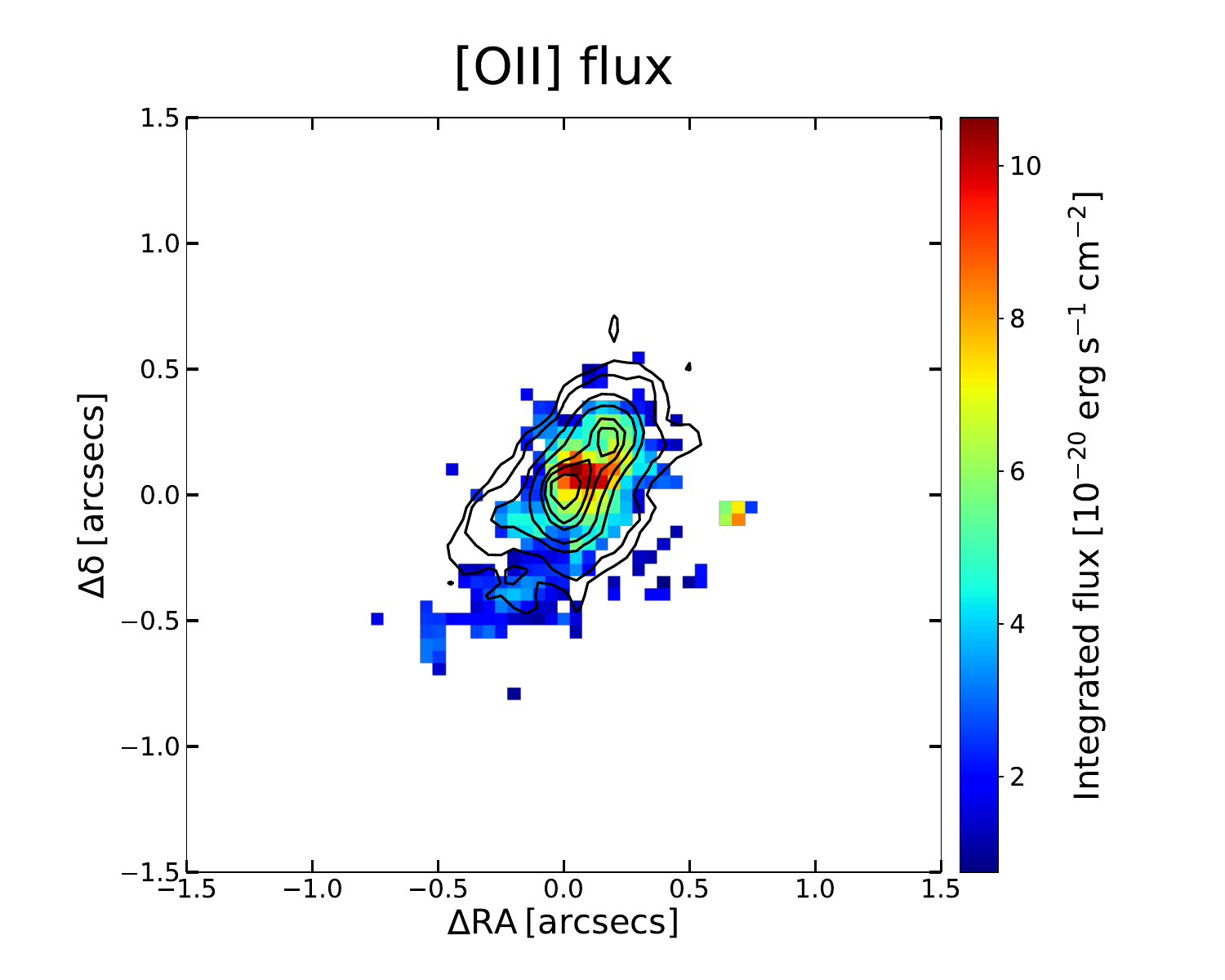}\\
    \includegraphics[scale=0.27,trim={1.5cm 0 1.5cm 0},clip]{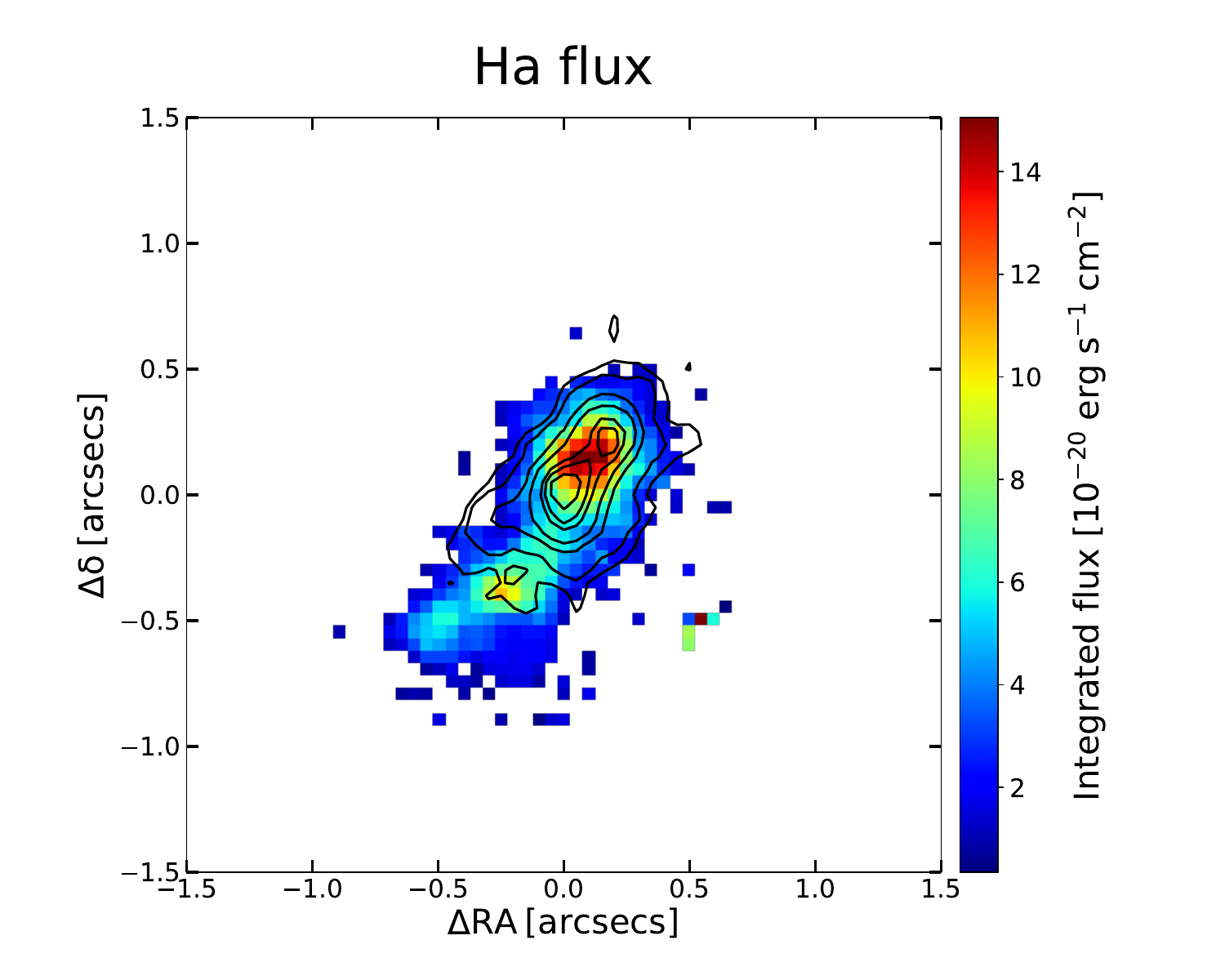}
    \includegraphics[scale=0.27,trim={1.5cm 0 1.5cm 0},clip]{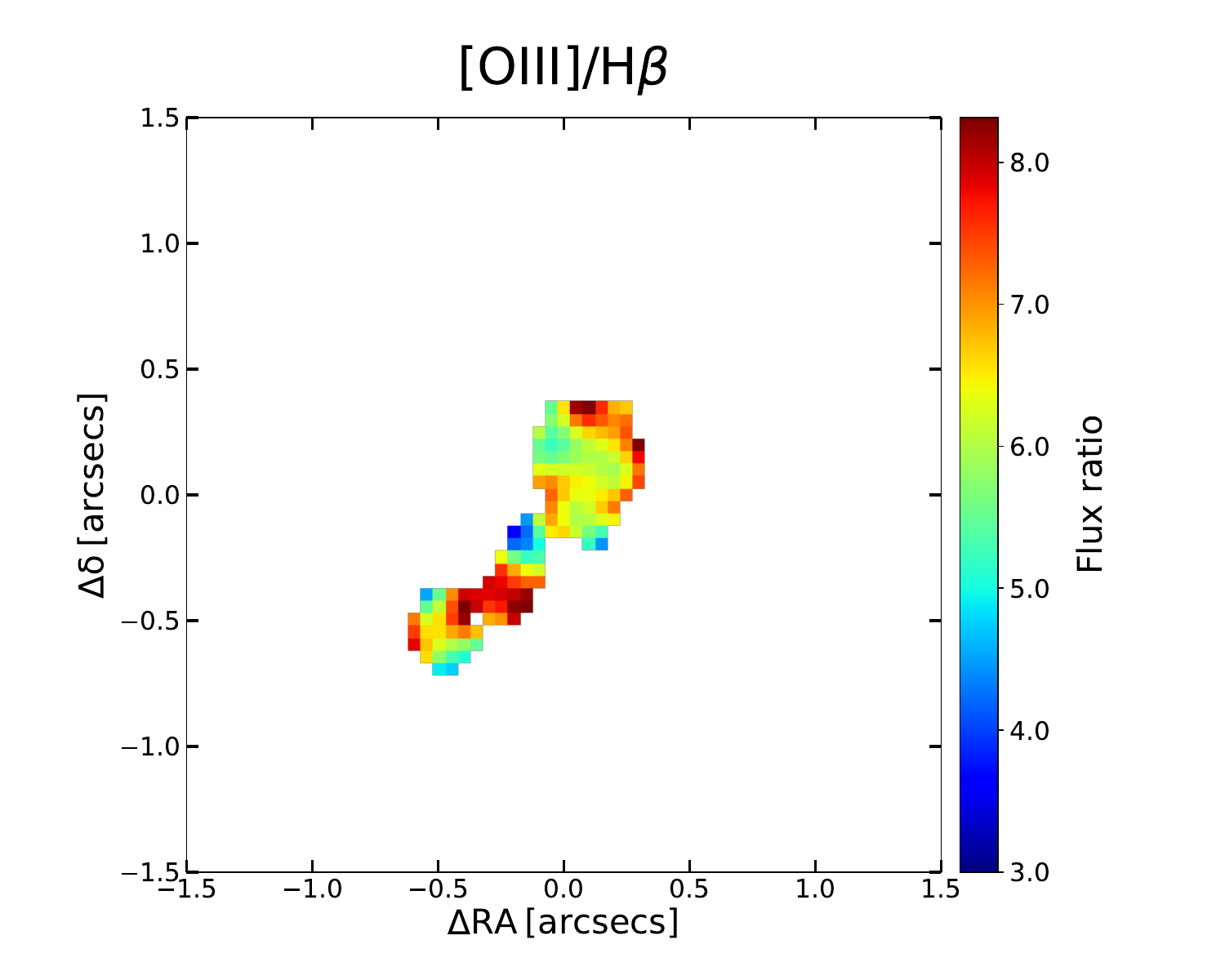}
    \includegraphics[scale=0.27,trim={1.5cm 0 1.5cm 0},clip]{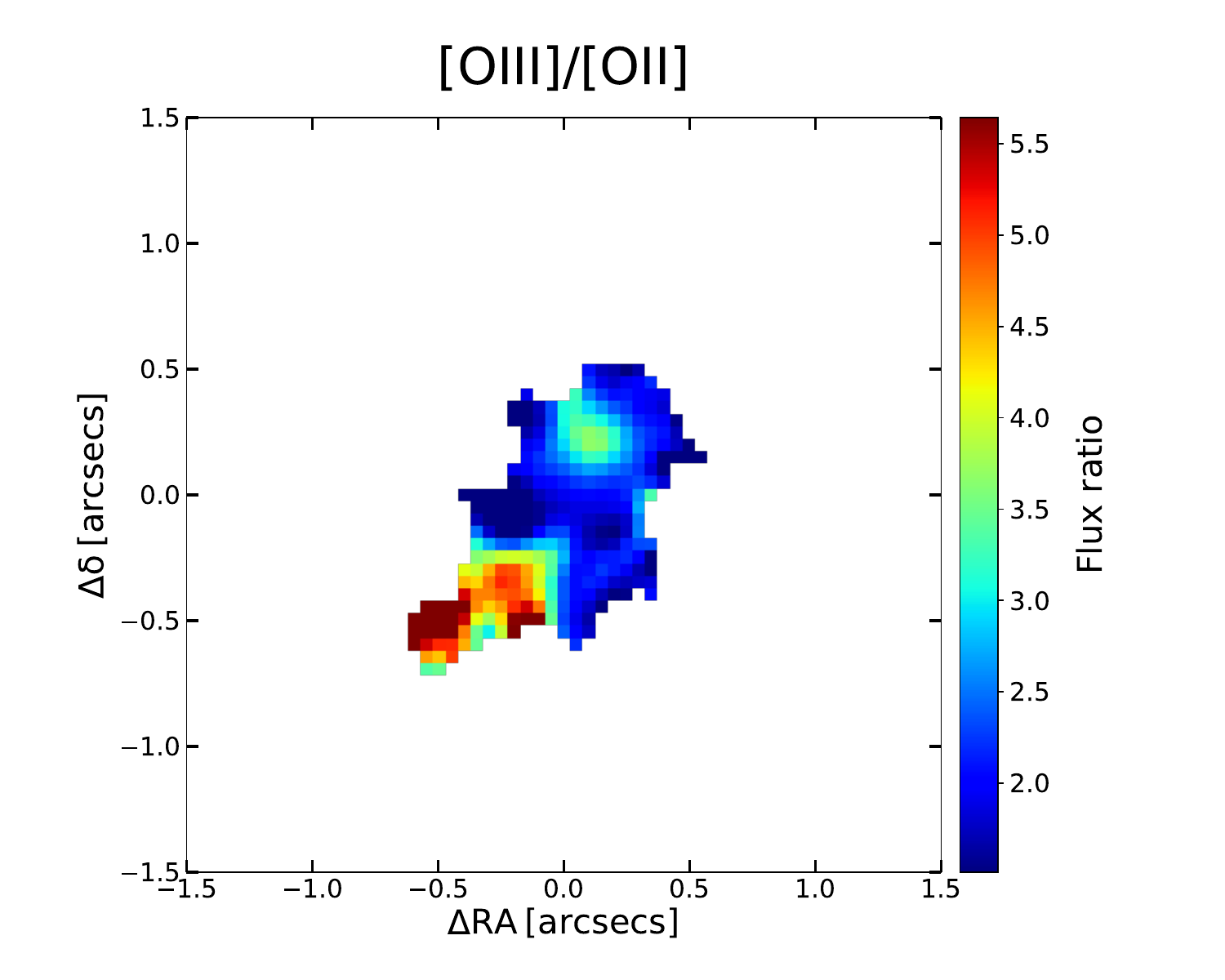}
    \captionof{figure}{Additional maps for \cosmos. As in Fig. \ref{fig:cosmos_maps}, the flux maps are obtained from the original data cube, while the line ratio maps from the spatially smoothed data cube (details in Sect. \ref{ssec:data_anal}). Top row: median continuum map in the observed spectral range 3--3.5 $\mu$m ($\sim$0.39--0.46 $\mu$m rest-frame; left), with the contours of the continuum in the observed spectral range 1--2 $\mu$m (same as in Fig. \ref{fig:cosmos_maps}, left) superimposed; \hb (centre-left), \oii (centre-right), and \ha (right) flux maps. Middle row: \oiii/\hb (left) and \oiii/\oii (centre-left) line ratio maps.
    }
    \label{fig:cosmos_maps_appdx}
\end{minipage}

\clearpage\noindent\begin{minipage}{\textwidth}
{\centering
    \includegraphics[scale=0.265,trim={1cm 0 1cm 0},clip]{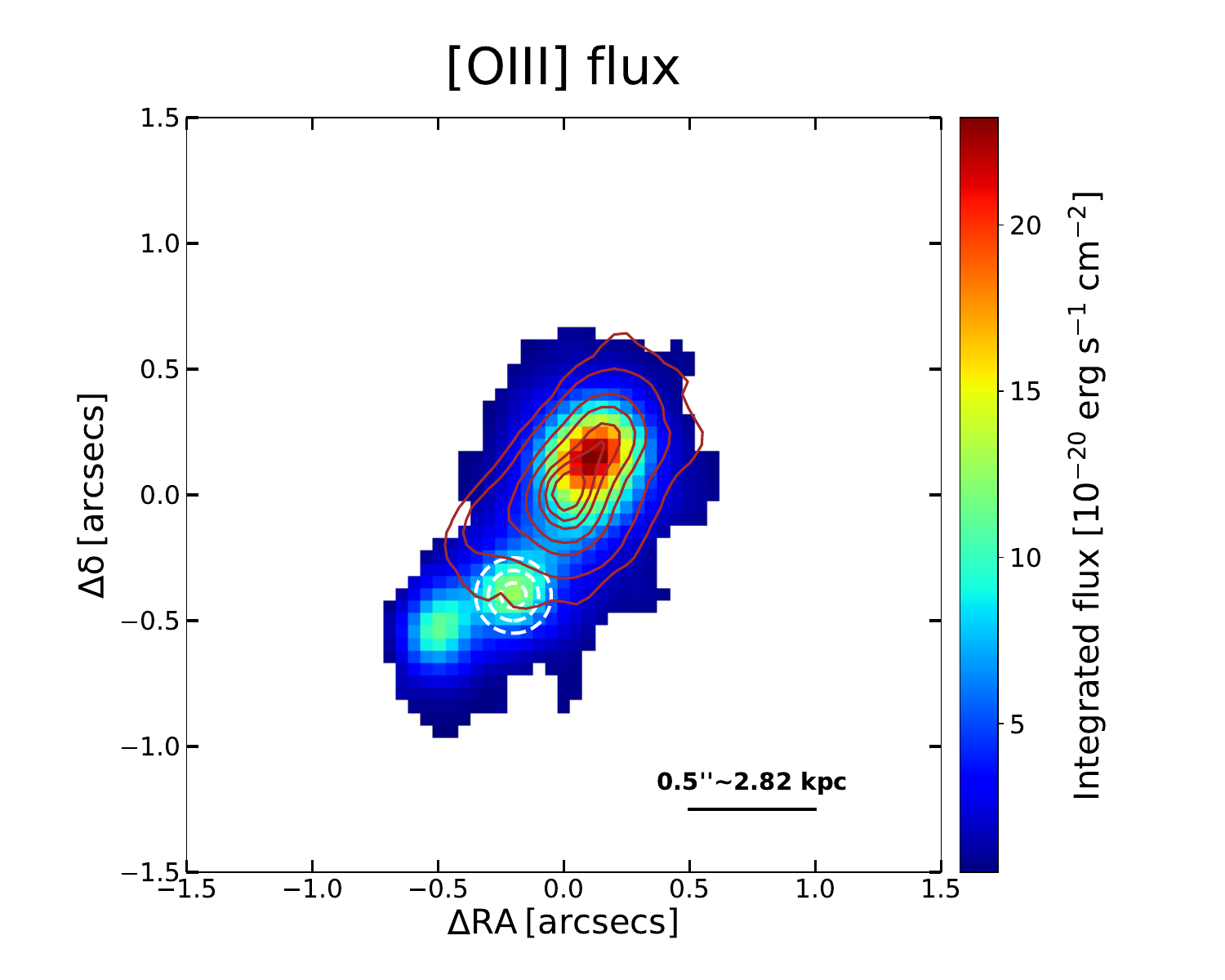}
    \includegraphics[scale=0.29,trim={1cm 0 1cm 0},clip]{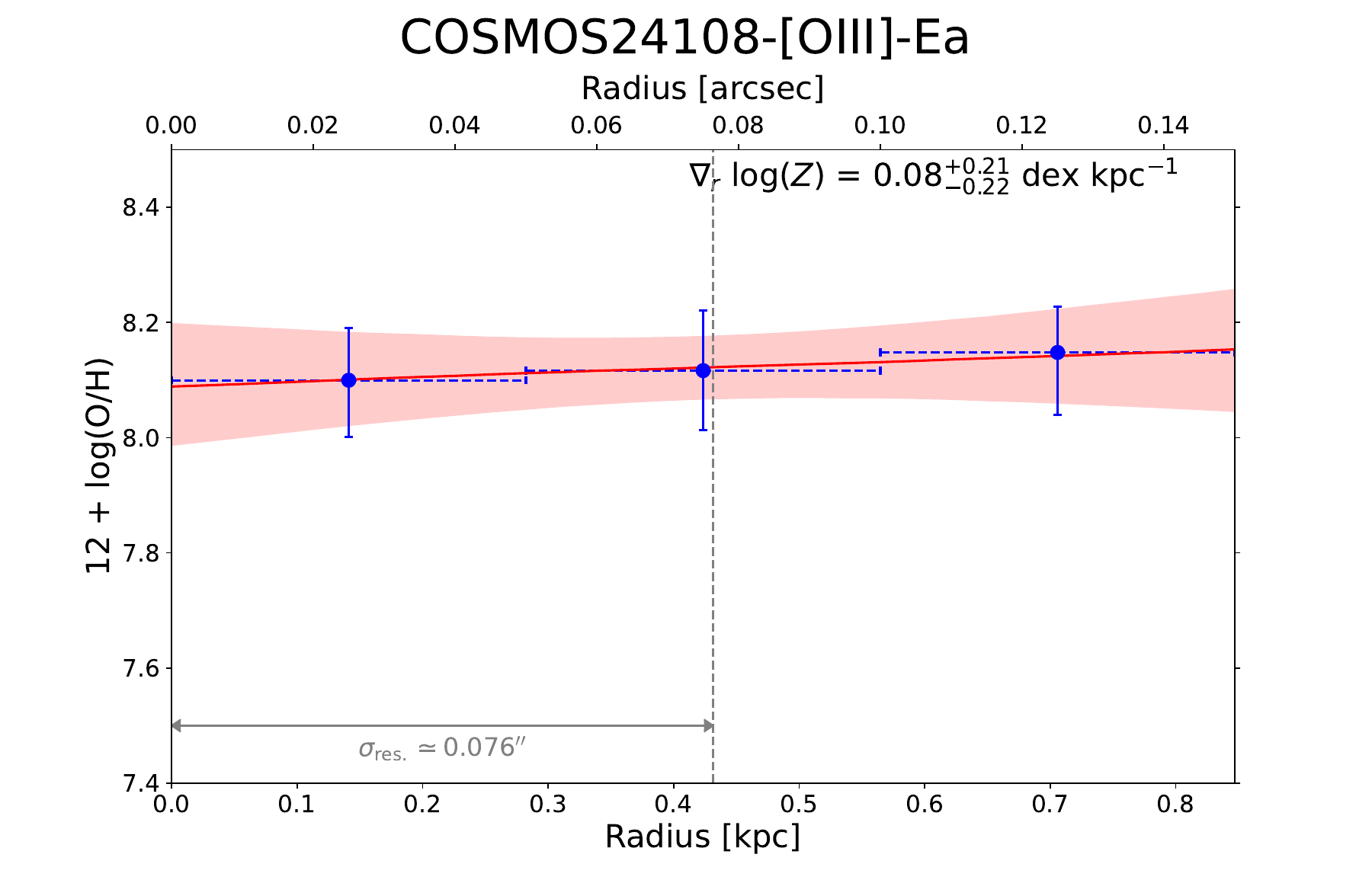}\\
    \includegraphics[scale=0.265,trim={1cm 0 1cm 0},clip]{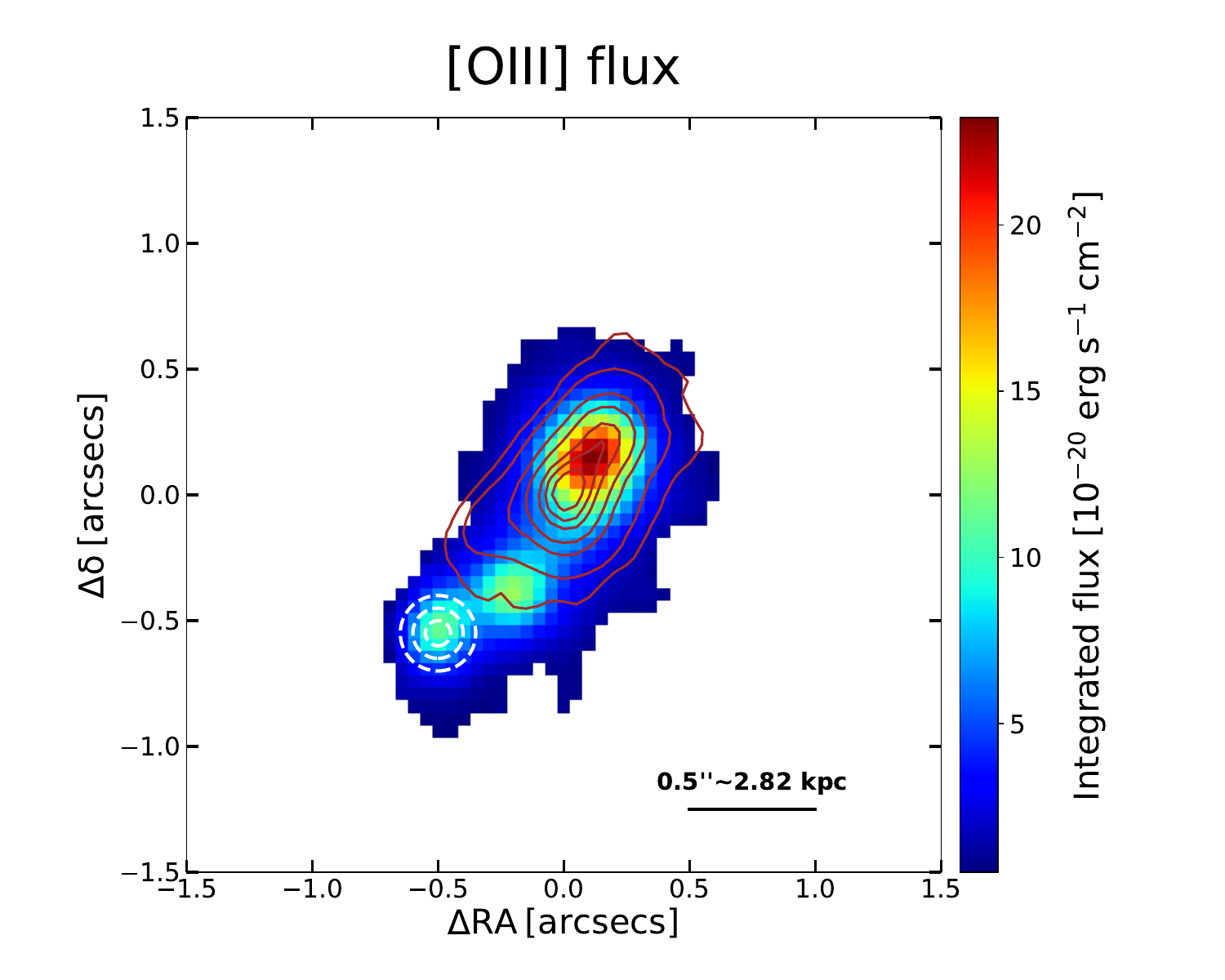}
    \includegraphics[scale=0.29,trim={1cm 0 1cm 0},clip]{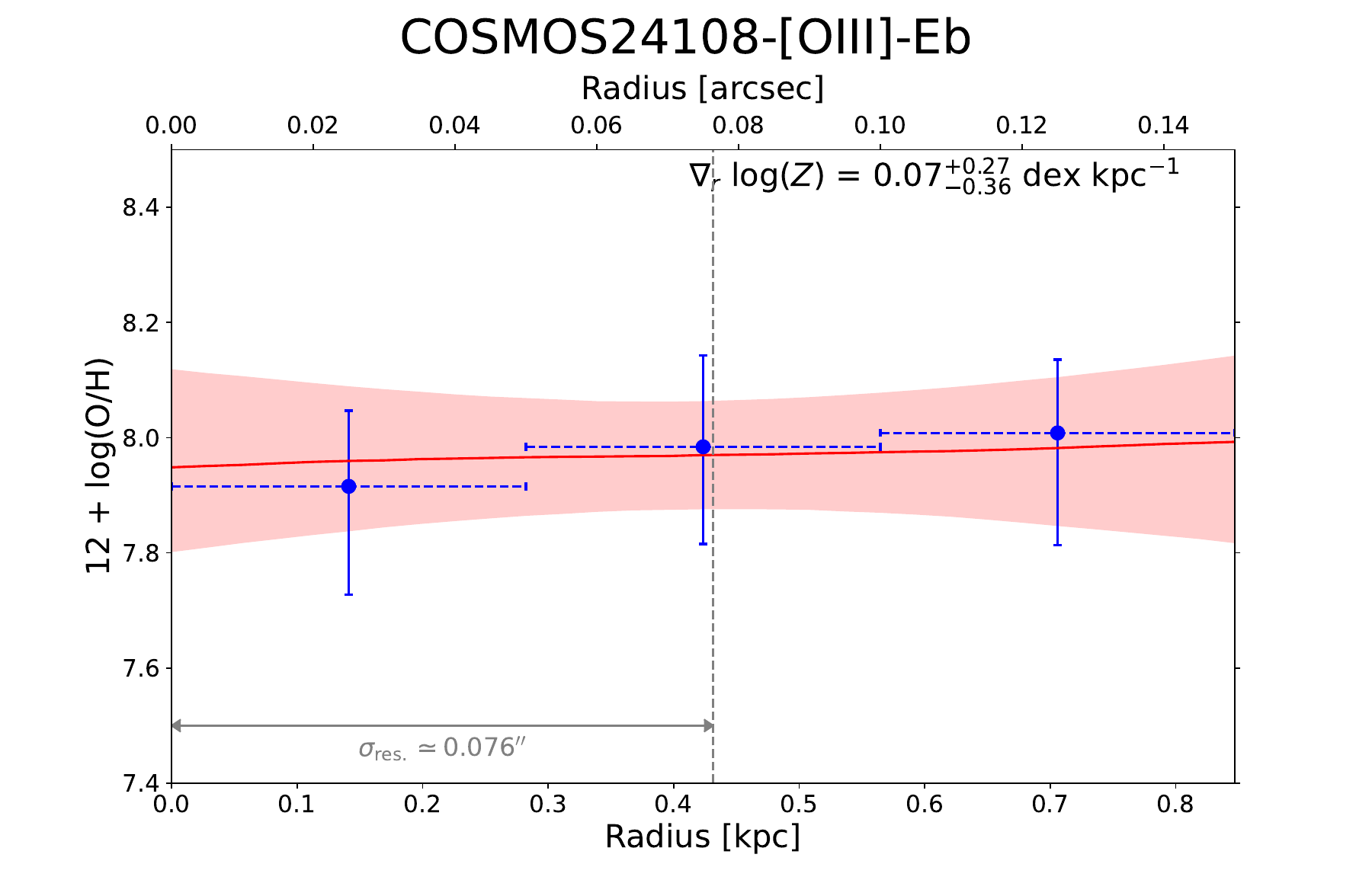}\\
    \includegraphics[scale=0.265,trim={1cm 0 1cm 0},clip]{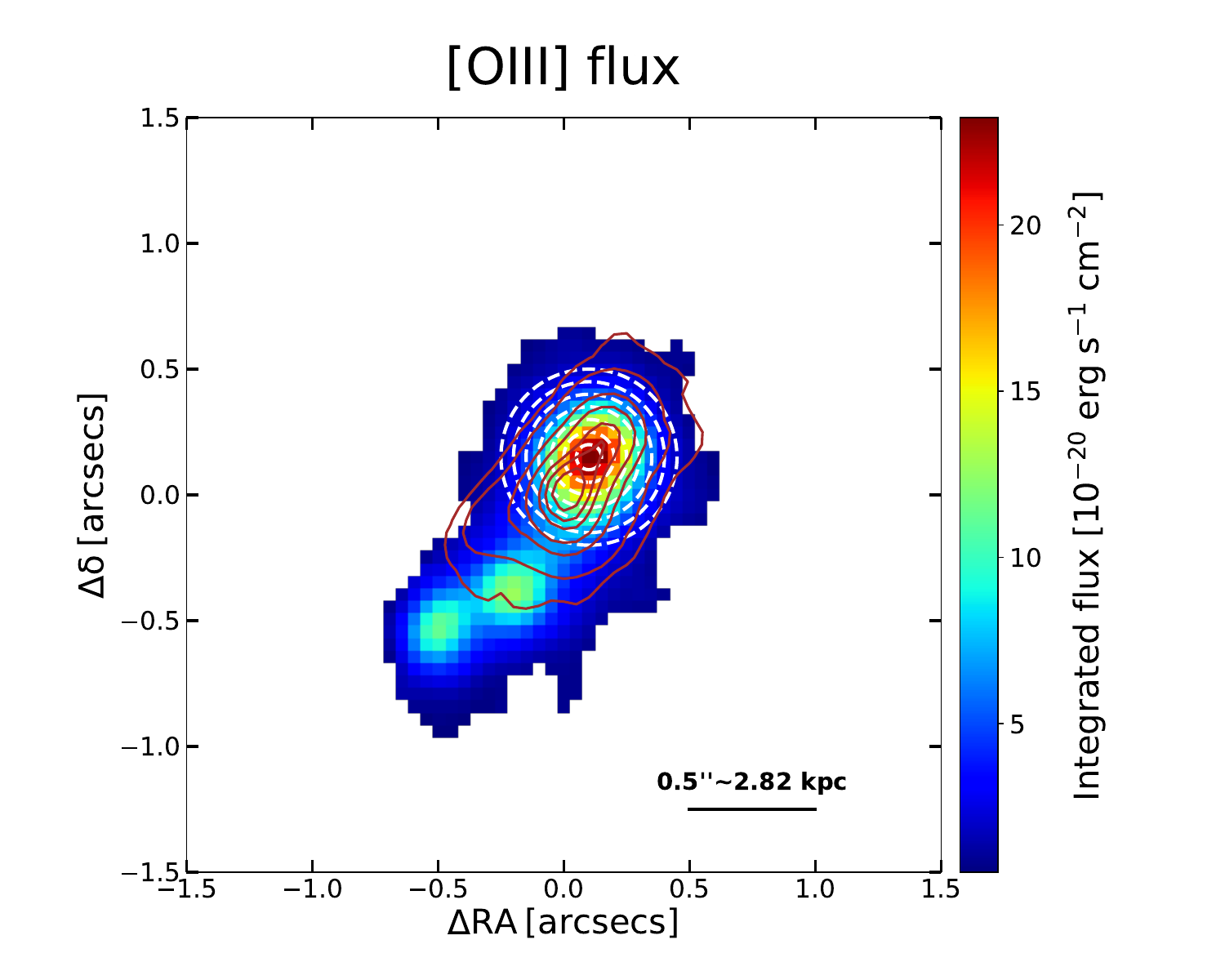}
    \includegraphics[scale=0.29,trim={1cm 0 1cm 0},clip]{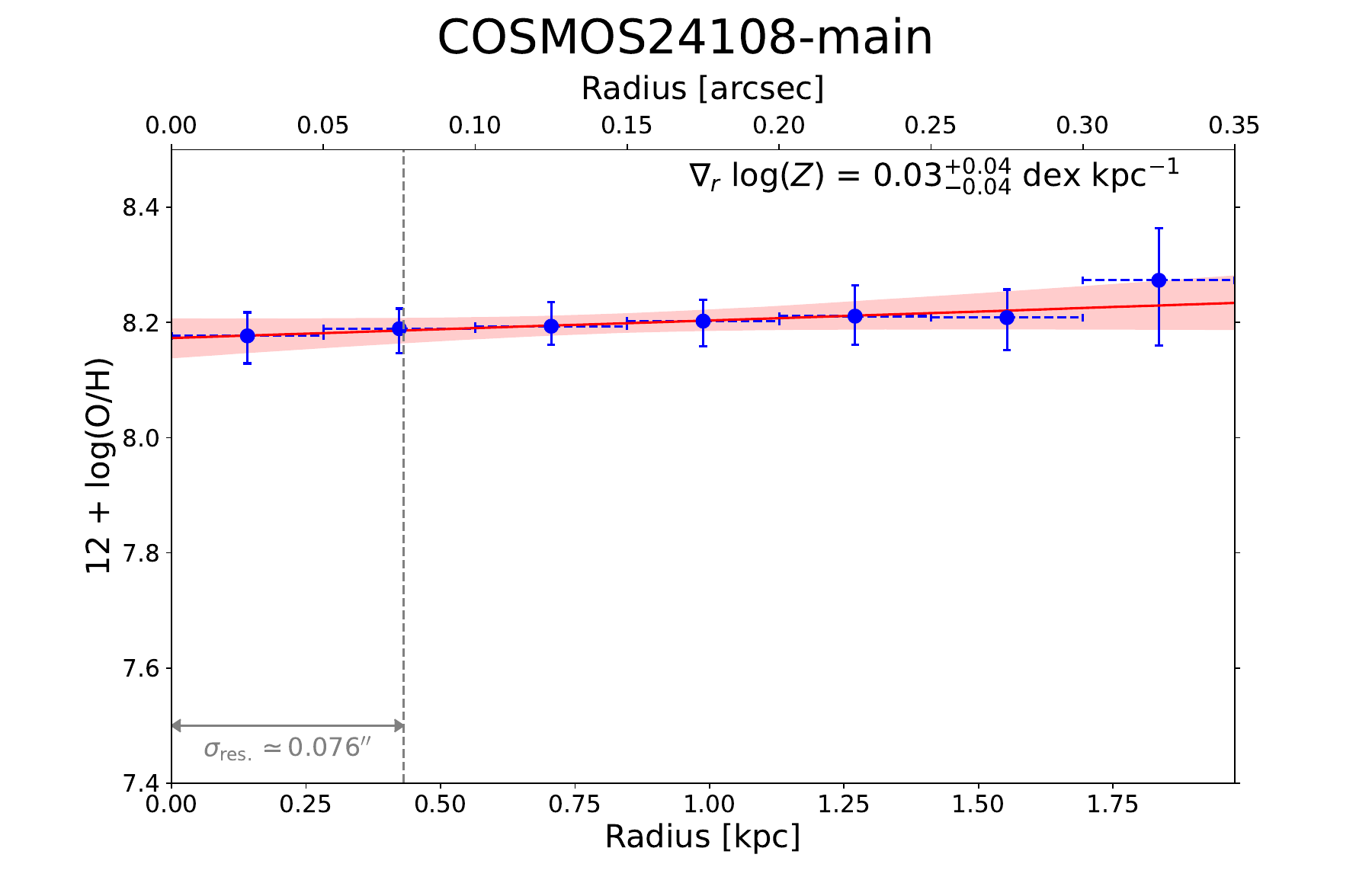}
    \caption{Metallicity gradients for the two minor sources in \cosmos, \oiii-Ea and \oiii-Eb (top and mid panels, respectively), and for the main system (bottom panel), when centring the gradient around the peak of the \oiii emission instead of around the continuum peak as in Fig.~\ref{fig:met_grad}.
    }
    \label{fig:cosmos_metgrad_clumps}
}
\end{minipage}



\end{appendix}

\end{document}